\documentclass[12pt]{article}
\usepackage{epsf}
\usepackage[dvips]{graphicx,psfrag}

\input colordvi

\renewcommand{\thefootnote}{\#\arabic{footnote}}
\renewcommand{\theequation}{\thesection.\arabic{equation}}

\newcommand{\BG}{ {\rm BG} }
\newcommand{\D}{ {\rm D} }

\newcommand{\abs}[1]{{\left|{#1}\right|}}

\newcommand{\beseven}{{\rm ^7Be}}
\newcommand{\cm}{ {\rm cm} }

\newcommand{\daikakko}[1]{{\left[{#1}\right]}}

\newcommand{\ev}{ {\rm eV} }

\newcommand{\gev}{ {\rm GeV} }
\newcommand{\hethree}{{\rm ^3He}}
\newcommand{\hefour}{{\rm ^4He}}
\newcommand{\kakko}[1]{{\left({#1}\right)}}
\newcommand{\kev}{ {\rm keV} }
\newcommand{\lisix}{{\rm ^6Li}}
\newcommand{\liseven}{{\rm ^7Li}}

\newcommand{\mb}{ {\rm mb} }
\newcommand{\mev}{ {\rm MeV} }

\newcommand{\mpi}{ {m_{\pi}} }
\newcommand{\order}{{\cal O}}

\newcommand{\pbg}{ {{p}_{\rm BG}} }

\newcommand{\staave}[1]{{\langle{#1}\rangle}}
\newcommand{\T}{ {\rm T} }
\newcommand{\tenten}{ {\cdots} }

\renewcommand{\thefootnote}{\#\arabic{footnote}}
\newcommand{\vect}[1]{{\mbox{\boldmath {$#1$}}}}

\newcommand{\Y}{Y}

\begin{document}

\newcommand{\gtrsim}{ \mathop{}_{\textstyle \sim}^{\textstyle >} }
\newcommand{\lesssim}{ \mathop{}_{\textstyle \sim}^{\textstyle <} }

\renewcommand{\thefootnote}{\fnsymbol{footnote}}
\setcounter{footnote}{0}
\begin{titlepage}

\def\thefootnote{\fnsymbol{footnote}}

\begin{center}

\hfill ICRR-Report-508-2004-6\\
\hfill OU-TAP-234\\
\hfill TU-727\\
\hfill astro-ph/0408426\\
\hfill August, 2004\\

\vskip .4in

{\Large \bf

Big-Bang Nucleosynthesis and Hadronic Decay of Long-Lived Massive
Particles

}

\vskip .4in

{\large
 Masahiro Kawasaki$^{(a)}$,
 Kazunori Kohri$^{(b)}$ and
 Takeo Moroi$^{(c)}$
}

\vskip .3in

{\em $^{(a)}$Institute for Cosmic Ray Research,
University of Tokyo\\
Kashiwa, Chiba 277-8582, Japan}

\vskip .1in

{\em $^{(b)}$Department of Earth and Space Science, 
Graduate School of Science\\
Osaka University, Toyonaka, Osaka 560-0043, Japan}

\vskip .1in

{\em $^{(c)}$ Department of Physics, Tohoku University,  Sendai
980-8578, Japan}

\end{center}

\vskip .4in

\begin{abstract}

We study the big-bang nucleosynthesis (BBN) with the long-lived exotic
particle, called $X$.  If the lifetime of $X$ is longer than $\sim
0.1\ {\rm sec}$, its decay may cause non-thermal nuclear reactions
during or after the BBN, altering the predictions of the standard BBN
scenario.  We pay particular attention to its hadronic decay modes and
calculate the primordial abundances of the light elements.  Using the
result, we derive constraints on the primordial abundance of $X$.

Compared to the previous studies, we have improved the following
points in our analysis: The JETSET 7.4 Monte Carlo event generator is
used to calculate the spectrum of hadrons produced by the decay of
$X$; The evolution of the hadronic shower is studied taking account of
the details of the energy-loss processes of the nuclei in the thermal
bath; We have used the most recent observational constraints on the
primordial abundances of the light elements; In order to estimate the
uncertainties, we have performed the Monte Carlo simulation which
includes the experimental errors of the cross sections and transfered
energies.

We will see that the non-thermal productions of D, ${\rm ^{3}He}$,
${\rm ^{4}He}$ and ${\rm ^6Li}$ provide stringent upper bounds on the
primordial abundance of late-decaying particle, in particular when the
hadronic branching ratio of $X$ is sizable.  We apply our results to
the gravitino problem, and obtain upper bound on the reheating
temperature after inflation.

\end{abstract}


\end{titlepage}

\renewcommand{\thepage}{\arabic{page}}
\setcounter{page}{1}
\renewcommand{\thefootnote}{\#\arabic{footnote}}
\setcounter{footnote}{0}

\renewcommand{\theequation}{\thesection.\arabic{equation}}

\section{Introduction}
\setcounter{equation}{0}
\label{sec:Introduction}

In modern cosmology, big-bang nucleosynthesis (BBN) is one of the most
important subjects.  In the standard scenario, neutrons freeze out
from the thermal bath when the cosmic temperature is $\sim 0.7\ {\rm
MeV}$ and then the light elements (i.e., ${\rm D}$, ${\rm ^4He}$,
${\rm ^7Li}$, and so on) are synthesized subsequently.  As we will
discuss in the next section, prediction of the standard BBN (SBBN)
scenario is in a reasonable agreement with the observations.

Predicted abundances of the light elements are, however, very sensitive
to the cosmological scenarios.  In particular, if we consider exotic
cosmological scenarios based on physics beyond the standard model,
theoretical predictions on the light-element abundances may be too
much affected to be consistent with the observations.  Thus, the BBN
provides significant constraints on the new particles which change the
cosmological evolution at the cosmic time $t\sim10^{-2}-10^{12}\ {\rm
sec}$. If we consider physics beyond the standard model, there exist
various candidates of such exotic particles.  (Hereafter, we call such
particle as $X$.)

One example of the long-lived particles is the gravitino in
supergravity theory \cite{Weinberg:zq}.  Gravitino acquires mass from
the effect of the supersymmetry breaking.  In addition, its
interactions are suppressed by inverse powers of the gravitational
scale and hence its lifetime becomes very long (if it is unstable).
In particular, for supersymmetric models with the gravitino mass
$m_{3/2}\sim {\cal O}(10^{2-3})\ {\rm GeV}$, lifetime of the gravitino
becomes much longer than $1\ {\rm sec}$ and its decay may
significantly affect the light-element abundances.  (This is called
the ``gravitino problem.'')  Thus, the BBN provides substantial
constraints on the properties of the gravitino and also on the
cosmological scenarios.  (For more details of the gravitino problem,
see \cite{Moroi:1995fs} and references therein.)

In addition, even for the case where the gravitino is the lightest
superparticle (LSP), the next-to-the lightest superparticle (NLSP) has
long lifetime since the NLSP decays into its superpartner and the
gravitino.  The BBN imposes significant constraints on the case where
the NLSP is the neutralino or the scalar-$\tau$
\cite{Moroi:1993mb,sWIMP}.  Furthermore, moduli fields in the
superstring theory are another candidates of $X$.  Some of the moduli
fields may acquire non-vanishing amplitude in the early universe.  If
so, their coherent oscillation may decay at a very late stage of the
evolution of the universe.

The exotic particles listed above are some of the famous examples and,
if one considers particle-physics models beyond the standard model,
there may exist long-lived particles which affect the BBN.  Thus, it
is important to study the BBN scenario with late-decaying particles.
Such studies have been done by various groups
\cite{BBNwX_OLD,Dominguez:1987,Reno:1987qw,DimEsmHalSta,KawMor,
Kawasaki:1994bs,Protheroe:dt,Holtmann:1998gd,
Jedamzik:1999di,Kawasaki:2000qr,Kohri:2001jx,Cyburt:2002uv,
Jedamzik04a,Kawasaki:2004yh}.

In most of the previous studies (except
\cite{Dominguez:1987,Reno:1987qw,DimEsmHalSta,Kohri:2001jx}), however,
hadronic decay modes of $X$ were ignored although, for many of the
candidates of the long-lived exotic particles, it is expected that the
hadronic branching ratio is sizable.  For example, even if $X$
dominantly decays into photon (and something else), the hadronic
branching ratio is expected to be non-vanishing since the (virtual)
photon can be converted to the quark anti-quark pair.  In this case,
the hadronic branching ratio is estimated to be at least $\sim
\alpha_{\rm em}/4\pi \sim 10^{-(2-3)}$ (with $\alpha_{\rm em}$ being
the fine structure constant), unless the hadronic decay mode is
kinematically suppressed.  Of course, if $X$ directly decays into
quarks and/or gluons, hadronic branching ratio can be close to $1$.
If the massive particles decay into quarks or gluons during/after the
BBN epoch, many mesons (mostly pions) and nucleons are produced. The
emitted hadrons lose their energy via the electromagnetic interactions
and scatter off the background nuclei.  The emitted hadrons affect the
BBN via two effects.  One is the ``inter-conversion'' effect; if $X$
decays at relatively early stage of the BBN (i.e., $t\lesssim
10^{2}~{\rm sec}$), emitted hadrons may change the neutron-to-proton
ratio.  On the contrary, at the later stage of the BBN (i.e.,
$t\gtrsim 10^{2}\ {\rm sec}$), hadrodissociation processes are caused
by energetic hadrons generated by the decay of $X$.  Due to these
effects, light-element abundances can be significantly affected.

The purpose of this paper is to investigate the BBN scenario with the
long-lived exotic particle using the currently available best
knowledges on particle physics, nuclear physics, and astrophysics,
paying particular attention to hadronic decay modes of $X$.  We
calculate the abundances of the light elements including relevant
hadronic scattering processes (as well as photodissociation
processes).  Then, in order not to spoil the success of the BBN, we
derive upper bounds on the primordial abundance of $X$.

This paper is the full-length version of our recent letter
\cite{Kawasaki:2004yh}. This paper is organized as follows.  In the
next section, we briefly review the current status of the observations
and SBBN.  Then, in Section \ref{sec:overview}, we give an overview of
the decay of massive particles and its cosmological effects.  In
Section \ref{sec:photodissociation}, we give a brief overview of the
photodissociation process. We outline the hadronic decay scenarios in
Section \ref{sec:hadonicdecay}.  In Section \ref{sec:interconversion}
we introduce the formulations and computations of inter-conversion
effects between background $p$ and $n$ by hadrons at earlier epochs in
the hadron injection scenario. In Section \ref{sec:hadrodis} we
discuss the destruction and production processes of light elements in
hadrodissociation scenario. We also consider the non-thermal
production processes of Lithium and Beryllium in hadronic decay
scenario in Section \ref{sec:ntlibe}.  In Section \ref{sec:results} we
compare the theoretical predictions with the observations in hadronic
and radiative decay scenario for general massive particles. Our main
results are shown in this section; if the reader is mostly interested
in the resultant constraints, see this section (in particular, see
Figs.\ \ref{fig:myx100gev} $-$ \ref{fig:myxphotodis}).  Then, in
Section \ref{sec:grav} we apply our results to the case of decaying
gravitinos in supergravity. Section \ref{sec:conclusion} is devoted to
the conclusions and discussion.

\section{Current Status: Observation and SBBN}
\setcounter{equation}{0}
\label{sec:obs_status}

\subsection{Current status of observations}

First we briefly summarize the current status of the observational
light element abundances.  The errors of the following observational
values are at 1$\sigma$ level unless otherwise stated.

The primordial value of the ratio $n_{\rm D}/n_{\rm H}$ is measured in
the high redshift QSO absorption systems.  (Here and hereafter,
$n_{A_i}$ denotes the number density of the nucleus $A_i$.)  Recently
a new deuterium data was obtained from observation of the absorption
system at the redshift $z=2.525659$ towards
Q1243+3047~\cite{Kirkman:2003uv} including improved modeling of the
continuum level, the Ly-$\alpha$ forest and the velocity structure of
the absorption systems.  The reported value of the deuterium abundance
by using Keck-I HIRES, was relatively low, $(n_{\rm D}/n_{\rm H})^{\rm
obs}=(2.42^{+0.35}_{-0.25})\times 10^{-5}$.  Combined with the
previous data
\cite{Tytler:1996eg,Burles:1998mk,O'Meara:2000dh,Pettini:2001yu}, it
is reported that the primordial abundance is given by~\footnote{
Note that higher deuterium abundance in relatively low redshift
absorption systems at z = 0.701 was also reported:  $n_{\rm D}/n_{\rm
H}=(2.0 \pm 0.5)\times 10^{-4}$ \cite{webb}.  Based on another
independent observation of the clouds, however, it is claimed that the
observed absorption is not due to ${\rm D}$ although there are still
some uncertainties.  Thus, we do not adopt the ``High D'' primordial
abundance in this paper.
}
\begin{equation}
      \label{lowd}
      (n_{\rm D}/n_{\rm H})^{\rm obs} 
      = (2.78^{+0.44}_{-0.38}) \times 10^{-5},
\end{equation}
which we adopt in this paper.  (Here and hereafter, the superscript
``obs'' is used for the primordial values inferred by the
observations.)  We should make a comment on taking mean of the data.
Although we think that this is reasonable as treatment of experimental
data, it was pointed out that the five measurements have a large
dispersion than expected~\cite{Kirkman:2003uv}.  Since the purpose of
our paper is to derive a conservative constraint, we have to care
about possible systematic errors.  Since non-thermal production of D
leads to a severe constraint on the abundance of massive particles,
the upper bound of D/H is important for us.  The highest value among
the five measurements
is~\cite{Tytler:1996jf,Burles:1998mk,Kirkman:2003uv}
\begin{equation}
  \label{highd}
   (n_{\rm D}/n_{\rm H})^{\rm obs} 
      = (3.98^{+0.59}_{-0.67}) \times 10^{-5}.
\end{equation}
Thus, when we derive the constraint on generic massive particles in
Section \ref{sec:results}, we will also show the result with adopting
Eq.\ (\ref{highd}).

The primordial abundance of $^4$He is inferred from the recombination
lines from the low metallicity extragalactic HII regions.  One
obtains the primordial value of the $^4$He mass fraction $Y$ by
regressing to the zero metallicity O/H $\rightarrow 0$ for the
observational data.  Based on the reanalysis of 
Fields and Olive~\cite{FieOLi},
which takes account of the effect of the HeI absorption, the
primordial mass fraction is given by 
\begin{equation}
    \label{FieOLi}
    \Y^{\rm obs}({\rm FO}) = 0.238 \pm (0.002)_{\rm stat} \pm
    (0.005)_{\rm syst}, 
\end{equation}
where the first and second errors are the statistical and systematic
ones, respectively.  On the contrary, Izotov and 
Thuan~\cite{Izotov:2003xn} reported a slightly higher value :
\begin{equation}
    \label{highY}
    \Y^{\rm obs}({\rm IT}) = 0.242 \pm (0.002)_{\rm stat} (\pm
    (0.005)_{\rm syst}),
\end{equation}
where we have added the systematic errors following Refs.\
\cite{Olive:1994fe,Olive:1996zu,Izotov:mi}.  Since there exists
sizable difference between the results of two groups, we use two
values $\Y^{\rm obs}({\rm FO})$ and $\Y^{\rm obs}({\rm
IT})$.~\footnote{
Recently Olive and Skillman~\cite{Olive:2004kq} re-analyzed the
Izotov-Thuan data of 1998~\cite{Izotov:1998} (and
2004~\cite{Izotov:2003xn}) and obtained $^{4}$He abundance with larger
uncertainties. They used only $^{4}$He lines to estimate the $^{4}$He
abundance, electron density and temperature. This "self-consistent"
approach has some merit but cannot determine the temperature and
density precisely as the authors admit. The standard approach adopted
by Izotov and Thuan (and others) is to use the OIII  lines to estimate
the temperature because it leads to much more precise determination.
At present we think that it is premature to judge which approach is
more adequate to estimate the $^{4}$He abundance. Therefore, we did
not adopt it in this paper. For reference's sake, if we adopted the
value given in Ref.~\cite{Olive:2004kq}, the constraint on the
abundance of the massive particle (or the reheating temperature after
the inflation) would become milder by a factor of 4 -- 5 compared to
the result with Izotov and Thuan's value at short lifetime ($\lesssim
10^{2}$ sec).}

As for $\liseven$, it is widely believed that the primordial abundance
of $^7$Li can be determined using Pop II old halo stars with
temperature higher than $\sim 6000\ {\rm K}$ and with low metallicity.
We use the most recent measurements by Ref.\ \cite{Bonifacio:2002yx}:
$\log_{10}\left[ \left(n_{\liseven}/n_{\rm H}\right)^{\rm obs}
\right]=-9.66\pm (0.056)_{\rm stat} \pm (0.06)_{\rm sys}$, which
corresponds to $n_{\liseven}/n_{\rm H}=(2.19^{+0.46}_{-0.38})\times
10^{-10}$.  It was claimed that there can be a significant dependence
of $n_{\rm ^7Li}$ on Fe abundance in the low metallicity region
\cite{RNB}.  In addition, assuming that this trend is due to the
cosmic ray interactions, Ref.\ \cite{RBOFN} inferred that the
primordial value is $n_{\liseven}/n_{\rm H}=(1.23^{+0.68}_{-0.32})
\times 10^{-10}$.  This differs by a factor of two from the result
given in \cite{Bonifacio:2002yx}.  This suggests that the systematic
errors in both observations may be underestimated.  We are afraid that
more $\liseven$ in the halo stars might have been supplemented (by
production in cosmic-ray interactions) or depleted (in
stars)~\cite{factor-of-two}.  Since the precise determination of the
primordial abundance from the observations is out of the scope of this
paper, we conservatively adopt the center value given by
\cite{Bonifacio:2002yx} with larger uncertainties in this paper:
\begin{equation}
    \label{eq:li7}
    \log_{10}
    \left[ \left(n_{\liseven}/n_{\rm H}\right)^{\rm obs} \right]
    =-9.66 \pm (0.056)_{\rm stat} \pm (0.3)_{\rm add},
\end{equation}
which corresponds to $n_{\liseven}/n_{\rm H}= 
(2.19^{+2.2}_{-1.1})\times 10^{-10}$.
This is also justified from the point of the view of deriving
conservative constraints; with the primordial abundance of ${\rm
^7Li}$ given by \cite{RBOFN}, discrepancy between the values of the
baryon-to-photon ratio $\eta$ determined by the SBBN and that by the
observations of the cosmic microwave background (CMB) anisotropies
becomes worse.  (See the next subsection.)

For $^6$Li, it is much more difficult to determine its primordial
abundance since $^6$Li is much rarer than $^7$Li.  Unfortunately, data
is insufficient and $^6$Li abundance cannot be reliably determined.
However, because it is generally believed that the evolution of $^6$Li
is dominated by the production through the cosmic ray spallation
(i.e., reactions of cosmic rays with the interstellar medium), we can
set an upper bound on the ratio $n_{\rm ^6Li}/n_{\rm ^7Li}$.  The
models of the nucleosynthesis through the cosmic ray spallation were
intrinsically required to simultaneously explain the whole
observational Li-Be-B abundances~\cite{li6metal,li6LSTC,li6FO}.  On
the other hand, recently it was claimed that the observational $^6$Li
abundance in halo stars is too abundant from the point of view of the
cosmic ray energy if $^9$Be is fitted by the model of the cosmic-ray
metal~\cite{ramaty}.  Therefore, there seems to be some uncertainties
in the models of the
cosmic ray spallation.~\footnote{
Recently, Suzuki and Inoue \cite{suzuki:2002} pointed out other
possibility of producing $^6$Li independently of the abundance of
$^9$Be through $\alpha$-$\alpha$ reactions induced by cosmic-ray
$\alpha$ accelerated in structure formation shocks. However, it would
be difficult to precisely predict the abundance of $^6$Li in the
current version of their model. Therefore, it is premature to
quantitatively discuss the abundance by using their model.
}
In this situation, at least it would be safe to assume that $^6$Li
abundance increases as the metallicity increases.  Today we observe
only the $^6$Li to $^7$Li ratio in low-metallicity ($[{\rm Fe/H}]\leq -
2.0$) halo stars~\cite{li6_obs},
\begin{equation}
    \label{eq:obs6}
    (n_{\lisix} / n_{\liseven})^{\rm halo} 
    = 0.05 \pm 0.02 ~~ (2\sigma).
\end{equation}
We take this value as an upper bound on the primordial value of
$n_{\lisix}/n_{\liseven}$: $(n_{\lisix} / n_{\liseven})^{\rm obs}\leq
(n_{\lisix} / n_{\liseven})^{\rm halo}$.  In our statistical analysis,
we use the ratio $n_{\lisix}/n_{\rm H}$ in deriving the constraints
since, with the long-lived exotic particle $X$, this ratio can be
calculated more reliably than $n_{\lisix}/n_{\liseven}$.  In
particular, combining Eq.\ (\ref{eq:obs6}) with Eq.\ (\ref{eq:li7}),
we use the upper bound~\footnote{
Recently, using the Subaru Telescope, $\lisix/\liseven$ was measured
in the metal poor subgiant HD 140283 with the use of a high-S/N and
high-resolution spectrum: $(n_{\rm ^6Li}/ n_{\rm H})^{\rm halo} < 5.0
\times 10^{-12}$ (2$\sigma$) \cite{aoki:2004}.  The subgiant HD 140283
is the metal-poorest among all of the objects which have been used to
derive the bound on $\lisix/\liseven$ ($[{\rm Fe/H}]=-2.5$).  It is,
however, premature and beyond the scope of this paper to judge its
reliability.  In addition, our purpose is to obtain a conservative
constraints.  Thus, we do not use this constraint.}
\begin{eqnarray}
    (n_{\rm ^6Li} / n_{\rm H})^{\rm obs} \leq
    (n_{\rm ^6Li} / n_{\rm H})^{\rm halo} =
    ( 1.10^{+ 5.00}_{-0.92} ) \times 10^{-11} ~~ (2 \sigma).
    \label{eq:obs6_upp}
\end{eqnarray}

For the constraint on $^3$He, we adopt the observational $\hethree$
abundance of the pre-solar measurements.  In this paper, we do not
rely upon any detailed models of galactic and stellar chemical
evolution because there are large uncertainties in extrapolating back
to the primordial abundance. According to such theories of the
chemical evolution, the $\hethree$ abundance can decrease or increase
after the BBN epoch. Therefore, ``$\hethree$-to-H ratio'' in itself
can not be solely used for a constraint. Instead we adopt the present
ratio of $\hethree$ to ${\rm D}$, $r_{3,2}$, as the upper bound on the
primordial value.  This is based on the following simple argument of
the chemical evolution. 
Suppose that some astrophysical process destroys
${\rm D}$ and/or $\hethree$ as
\begin{eqnarray}
    \Delta n_{\hethree} &=& - R_{3} n_{\hethree}, 
    \\
    \Delta n_{\rm D} &=& -R_{2} n_{\rm D},
\end{eqnarray}
where $\Delta n_{\hethree}$ ($\Delta n_{\rm D}$) is the change of the
$\hethree$ (D) abundance and $R_{3}$ ($R_2$) is the probability of
destruction of $\hethree$ (${\rm D}$). Then, the change of the ratio
$r_{3,2}\equiv n_{\rm ^3He}/n_{\rm D}$ is
\begin{equation}
    \Delta r_{3,2} 
    \equiv 
    \frac{n_{\hethree} + \Delta n_{\hethree}}
    {n_{\rm D} + \Delta n_{\rm D}}
    - \frac{n_{\hethree}}{n_{\rm D}}
    = \frac{R_{2}-R_{3}}{1-R_{2}} r_{3,2}.
\end{equation}
Since D is more easily destroyed than $\hethree$,~\footnote{
The binding energy of D is 2.2 MeV while the threshold energies of any
destruction processes of ${\rm ^3He}$ are larger.
}
it is quite reasonable to assume
\begin{equation}
    \label{eq:D-he3-chem}
    R_{2}\ge R_{3},
\end{equation}
which leads to $\Delta r_{3,2}\ge 0$. Thus, the ratio $r_{3,2}$ is
monotonically increasing function of the cosmic time.  Note that, in
order to derive this result, we only rely on the fact that D is more
fragile than $^{3}$He. Therefore the present ratio gives us an upper
bound on the primordial value of $r_{3,2}$.  When we adopt the
solar-system data \cite{Geiss93}, the $^3$He to D ratio is given by
\begin{equation}
    r_{3,2}^{\odot} = 0.59 \pm 0.54 ~~ (2\sigma).
    \label{He3/D}
\end{equation}
We take this to be an upper bound on the primordial $^3$He to D ratio
\begin{equation}
    \label{upper_bound32}
    r_{3,2}^{\rm obs} \le r_{3,2}^{\odot}.
\end{equation}
Naively it means the upper bound $r_{3,2}^{\rm obs} \le 1.13$
(2$\sigma$).

\subsection{Current status of SBBN}
\label{subsec:SBBN}

In this subsection, we briefly discuss the current status of the SBBN.
In the recent years, there have been great progresses in the
experiments of the low energy cross sections for 86 charged-particle
reactions by the NACRE collaboration~\cite{NACRE}. In the compilation,
22 reactions are relevant to the primordial nucleosynthesis, and the
old data were revised. In particular, 7 reactions of them are
important for the most elementary processes generating nuclei with
atomic number up to $7$.  Cyburt, Fields and Olive reanalyzed the
NACRE data~\cite{Cyburt:2001pp,Cyburt:2004cq}. They properly derived
the $1\sigma$ uncertainty and the normalization of the center value
for each reaction.  In addition, they also reanalyzed the four
remaining reactions, using the existing data~\cite{SKM,kawano,Brune}
and the theoretical prediction (for one reaction)~\cite{Hale}. Their
efforts are quite useful for the study of the Monte Carlo simulation
in BBN, and it was shown that their treatment is consistent with the
other earlier studies adopting the results of NACRE~\cite{NB,VCC}.

In our numerical study of the SBBN, we used the Kawano Code (Version
4.1) \cite{kawano} with some updates of the cross sections of the
nuclear reactions.  We use the center values and errors of the cross
sections for 11 elementary nuclear reactions given in
Refs.~\cite{Cyburt:2001pp,Cyburt:2004cq}. (For the neutron lifetime,
see Eq.\ (\ref{eq:tau_n}).)  To systematically take into account the
uncertainties, we perform the $\chi^{2}$ fitting including both the
observational and theoretical errors which are obtained in Monte Carlo
simulation.  (See the Appendix in \cite{Holtmann:1998gd}.)  In our
analysis, we assume that the theoretical predictions of $n_{\rm
D}/n_{\rm H}$, $Y$, and $\log_{10}[(n_\liseven/n_{\rm H})]$ obey the
Gaussian probability distribution functions with the widths given by
the 1$\sigma$ errors.  Concerning the observational values, they are
also assumed to obey the Gaussian probability distribution functions.
Note that we consider two cases for $Y^{\rm obs}$, i.e., Fields and
Olive (FO) given in Eq.\ (\ref{FieOLi}) and Izotov and Thuan (IT)
given in Eq.\ (\ref{highY}).

\begin{figure}
    \centering
    \centerline{{\vbox{\epsfxsize=0.75\textwidth
    \epsfbox{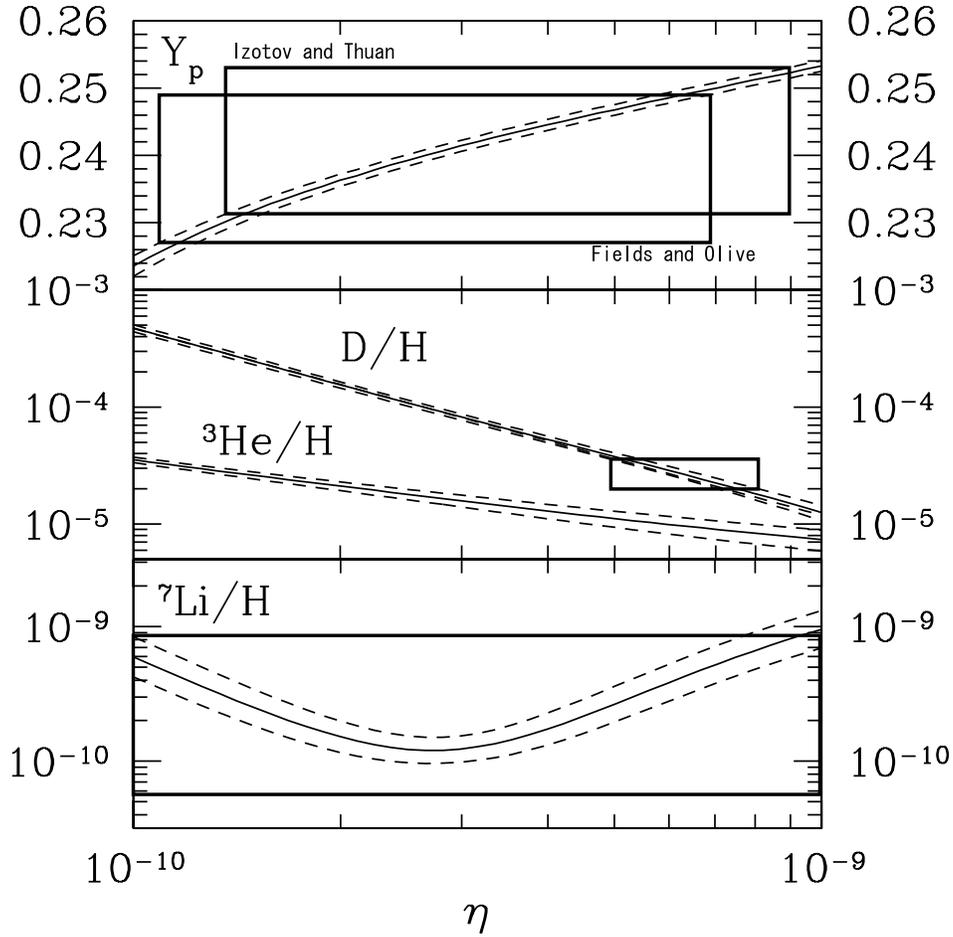}}}}
    \caption{Abundances of the light elements as functions of $\eta$.
    The solid lines are the center value while the dotted lines show
    the theoretical uncertainties.  Observational constraints are also
    shown.}
    \label{fig:yield_2sigma_2}
\end{figure}

We calculated the abundances of the light elements as functions of
baryon-to-photon ratio:
\begin{eqnarray}
    \eta\equiv \frac{n_B}{n_\gamma},
\end{eqnarray}
where $n_B$ and $n_\gamma$ are number densities of the baryon and
photon, respectively.  The results are plotted in Fig.\
\ref{fig:yield_2sigma_2}.  As one can see, theoretical predictions
become more or less consistent with the observational constraints when
$\eta\sim 6 \times 10^{-10}$.

\begin{figure}[t]
    \centering
    \centerline{{\vbox{\epsfxsize=8.0cm\epsfbox{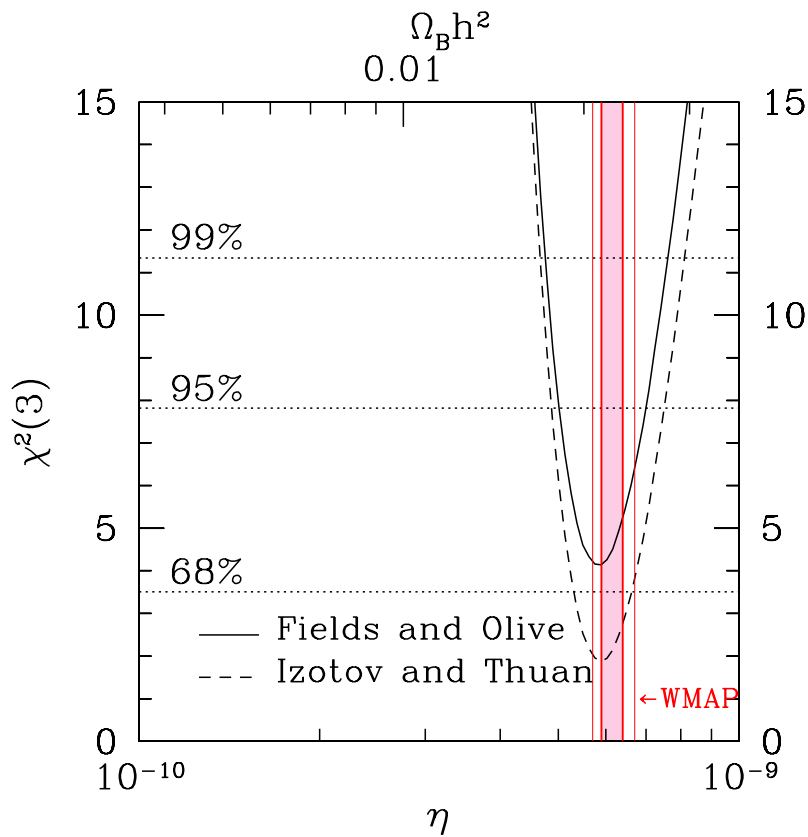}}}}
    \caption{$\chi^2$ variable as a function of $\eta$ for SBBN 
    with three degrees of freedom.  For the constraint on $Y$, we used
    Fields and Olive's result (solid) and Izotov and Thuan's (dashed),
    which are given in Eqs.\ (\ref{FieOLi}) and (\ref{highY}),
    respectively.  The shaded band (vertical solid lines) shows the
    baryon-to-photon ratio suggested by the WMAP at the 1$\sigma$
    (2$\sigma$) level.}
    \label{fig:chi_2003}
\end{figure}

For a more precise determination of $\eta$, we calculate the
$\chi^2$-variable as a function of $\eta$, and the result is shown in
Fig.~\ref{fig:chi_2003}.  The solid line (dashed line) is for the case
of Fields and Olive (Izotov and Thuan).  From this figure, we see that
the prediction of the SBBN agrees with the observation of $\hefour$,
D, and $\liseven$ at 95 $\%$ C.L.  In addition, we obtain the baryon
to photon ratio at 95 $\%$ C.L.\ as $\eta^{\rm
(SBBN)}=5.85^{+1.15}_{-0.85}\times 10^{-10}$
($5.90^{+1.63}_{-1.02}\times 10^{-10}$) using the value of $Y$ in
Fields and Olive (Izotov and Thuan).  Since the baryon to photon ratio
is related to the baryon density parameter as
$\Omega_Bh^2=3.67\times 10^7\eta$, we obtain, at 95 $\%$ C.L.,
\begin{eqnarray}
    \label{eq:omegab}
    \Omega_B h^2 = \left\{
        \begin{array}{ll}
            0.0212^{+0.0043}_{-0.0031}
            ~~ \mbox{(Fields and Olive)}
            \\
            0.0214^{+0.0059}_{-0.0037}
            ~~ \mbox{(Izotov and Thuan)}
        \end{array}
    \right. ,
\end{eqnarray}
where $h$ is Hubble parameter in units of 100\ km/sec/Mpc.

We also plot the value of $\eta$ reported by the WMAP collaborations
in observations of the cosmic microwave background (CMB)
anisotropies~\cite{Spergel:2003cb}, which is approximately given by
\begin{eqnarray}
    \label{eq:eta_WMAP}
    \eta = (6.1\pm 0.3) \times 10^{-10},
\end{eqnarray}
where we have adopted the slightly larger error for the lower
bound. The shadowed band in Fig.~\ref{fig:chi_2003} represents the
baryon to photon ratio at 1$\sigma$. The vertical solid lines are
their constraints at 2$\sigma$. From this figure, we find that SBBN
is consistent with the CMB observation.

Under these circumstances, comparing the predictions of the BBN
computations with observations, we can constrain the non-standard
scenario such as the radiative decay or the hadronic decay of
long-lived massive particles.

Here we should mention that the consistency between theoretical
predictions and observed abundances in SBBN or between CMB and SBBN is
partly because we have adopted rather large systematic errors for the
observed abundances of $^{4}$He and $^{7}$Li. In fact, if we adopted
smaller systematic errors reported in the original papers, we would be
confronted with difficulty that the $\eta$ inferred from D and CMB
disagree with that from $^{4}$He and $^{7}$Li (e.g. see,
Ref.~\cite{Ichikawa:2004pb}). But here we assume large systematic
errors and that SBBN is consistent because the purpose of the present
paper is to derive conservative constraints on the massive particles
with hadronic decay mode.

\section{Overview}
\setcounter{equation}{0}
\label{sec:overview}

Before going into the detailed discussion of the BBN with long-lived
particle $X$, we give an overview of the cosmological scenario we
consider, and define parameters which are used in our analysis.

\subsection{Production}

In this paper, we consider a scenario where a massive particle $X$,
with mass $m_X$ and decay rate $\Gamma_X$, has non-vanishing number
density at the early universe.  First, we consider the production of
$X$ in the early universe.  Throughout this paper, we consider a
situation where the $X$ particle is somehow produced in the early
universe.  Production mechanism depends on the property of $X$.  For
example, if $X$ is a particle, like gravitino, it can be produced by
scattering processes of the thermal particles.  In addition, it may be
also produced by the decay of other particles.  Moreover,
condensation of some (exotic) scalar field may play the role of $X$.
In such a case, non-vanishing initial amplitude of the scalar field
provides non-vanishing number density of $X$ at the late stage of the
evolution of the universe.

In order to perform our analysis as model-independent as possible, we
do not specify the production mechanism of $X$.  Indeed, constraints
we will obtain depends only on the relic density of $X$ (before it
decays).  In order to parameterize the number density of $X$ in the
early universe, we define the ``yield variable''
\begin{eqnarray}
    Y_X \equiv \frac{n_X}{s},
\end{eqnarray}
which is defined at the time $t\ll\Gamma_X^{-1}$.  Here, $n_X$ is the
number density of $X$ while $s$ is the total entropy density of the
universe.  Notice that, as far as we can neglect the entropy
production, $Y_X$ is a constant when $t\ll \Gamma_X^{-1}$.

\subsection{Decay}

\begin{figure}
    \centering
    \includegraphics[width=13cm]{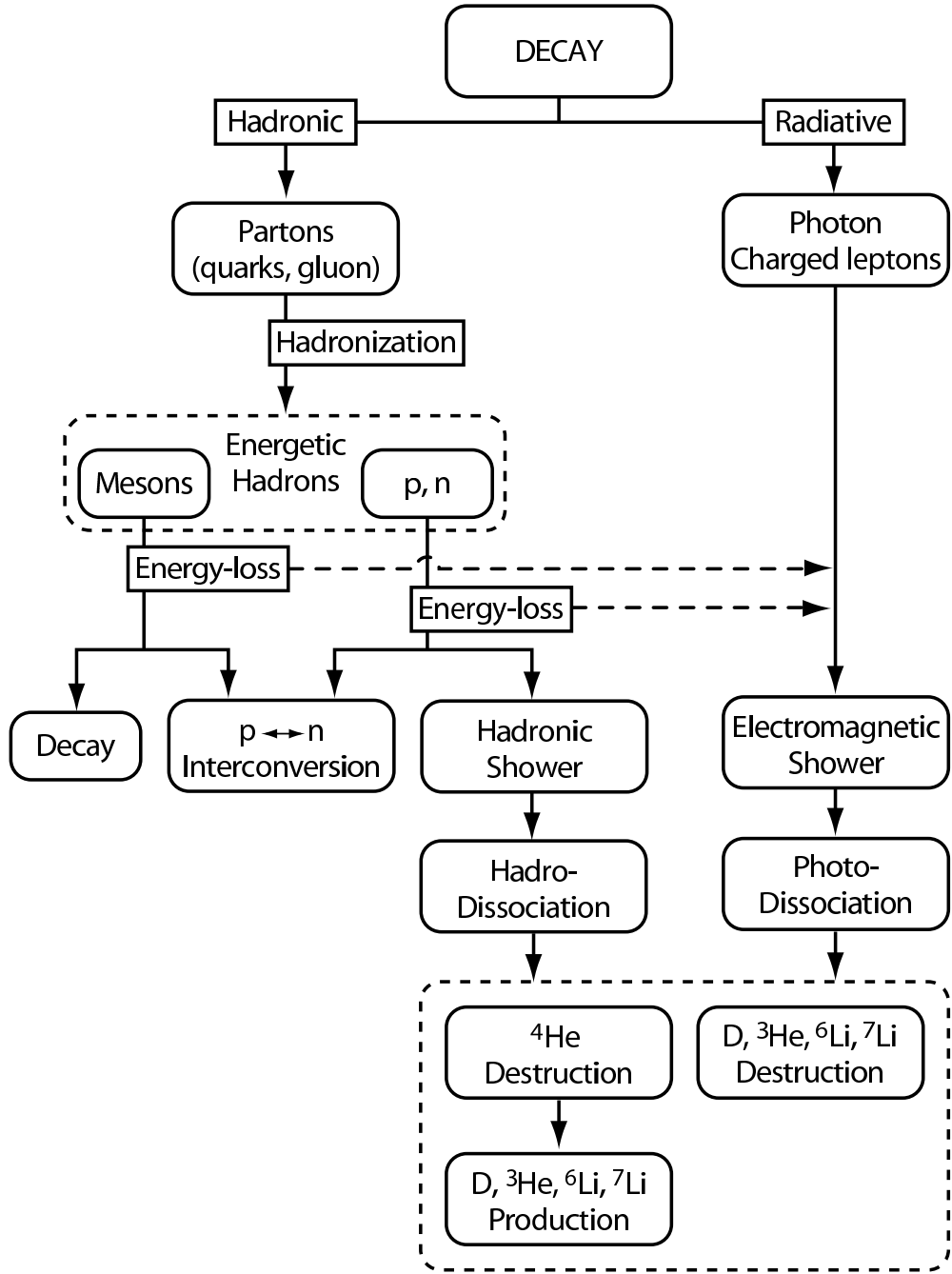}
    \caption{Flow-chart of the hadronic decay of massive particles.}
    \label{fig:overview}
\end{figure}

In this subsection we discuss the decay of massive particles and its
cosmological effects.  The overview is schematically presented in
Fig.~\ref{fig:overview}.

\subsubsection{Decay processes}

In studying the effects of $X$ on the BBN, we classify the the decay
process of $X$ into two categories; radiative and hadronic decays.
These decay processes cause different types of reactions and it is
necessary to take account of both processes.  Branching ratios
decaying into radiative and hadronic particles depend on the model.
In order to perform our analysis as model-independent as possible, we
define the ``hadronic branching ratio''
\begin{eqnarray}
  B_h = \frac{\Gamma_{X\rightarrow {\rm hadrons}}}{\Gamma_X},
\end{eqnarray}
where $\Gamma_X$ is the decay rate of $X$ and $\Gamma_{X\rightarrow
{\rm hadrons}}$ is the hadronic decay width of $X$.

If $X$ may directly decay into colored particles, $B_h$ may become
close to 1.  In addition, even when $X$ primarily decays into photon
(and other non-hadronic particles), $B_h$ is expected to be
non-vanishing since the quark-anti-quark pair can be attached at the
end of the (virtual) photon line.

\begin{figure}
    \begin{center}
        \centerline{{\vbox{\epsfxsize=0.6\textwidth
        \epsfbox{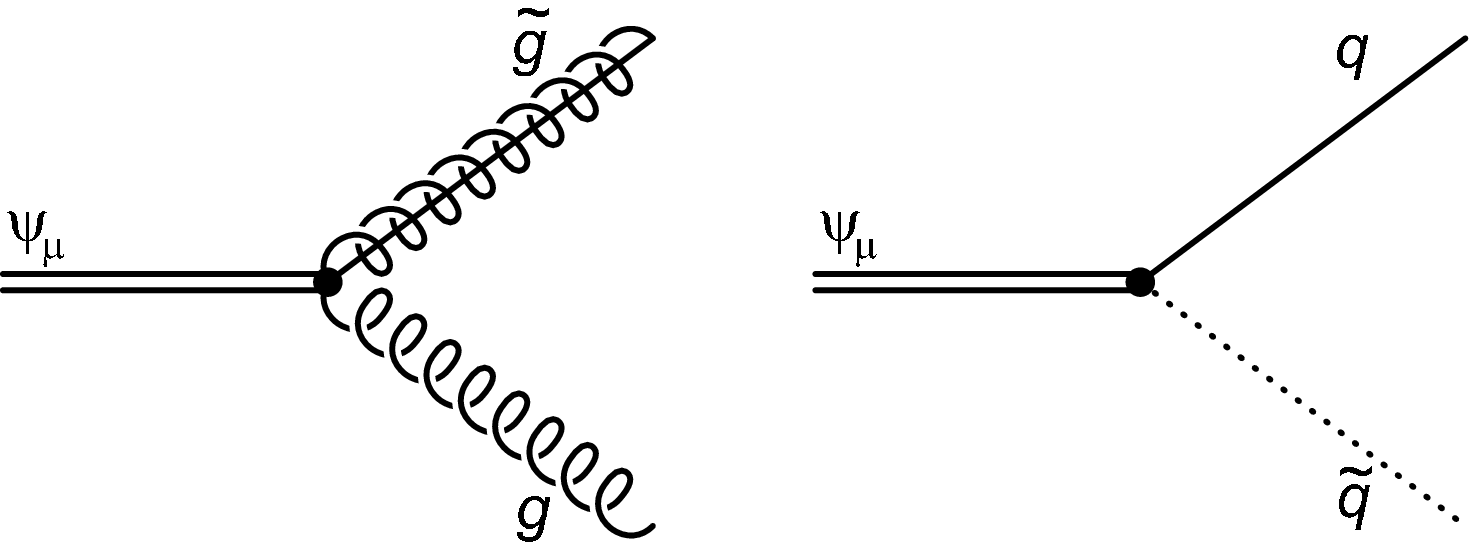}}}}
        \caption{Feynman diagrams for the decay processes
        $\psi_\mu\rightarrow g+\tilde{g}$ and $\psi_\mu\rightarrow
        q+\tilde{q}$, where $\psi_\mu$, $g$, $\tilde{g}$, $q$, and
        $\tilde{q}$ are the gravitino, gluon, gluino, quark, and
        squark, respectively.  Here, the black blob represents the
        vertex originating from the gravitino-supercurrent
        interaction.}
        \label{fig:feyngravhad}
    \end{center}
\end{figure}
%
\begin{figure}
    \begin{center}
        \centerline{{\vbox{\epsfxsize=0.6\textwidth
        \epsfbox{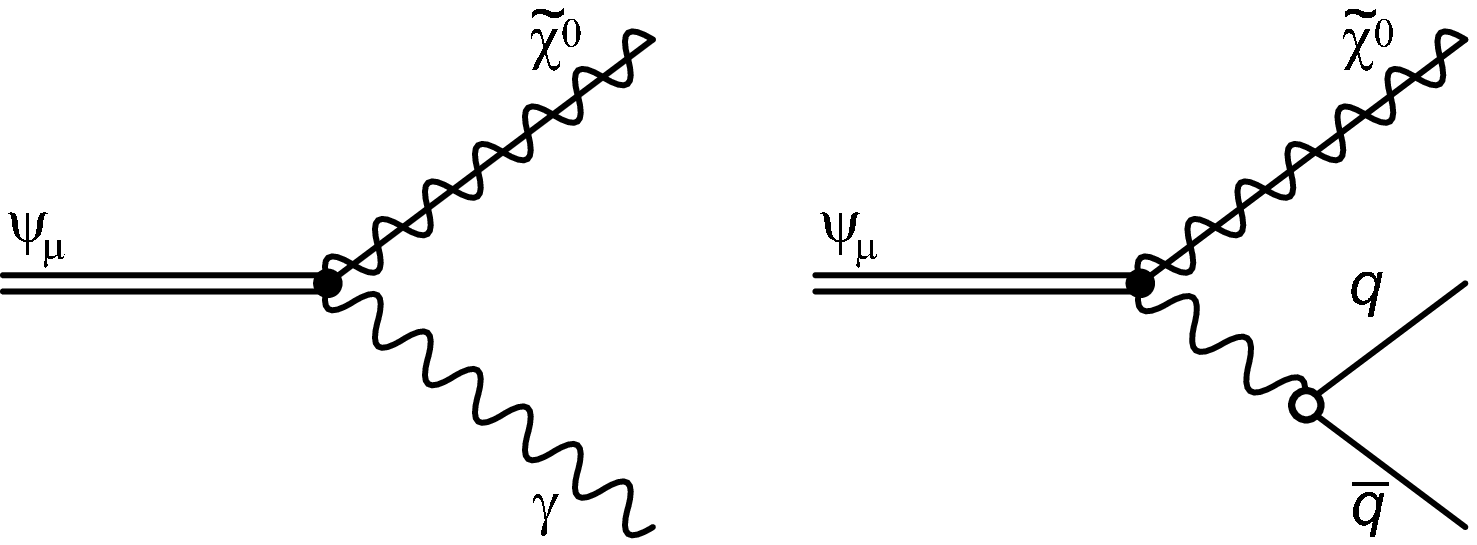}}}}
        \caption{Same as Fig.\ \ref{fig:feyngravhad}, but for the 
        radiative decay modes.  (Here, $\gamma$ is the photon while 
        $\tilde{\chi}^0$ is the neutralino.)}
        \label{fig:feyngravrad}
    \end{center}
\end{figure}

For example, for the case where the unstable gravitino $\psi_\mu$
plays the role of $X$, which is one of the most well-motivated cases,
the gravitino may directly decay into gluon-gluino and/or quark-squark
pairs.  (See Fig.\ \ref{fig:feyngravhad}.)  If the decay rate of these
modes are sizable, $B_h$ becomes close to 1.  If these hadronic decay
modes are kinematically blocked, however, the gravitino may primarily
decay into photon and the neutralino $\tilde{\chi}^0$:
$\psi_\mu\rightarrow\gamma+\tilde{\chi}^0$.  If this is the only
kinematically allowed two-body decay process of the gravitino, $B_h$
becomes much smaller than $1$.  However, even in this case, it is also
expected that $B_h$ is non-vanishing since the decay mode
$\psi_\mu\rightarrow q+\bar{q}+\tilde{\chi}^0$ (with $q$ and $\bar{q}$
being the quark and anti-quark, respectively) exists.  (See Fig.\ 
\ref{fig:feyngravrad}.)  In this case, $B_h$ is expected to be
$B_h\sim 10^{-(2-3)}$, since the process is three body process so the
branching ratio is suppressed by the factor $\alpha_{\rm em}/4\pi$.
To be more quantitative, we considered the case where the lightest
neutralino is purely the photino $\tilde{\gamma}$, the superpartner of
the photon, as an example.  We calculated the branching ratio for the
process $\psi_\mu\rightarrow q+\bar{q}+\tilde{\gamma}$ (where $q$ here
corresponds to $u$, $d$, $s$, $c$, and $b$).  In our calculation, the
Feynman amplitude is calculated by using the HELAS package
\cite{HELAS}, then the phase-space integration for the final-state
particles is done with BASES package \cite{BASES}.  Results for
several cases are shown in Table \ref{table:Bh(grav)}.  As one can
see, $B_h$ in this case is indeed $10^{-(2-3)}$.  (We have checked
that $B_h$ is insensitive to the gravitino mass as far as the mass
difference $m_{3/2}-m_{\tilde{\gamma}}$ is fixed.)

\begin{table}[t]
    \begin{center}
        \begin{tabular}{lll}
            \hline\hline
            $m_{3/2}$ & $m_{3/2}-m_{\tilde{\gamma}}$ & $B_h$ \\
            \hline
            100 GeV
            & 20 GeV
            & 0.004  \\
            100 GeV
            & 40 GeV
            & 0.008 \\
            100 GeV
            & 60 GeV
            & 0.010 \\
            300 GeV
            & 20 GeV
            & 0.003 \\
            300 GeV
            & 40 GeV
            & 0.008 \\
            300 GeV
            & 60 GeV
            & 0.010 \\
            \hline\hline
        \end{tabular}
        \caption{Hadronic branching ratio of the gravitino for several 
        values of the gravitino mass $m_{3/2}$ and the mass difference
        between the gravitino and the photino
        $m_{3/2}-m_{\tilde{\gamma}}$.  Here, the lightest neutralino
        is assumed to be the pure photino and all the superparticles
        other than the photino are assumed to be heavier than the
        gravitino.  The minimal possible value of the invariant mass
        of the quark anti-quark system is set to be 10 GeV.}
        \label{table:Bh(grav)}
    \end{center}
\end{table}

We also introduce several parameters in order to characterize the
decay of $X$.  First, we assume that {\sl each} primary parton jet has
the energy $E_{\rm jet}$.  (For example, when $X$ decays into
$q$-$\bar{q}$ pair, $E_{\rm jet}=\frac{1}{2}m_X$.)  This parameter is
used when we study the hadronization processes with JETSET event
generator.  On the contrary, for the decay process of $X$ with
energetic photon in the final state, we define $E_{\gamma}^{(0)}$
which is the energy of the emitted photon.  In addition, in some case,
invisible particle may be also emitted.  For example, when the
gravitino plays the role of $X$, some fraction of the energy is
carried away by the LSP which we assume is the neutralino.  Thus, we
define $E_{\rm vis}$, which is the (averaged) energy emitted in the
form of the ``visible'' particles.  As we will discuss, number of
high-energy photons produced in the electromagnetic cascade process is
proportional to $E_{\rm vis}$.

\subsubsection{Radiative decay}

In the radiative decay the massive particles decay into photons
(and/or electrons, and so on).  The energy of the emitted particles
can be as large as the mass of the parent particles; with the high
energy primary particles, electromagnetic showers are induced and
energetic photons are recursively produced in the shower (see the
right branch in Fig.~\ref{fig:overview}).  Some of soft photons
produced secondarily in the shower induce destruction and production
processes of the light elements (D, $^{3}$He, $^{4}$He, $^{6}$Li and
$^{7}$Li).  For the details, see, for example, Ref.\ 
\cite{KawMor,Holtmann:1998gd,Kawasaki:2000qr}.

In the case of radiative decay, the most important process that
determines the destruction rate of the light elements is
``photon-photon process'' where the high energy photons scatter off
the background photons into electrons and positrons. Since the number
of the background photons is about $10^{10}$ times larger than that of
electrons, the photon-photon process, if it occurs, quickly
thermalizes the high energy photons and the destruction of light
elements does not take place frequently. However, since there exists
the threshold of the photon energy ($E_{\rm th}\sim m_{e}/22T$), soft
photons with energy less than the threshold can destroy the light
elements.  Comparing the binding energies of the light elements with
$E_{\rm th}$, we can see that the photodissociation processes of D and
$^{4}$He may become effective when the temperature becomes lower than
$T\lesssim 0.01$~MeV and $0.001$~MeV, respectively.  Therefore,
roughly speaking, the constraint on the radiatively decaying particle
comes from the destruction of D for lifetime less than $10^{6}$~sec
and from the overproduction of D and $^{3}$He due to the destruction
of $^{4}$He for lifetime longer than $10^{6}$~sec.

\subsubsection{Hadronic decay}

When quarks or gluons are emitted in the decay of the massive
particles, they firstly fragment into a lot of hadrons and form
hadronic jets. As a result, many high energy mesons and nucleons are
injected into the cosmic plasma.

At earlier epochs ($t \lesssim 100$~sec) the high energy mesons and
nucleons lose their energy very quickly through electromagnetic
interaction. Thus, they are completely stopped and reach to the
kinetic equilibrium. Thus, the emitted hadrons do not directly destroy
the light elements.  After energy loss, they scatter off the
background $p$ or $n$ through the strong interaction with their
threshold cross sections. Then, they inter-convert the background $p$
and $n$ each other, even after the normal freeze-out time of the $n/p$
ratio of the weak interaction. Since the typical mean free time for
strong interaction is ${\cal O} (10^{-8})$~sec, only mesons with
relatively long lifetimes such as $\pi^{\pm}$ and $K^{0,\pm}$ and
nucleons ($p$, $n$, $\bar{p}$, $\bar{n}$) can cause the $p$-$n$
inter-conversion.  Since, at $T\lesssim 1\ {\rm MeV}$, the proton is
more abundant than the neutron, the conversion from $p$ to $n$ takes
place more frequently than its inverse process, and hence the hadron
injection extraordinarily tends to increase the ratio $n/p$.  As a
result, the produced $\hefour$ and D would increase in the hadron
injection scenario compared to the SBBN case.

At later epoch ($t \gtrsim 100$~sec), mesons decay before they
interact the background nucleons, and hence they become cosmologically
irrelevant. On the other hand the emitted high energy protons and
neutrons can scatter off the background $p$, $n$ and $^{4}$He (which
is synthesized in BBN). Since the energy loss due to the
electromagnetic interaction is insufficient, the high energy $p$ and
$n$ interact with background hadrons before they lose energy, and
produce secondary hadrons through elastic and inelastic
collisions. Such hadronic interactions occur successively and evolve
hadronic showers. During evolution of the hadronic shower, a lot of
$^{4}$He's are destroyed by the inelastic collisions, and D, T and
$^{3}$He are produced from the $^{4}$He dissociation. Then the
energetic T and $^{3}$He scatter off the background $^{4}$He, and
produce $^{6}$Li and $^{7}$Li. Since $^{4}$He is much more abundant
than the other light elements, those non-thermal production of D,
$^{3}$He, $^{6}$Li and $^{7}$Li drastically changes the prediction of
SBBN. Thus, the severe constraint is imposed on the hadronic
decay. Contrary to the radiative decay, non-thermal production due to
$^{4}$He dissociation is even important at $t\lesssim 10^{6}\ {\rm
sec}$ when the high energy photons quickly lose their energy by the
photon-photon process and cannot destroy $^{4}$He.

Here, we remark that the almost all of the energy of the primary
hadrons are transferred to the electrons (positrons) and photons
through the electromagnetic energy loss processes, and decays of
mesons and heavy charged leptons. Then the energetic photons and
electrons cause electromagnetic showers. In this sense, the hadronic
decay also has the same effect as the radiative decay. This is
indicated by dashed arrows in Fig.~\ref{fig:overview}.

In order to take account of the non-standard processes, we have
modified the Kawano Code (Version 4.1, with the updates of the nuclear
cross sections).  In particular, we have developed new subroutines
which deal with the photodissociation and hadrodissociation processes
and implemented them into the Kawano Code.  Details will be explained
in the following sections.

\subsection{Comparison with previous works}

Before going into the details of our analysis, it would be relevant to
compare our analysis with previous ones.  As we mentioned, once the
hadronic decay occurs, the BBN processes are affected by the
inter-conversion and hadrodissociation processes.  

The effects of the inter-conversion was first studied in
\cite{Reno:1987qw}, whose results were also applied to some of the
topics in the early universe in \cite{Kawasaki:2000en,Kohri:2000ex}.
In these works, however, it is simply assumed that all of emitted
hadrons are effectively stopped even when the cosmic time is longer
than $t \sim 10^{2}\ {\rm sec}$.  Because the energy loss process
becomes inefficient at $t\gtrsim 10^{2}\ {\rm sec}$ for proton and
$t\gtrsim 10^{3}\ {\rm sec}$ for neutron, as we will show in detail in
Section \ref{sec:hadonicdecay}, their assumption becomes inappropriate
at low temperature.  In our analysis, we carefully reconsider
study the stopping of the hadrons through the electromagnetic
interactions with background plasma.  (Such stopping processes are
also important for the study of the hadrodissociations.)  On the
other hand, effects of the hadrodissociation processes were studied in
\cite{DimEsmHalSta}.  This study is, however, based on theoretical and
experimental information which can be improved with our current
knowledge.  Thus, in our study of the evolution of the hadronic
showers, the basic framework is the same as that used in
\cite{DimEsmHalSta}, but there are several important modifications
(see the following sections).

After these pioneering works, there have been various theoretical,
experimental and observational progresses to study the BBN scenario
with long-lived exotic particles.  First, with the progresses in the
high-energy experiments, now we have better information on the hadron
fragmentation processes.  In particular, we use the JETSET 7.4 Monte
Carlo event generator~\cite{Sjostrand:1994yb} to estimate the
distributions of the nucleons and mesons produced by the hadronic
decay of $X$.  In addition, we have more experimental data of the
hadron-nucleon cross sections and energy distributions of the hadronic
particles generated by the hadrodissociation processes.  With these
improvements, we can perform a better study of the evolution of the
hadronic shower.  Furthermore, observational constraints on the
primordial abundances of the light elements are also improved.

In summary, the most important improvements which are made in this
paper are as follows.  (i) We carefully take into account the energy
loss processes for high-energy nuclei through the scattering with
background photons or electrons. In particular, dependence on the
cosmic temperature, the initial energies of nuclei, and the background
$\hefour$ abundance are considered. (ii) We use available data of
cross sections and transfered energies of elastic and inelastic
hadron-hadron scattering processes as much as possible. (iii) The time
evolution of the energy distribution functions of high-energy nuclei
are computed with proper energy resolution. (iv) The JETSET 7.4 Monte
Carlo event generator \cite{Sjostrand:1994yb} is used to obtain the
initial spectrum of hadrons produced by the decay of $X$. (v) The most
resent data of observational light element abundances are
adopted. (vi) We estimate uncertainties with Monte Carlo simulation
which includes the experimental errors of the cross sections and
transfered energies, and uncertainty of the baryon to photon ratio
\cite{Spergel:2003cb}.

\section{Photodissociation}
\label{sec:photodissociation}
\setcounter{equation}{0}

We are at the position to discuss various processes induced by the
hadronic (as well as the radiative) decay of the late-decaying
particle $X$.  In calculating of the abundances of the light elements,
we take account of two types of dissociation processes of the light
elements; one is the photodissociations induced by the energetic
photons and the other is the hadrodissociations by the hadrons.
Importantly, even if we consider hadronic decay modes of $X$, kinetic
energy of the hadrons are eventually converted to radiation via the
scattering processes.  Thus, even in the case of the hadronic decay
mode, it is important to consider the photodissociation processes.  In
this section, we first discuss the simple photodissociation reactions
of the light elements.

Once the late-decaying particle $X$ decays in the thermal bath, most
of the (visible) energy released by the decay of $X$ is eventually
converted to the form of the photon for the situation we are
interested in.  Then, the electromagnetic cascade processes are
induced.  In order to study the photodissociation processes of the
light elements, it is necessary to understand the spectrum of the high
energy photon generated by the cascade process.  In our study, we have
calculated the photon spectrum taking account of effects of the
following processes:
\begin{itemize}
\item Injection of the high energy photon from the radiative decay of
    $X$
\item Double photon pair creation ($\gamma+\gamma_{\rm BG}\rightarrow
    e^{+}+e^{-}$)
\item Photon-photon scattering
    ($\gamma+\gamma_{\rm BG}\rightarrow\gamma+\gamma$)
\item Compton scattering off thermal electron
    ($\gamma+e^{-}_{\rm BG}\rightarrow\gamma +e^{-}$)
\item Inverse Compton scattering off background photon
    ($e^{\pm}+\gamma_{\rm BG}\rightarrow e^{\pm}+\gamma$)
\item Pair creation in background proton (and $\alpha_{\rm BG}$)
    ($\gamma+p_{\rm BG}\rightarrow e^{+}+e^{-}+p$)
\end{itemize}
For details of the calculation of the photon spectrum, see Appendix
\ref{app:photon}.

One important point is that the energy distribution of the photon in
the electromagnetic shower is mostly determined by the total amount of
the injected energy and is insensitive to the detail of the primary
spectrum of the injected high-energy particles.  Thus, the photon
spectrum depends on the temperature, number density of $X$, decay rate
of $X$, and the total visible energy released by the single decay of
$X$ which we call $E_{\rm vis}$; once these parameters are fixed, the
photon spectrum $f_\gamma$ is determined.

In addition, if the temperature is high enough, pairs of charged
particles like $\mu^+\mu^-$, $\pi^+\pi^-$, and so on, may be produced
by the photon-photon scattering.  Such pair-production processes,
however, do not significantly change the photodissociation rates since
photon spectrum for the photon energy relevant for the
photodissociation processes is determined by the Compton scattering
and the $e^+e^-$ pair creation in the nuclei.  Thus, for the study of
the photodissociation, we can neglect the pair production of charged
particles heavier than the electron. If the hadronic particles are
pair-produced, however, it may provide new sources of the hadronic
particles.  Such effects will be considered in the next section.

Effects of the photodissociation is taken into account by including
the following terms into the Boltzmann equations describing the
evolutions of the light elements:
\begin{eqnarray}
    \left[ \frac{d n_{A_i}}{d t} \right]_{\rm photodiss} &=&
    - n_{A_i} \sum_j \int_{E_\gamma^{\rm (th)}} d E_\gamma
    \sigma_{A_i \rightarrow A_j} (E_\gamma)
    f_{\gamma} (E_\gamma)
    \nonumber \\ &&
    + \sum_j  n_{A_j} \int_{E_\gamma^{\rm (th)}} d E_\gamma
    \sigma_{A_j \rightarrow A_i} (E_\gamma)
    f_{\gamma} (E_\gamma),
\end{eqnarray}
where $\sigma_{A_i \rightarrow A_j}$ is the cross section for the
process $A_i+\gamma\rightarrow A_j+\cdots$ with the threshold energy
$E_\gamma^{\rm (th)}$, and the summations are over all the possible
reactions.  (In this paper, $A_i$ is used for light elements, i.e.,
$n$, $p$, ${\rm D}$, ${\rm T}$, ${\rm ^3He}$, ${\rm ^4He}$, ${\rm
^6Li}$, ${\rm ^7Li}$, and ${\rm ^7Be}$.)  The above equation for
$i$-th nucleus can be also written as
\begin{eqnarray}
    \left[ \frac{d n_{A_i}}{d t} \right]_{\rm photodiss}
    = - \sum_j \Gamma_{A_i\rightarrow A_j}^{\rm (photodiss)} n_{A_i} 
    + 
    \sum_j \Gamma_{A_j\rightarrow A_i}^{\rm (photodiss)} n_{A_j},
\end{eqnarray}
where 
\begin{eqnarray}
    \Gamma_{A_j\rightarrow A_i}^{\rm (photodiss)} \equiv
    \int_{E_{\gamma}^{\rm (th)}} d E_\gamma
    \sigma_{A_i \rightarrow A_j} (E_\gamma)
    f_{\gamma} (E_\gamma).
\end{eqnarray}
Thus, the BBN reactions with the radiatively decaying particle can be
followed once the quantities $\Gamma_{A_j\rightarrow A_i}^{\rm
(photodiss)}$ are known.

\begin{table}[t]
    \begin{center}
        \begin{tabular}{lll} 
            \hline\hline
            Reaction & Error& Reference \\ 
            \hline
            $\gamma + {\rm D}\rightarrow n + p$
            & 6\ \%
            & \cite{Evans} \\
            $\gamma + {\rm T} \rightarrow n + {\rm D}$
            & 14\%
            & \cite{ZP208-129,PRL44-129}\\
            $\gamma + {\rm T} \rightarrow  p + n + n$
            &  7\% 
            & \cite{PRL44-129}\\
            $\gamma + {\rm ^3He} \rightarrow p + {\rm D}$
            & 10\% 
            & \cite{PL11-137}\\
            $\gamma + {\rm ^3He} \rightarrow p + p + n$
            & 15\%
            & \cite{PL11-137}\\
            $\gamma + {\rm ^4He} \rightarrow p + {\rm T}$
            &  4\%
            &\cite{SJNP19-589} \\
            $\gamma + {\rm ^4He} \rightarrow n + {^3{\rm He}}$
            &  5\%
            & \cite{CJP53-802,PLB47-433} \\
            $\gamma + {\rm ^4He} \rightarrow p + n + {\rm D}$
            & 14\%
            & \cite{SJNP19-589} \\ 
            $\gamma + {\rm ^6Li} \rightarrow {\rm anything}$
            &  4\%
            & \cite{SJNP5-349} \\ 
            $\gamma + {\rm ^7Li} \rightarrow n + {^6{\rm Li}}$
            &  4\%
            & \cite{Berman} \\
            $\gamma + {\rm ^7Li} \rightarrow {\rm anything}$
            &  9\%
            & \cite{SJNP5-344} \\ 
            $\gamma + {\rm ^7Be} \rightarrow p + {^6{\rm Li}}$
            & \\ 
            $\gamma + {\rm ^7Be} \rightarrow {\rm anything}$
            & \\
            \hline\hline
        \end{tabular}
        \caption{Photodissociation reactions included in our
        analysis.  We also write down the errors of each cross
        sections which are used in our Monte Carlo analysis.}
        \label{table:photodis}
    \end{center}
\end{table}

In our analysis, we have calculated $\Gamma_{A_i\rightarrow A_j}^{\rm
(photodiss)}$ for all the relevant processes.  In order to derive the
constraint on the primordial abundance of $X$, we calculate the
abundances of the light elements up to $^7$Li.  All the processes
included in our analysis are listed in Table~\ref{table:photodis}.
For the photodissociation of ${\rm D}$, we use the cross section in
the analytic form.  For the other processes, the cross sections are
taken from experimental data.  For the photodissociation cross
sections of $^7{\rm Be}$, we could not find experimental data.  Thus,
we use the photodissociation rates of $^7{\rm Li}$ for corresponding
dissociation processes of $^7{\rm Be}$; $\Gamma_{^7{\rm Be}\rightarrow
A}^{\rm (photodiss)}\simeq \Gamma_{^7{\rm Li}\rightarrow A}^{\rm
(photodiss)}$.  We have checked that, in deriving the constraints on
the properties of $X$, $^7{\rm Be}$ does not play a significant role.
Thus, our results are not significantly affected by this assumption.

Since the abundance of $^4{\rm He}$ is large, the photodissociation of
$^4{\rm He}$ may significantly change the abundances of nuclei with
atomic number $A\le 4$.  Thus, in considering the photodissociation of
nuclei with $A\le 4$, we specify the final-state in order for precise
calculation of the abundances of the light elements.  
The dissociation of
${\rm Li}$, on the contrary, do not change the abundances of the
nuclei lighter than ${\rm Li}$ because of the smallness of the
abundances of $^6{\rm Li}$ and $^7{\rm Li}$.  Thus, for most of the
photodissociation processes of ${\rm Li}$, we do not specify the
final-state particles.  The only exception is the process $^7{\rm
Li}+\gamma\rightarrow n+{^6{\rm Li}}$; this process may be important
for the calculation of the $^6{\rm Li}$ abundance; In our analysis,
the effect of the $^6{\rm Li}$ production through this 
process is properly taken into account.

So far, we have discussed the scatterings of the photons with the
background nuclei.  Importantly, since the photons are energetic, the
final-state particles produced by the photodissociation processes may
acquire sizable energy and participate in other class of non-thermal
production processes of the light elements.  In particular, energetic
${\rm T}$ and ${\rm ^3He}$ may scatter off the background $\alpha$ to
produce $^6{\rm Li}$.  Since the observational upper bound on the
primordial abundance of $^6{\rm Li}$ is very stringent, non-thermal
production of $^6{\rm Li}$ may impose significant constraint on the
properties of $X$ \cite{Jedamzik:1999di,Kawasaki:2000qr}.  Such
non-thermal production of $^6{\rm Li}$ will be discussed in Section
\ref{sec:ntlibe} in detail.

\section{Hadronic Decay of Massive Particle}
\label{sec:hadonicdecay}
\setcounter{equation}{0}

Now, we consider hadronic decay of $X$ and its effects on the
light-element abundances.  For this purpose, it is important to
understand how the partons emitted by the decay of $X$ are hadronized
and how the hadrons propagate in the thermal bath.  Thus, let us
discuss these subjects in this section.

\subsection{Hadronization}
\label{subsec:hadronization}

We first consider the hadronization processes.  Since we mostly
consider the cases where the mass of $X$ is larger than the QCD scale,
$X$ primarily decays into the quarks and/or gluons for the hadronic
decay process.  For the cosmic temperature we consider, however, the
time scale for the hadronization is much shorter than that for the
cosmic expansion.  Thus, the partons emitted from the decay of $X$ are
instantly hadronized and fragment into the mesons and nuclei
($\pi^{\pm}$, $\pi^0$, $K^{\pm}$, $K^0_{L,S}$, $n$, $p$, $\Lambda^0$,
and so on).  In studying the evolution of the cascade processes, which
will be discussed in the next section, those energetic nuclei and
mesons (in particular, $p$ and $n$) become the primary particles which
cause the hadronic shower.

\begin{figure}[t]
    \centering
    \centerline{{\vbox{\epsfxsize=8.0cm\epsfbox{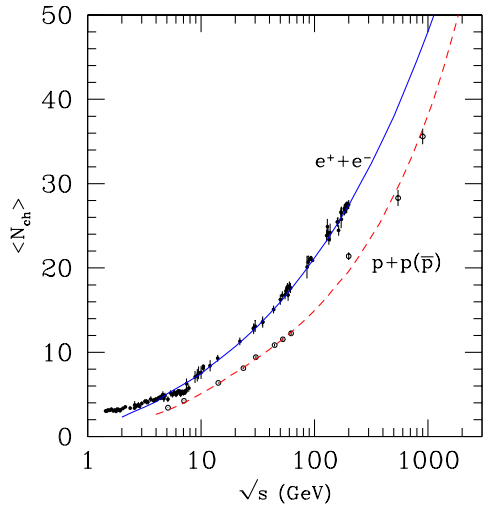}}}}
    \caption{Charged-particle multiplicity $\langle N_{\rm
    ch}\rangle$ per two hadronic jets as a function of the center of
    mass energy $\sqrt{s}$ \cite{Hagiwara:fs}.  The solid line is the
    case with $e^{+}e^{-}$ annihilation, while the dashed one is for
    the case with $p+p(\bar{p})$ collisions.  For the $p+p(\bar{p})$
    case, we use PYTHIA Monte Carlo event generator and do not include
    the single diffraction.}
    \label{fig:nch2003}
\end{figure}

In our study, the hadronization processes are followed by the JETSET
7.4 Monte Carlo event generator \cite{Sjostrand:1994yb}, which
computes the fragmentation of the hadrons from the partons.  For
  the decay process of $X$ in which colored particles are emitted, we
  used the JETSET event generator to estimate the spectrum of the
  proton, neutron, pion, and so on.  Importantly, predictions of the
JETSET package well agree with the experimental data.  Thus, in our
analysis, we expect that the uncertainties arising from the
hadronization processes are small enough to be neglected.

To demonstrate the agreements of the results from the JETSET package
with experiments, in Fig.~\ref{fig:nch2003}, we plot the averaged
charged-particle multiplicity $\langle N_{\rm ch}\rangle$ (which is
the total number of the charged hadrons) as a function of the
center-of-mass energy of the $q\bar{q}$ system
$\sqrt{s}$.~\footnote{
 Here $\langle N_{\rm ch}\rangle$ is defined
as the value after the decay of the $K_S$ and $\Lambda^0$ since their
lifetimes ($\tau_{K^0_S} = 0.89 \times 10^{-10}$ sec and
$\tau_{\Lambda^0} = 2.63 \times 10^{-10}$ sec,
respectively~\cite{Hagiwara:fs}) are much shorter than the
cosmological time scale we are interested in.  For details, see the
subsection \ref{subsec:timescale}.}
From Fig.~\ref{fig:nch2003} we find that the predicted $\langle N_{\rm
ch}\rangle$ well agrees with the experimental values.  In addition, as
we can see from the figure, $\langle N_{\rm ch}\rangle$ increases as
$\sqrt{s}$ increases.

\begin{figure}
    \centering
    \centerline{{\vbox{\epsfxsize=8.0cm\epsfbox{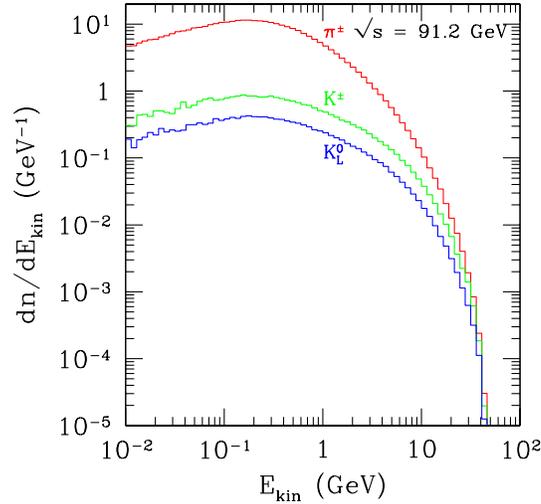}}}}
    \caption{Spectra of the mesons ($\pi^++\pi^-$, $K^++K^-$, and
    $K^0_L$) produced by the $e^{+}e^{-}$ annihilation process as
    functions of their kinetic energy $E_{\rm kin}$.  The
    center-of-mass energy is taken to be $\sqrt{s}=91.2$ GeV.}
    \label{fig:hadron_spec}
\end{figure}

\begin{figure}
    \centering
    \centerline{{\vbox{\epsfxsize=14.0cm\epsfbox{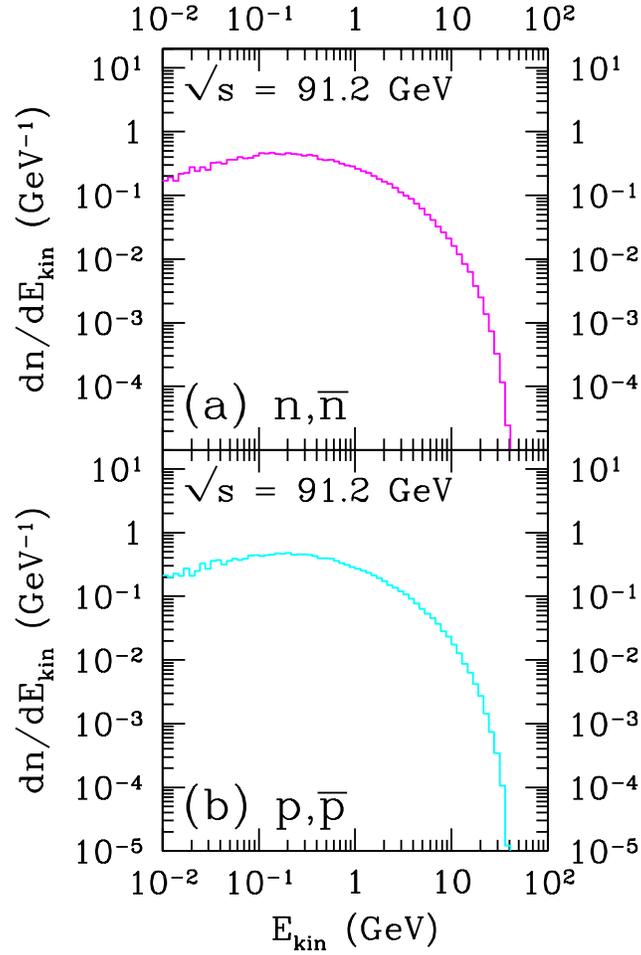}}}}
    \caption{Spectra of the baryons ((a) $n+\bar{n}$ and (b)
    $p+\bar{p}$) produced by the $e^{+}e^{-}$ annihilation process as
    functions of their kinetic energy $E_{\rm kin}$.  The
    center-of-mass energy is taken to be $\sqrt{s}=91.2$ GeV.}
    \label{fig:baryon_spec}
\end{figure}

In addition, in Figs.\ \ref{fig:hadron_spec} and
\ref{fig:baryon_spec}, we plot the spectra of the mesons
($\pi^++\pi^-$, $K^++K^-$, and $K^0_L$) and baryons ($n+\bar{n}$, and
$p+\bar{p}$), respectively, as functions of their kinetic energy in
the center-of-mass frame $E_{\rm kin}^{\rm (CM)}$.  Here we take the
total energy of the jets to be $\sqrt{s}=91.2$ GeV, for which we see
that the typical energy of the produced hadrons are ${\cal O}(10)~{\rm
GeV}$.  In studying the effects of the hadronic decay of $X$, we
calculate the spectra of the primary hadrons with the relevant total
energy of the jets which is determined by the mass of $X$ and the
decay mode.  For $m_X\sim \order(100)\ {\rm GeV} - \order(100)\ {\rm
TeV}$, we have found that the typical energy of the produced hadrons
is about $\order(1-100)\ {\rm GeV}$

\begin{figure}
    \centering
    \centerline{{\vbox{\epsfxsize=8.0cm
    \epsfbox{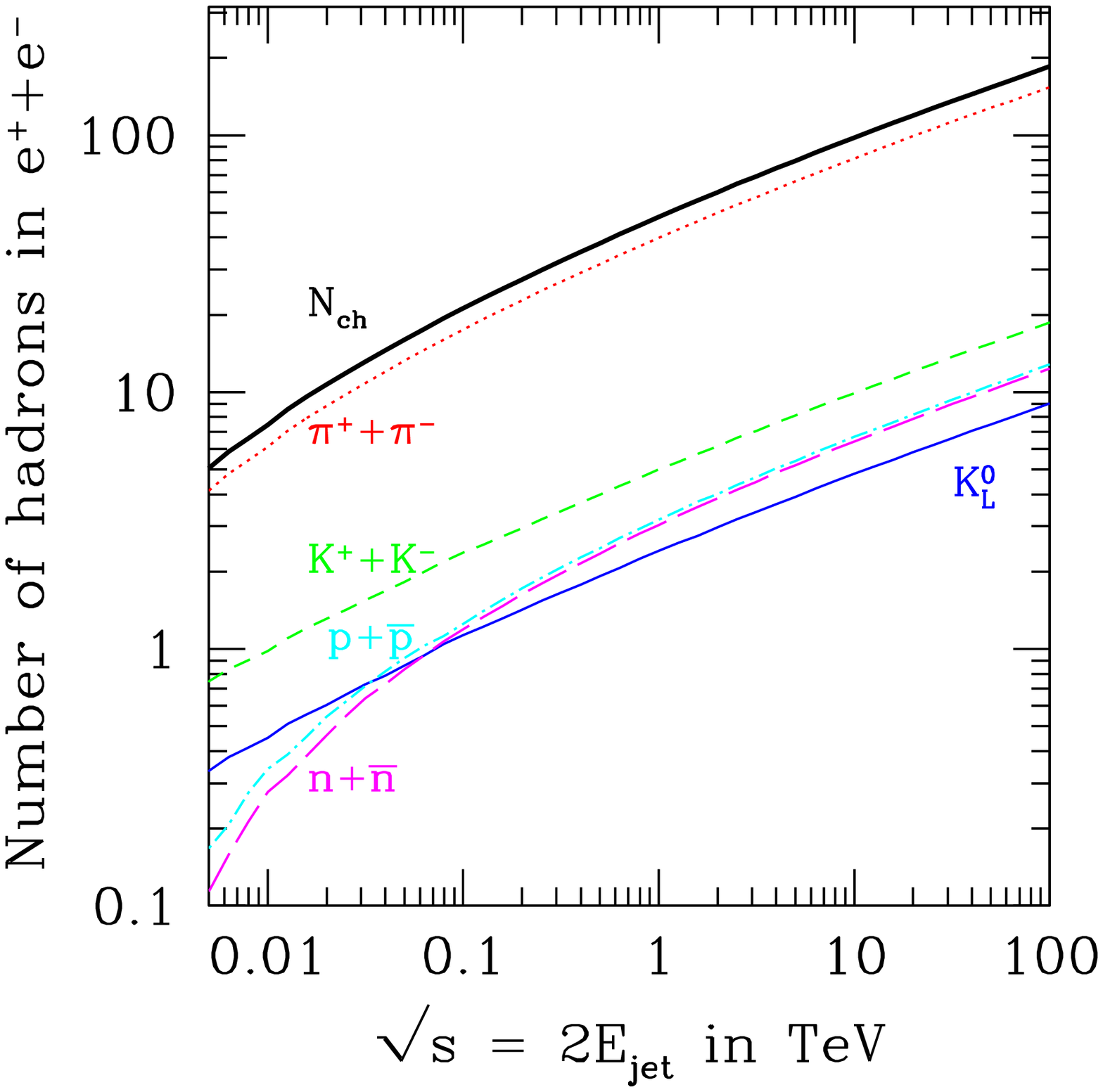}}}}
    \caption{Number of the hadrons produced by the $e^{+}e^{-}$
    annihilation process as functions of the center-of-mass energy
    $\sqrt{s}$.  The dotted, the short dashed, the thin solid, the
    dot-dashed, and the long dashed lines are $\pi^++\pi^-$,
    $K^++K^-$, $K^0_L$, $p+\bar{p}$, and $n+\bar{n}$, respectively.
    We also plot the charged-particle
    multiplicity $N_{\rm ch}$ by the thick solid line.}
    \label{fig:ee_nch2002_large}
\end{figure}

\begin{figure}
    \centering
    \centerline{{\vbox{\epsfxsize=8.0cm
    \epsfbox{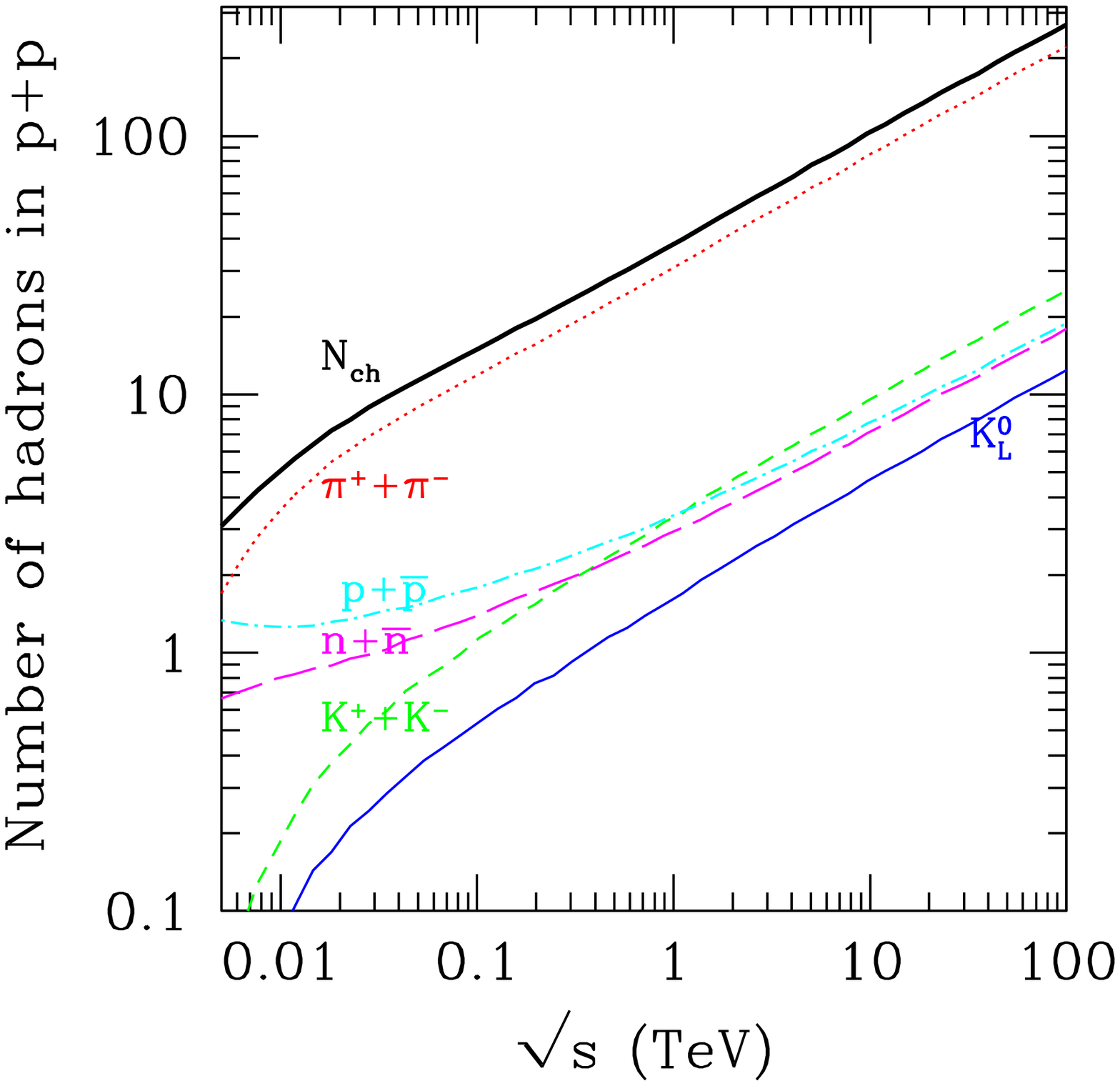}}}}
    \caption{Number of the hadrons produced by the $p+p(\bar{p})$
    scattering process.  Meanings of the lines are the same as Fig.\ 
    \ref{fig:ee_nch2002_large}.}
    \label{fig:pp_nch2003_large}
\end{figure}

In Fig.~\ref{fig:ee_nch2002_large} we also plot the averaged number of
the produced hadron per two hadronic jets as a function of the
center-of-mass energy $\sqrt{s}$.  From this figure, we can see that
the hadrons directly emitted from the decay of $X$ is mostly pions.
In addition, in Fig.~\ref{fig:pp_nch2003_large} we plot the averaged
numbers of the produced hadrons in $p+p(\bar{p})$ collision.  We see
that the number of the secondarily-produced nucleon-anti-nucleon pairs
is small for the center-of-mass relevant for our study $\sqrt{s}
\lesssim \order(10)\ {\rm GeV}$.  Therefore, we ignore them in this
paper.

In considering the hadronic processes in our analysis, we use the
hadron fragmentation obtained for the process
$e^++e^-\rightarrow\bar{q}+q$ with $\sqrt{s}=2E_{\rm jet}$ as a
primary spectrum of the hadrons generated from $X$.  Here, $E_{\rm
jet}$ is the energy of the primary jet, and will be related to $m_X$
later.  We have checked that our results are insensitive to the
Lorentz and color structure of the $q\bar{q}$ system as far as the
value of $\sqrt{s}$ is fixed.

As we mentioned in the previous section, there is another possible
production process of hadrons via the scattering of the high energy
photons emitted from $X$ with the background photons.  In particular,
the lightest charged mesons, $\pi^\pm$, can be generated from the
process $\gamma+\gamma_{\rm BG}\rightarrow\pi^++\pi^-$.  Such mesons
may contribute to the inter-conversion processes between the proton
and neutron, which will be discussed in Section\ 
\ref{sec:interconversion}.

For the high energy photon with energy $E_\gamma$, the center-of-mass
energy of such process is $\sqrt{s}\sim\sqrt{3 E_\gamma T}$.  Thus,
pair-creation rate of the pion becomes sizable only when the
temperature is high enough.  Importantly, if the pair creation of the
charged pions are effective, electron-positron (and other lepton) pair
can be also produced without kinematical suppression.  Thus, we
estimate the number of the pions produced by the radiative decay of a
single $X$ as
\begin{eqnarray}
    \xi_{\pi^\pm}^{\rm (rad)}
    = \left[
        \frac{\sigma_{\gamma+\gamma\rightarrow\pi^++\pi^-}}
        {\sum_l \sigma_{\gamma+\gamma\rightarrow l^++l^-}+
        \sigma_{\gamma+\gamma\rightarrow\pi^++\pi^-} }
    \right]_{s=3E_\gamma^{(0)}T},
\end{eqnarray}
where $\sigma_{\gamma+\gamma\rightarrow\pi^++\pi^-}$ and
$\sigma_{\gamma+\gamma\rightarrow l^++l^-}$ are cross sections of the
pion and charged lepton pair creation processes, respectively.  Those
cross sections are evaluated at $s=3E_\gamma^{(0)}T$.  For the
explicit formulae for these cross sections, see
\cite{Bijnens:1987dc,BerLifPit}.  Here, we only take account of the
charged pion production by the primary photons directly emitted from
the decay of $X$.  Since the energy of the photon is drastically
reduced after several steps of pair creations (and inverse Compton
scattering of electron and positron), the pair-creation processes are
mostly induced by the photons directly emitted from the decay process
of $X$.  In addition, if the energy of the primary photon become
large, hadrons heavier than the pions may be produced by the
photon-photon scattering processes.  However, heavier hadrons mostly
decay before inducing the inter-conversion processes.  (Exceptions are
kaons for some cases.  Since the effects of the kaons are not included
in discussing the inter-conversion effects, we do not consider the
pair creation of the charged kaons here.  For details, see Section
\ref{sec:interconversion}.)  In addition, productions of heavier
hadrons are kinematically suppressed.  Thus, we neglect the
productions of those heavier hadrons.

\subsection{Various time scales}
\label{subsec:timescale}

Once energetic hadrons are emitted into the thermal bath, which
consists of $\gamma$, $e^{\pm}$, and nucleons, hadrons scatter off
various background particles and induce cascade processes.  Thus, in
order to study the propagation of the hadronic particles in the
expanding universe, it is important to understand the time scales of
various processes; in particular, there are three important processes:
(i) hadronic scatterings, (ii) decay (for unstable particles), and
(iii) scatterings with thermal plasma through electromagnetic
interactions.  For the cosmic temperature we consider, the expansion
rate is much smaller than the rates for the above processes and hence
we can neglect the cosmic expansion in studying the effects of the
processes (i) $-$ (iii).  Time scales of the electromagnetic
processes will be discussed in the next subsection, and here we first
consider (i) and (ii).

First, we discuss the typical time scale of hadronic scattering
processes.  The interaction rate for the scattering processes between
the projectile hadron $H_i$ and the background nucleon $A_j$ through
$H_i+A_j\rightarrow A_k+\cdots$ is estimated as

\begin{eqnarray}
    \Gamma^{H_i}_{A_j\rightarrow A_k} &=& 
    n_{A_j} \sigma^{H_i}_{A_j \rightarrow A_k}
    \beta_{H_i}
    \nonumber \\ 
     &\simeq& 
    \left(4.4 \times 10^{-8} \sec \right)^{-1} f_{A_j}
    \left(\frac{\eta_{10}}{6}\right) 
    \left(\frac{\sigma^{H_i}_{A_j \rightarrow A_k} 
    \beta_{H_i} }{40 \mb}
    \right) 
    \left(\frac{T}{\mev}\right)^3,
    \label{eq:gamma^i_nn}
\end{eqnarray}
where, in this paper, $H_i$ is used for hadrons.  Here, $\beta_{H_i}$
is the velocity of $H_{i}$, $n_{A_j}$ is the number density of the
nucleon species $A_j$, $\eta_{10}\equiv\eta\times 10^{10}$, and
$f_{A_j}\equiv n_{A_j}/n_B$ with $n_B$ being the total baryon number
density.  For nucleon-nucleon collision processes, experimental data
suggest that the product of $\beta$ and $\sigma^{H_i}_{A_j\rightarrow
A_k}$ is approximately constant ($\sim 40$ mb) for large range of the
beam energies (see Figs. \ref{fig:ppsig} and \ref{fig:npsig}).  Thus,
in the following discussion, we sometimes use 40 mb as a typical value
of the cross section $\sigma^{H_i}_{A_j\rightarrow A_k}$, although we
use energy-dependent experimentally measured cross sections for our
numerical calculations.

The rate for the hadronic scattering process should be compared with
the decay rate and also with the stopping rate in the thermal plasma.
Among these two, we first consider the decay of the hadrons.
Eq.~(\ref{eq:gamma^i_nn}) shows that the typical time scale for the
hadronic scattering processes is longer than ${\cal O}(10^{-8})$ sec
for $T\lesssim 1\ {\rm MeV}$.  Thus, hadrons with lifetime longer than
$\sim 10^{-8}$ sec participate in the hadron-hadron collision
processes after the BBN starts.  Thus, hereafter, we only consider
such relatively long-lived mesons ($\pi^{\pm}$, $K^{\pm}$, and $K_L$)
and baryons ($p$, $\overline{p}$, $n$, and $\overline{n}$), whose
lifetimes are \cite{Hagiwara:fs}
\begin{eqnarray}
    \label{eq:lifetimes}
    \tau_{\pi^{\pm}} &=& (2.6033 \pm 0.0005) \times 10^{-8} \sec, \\
    \tau_{K^{\pm}} &=& (1.2384 \pm 0.0024) \times 10^{-8} \sec, \\
    \tau_{K^0_L} &=& (5.17\pm 0.04) \times 10^{-8} \sec, \\
    \label{eq:tau_n}
    \tau_n &=& 885.7 \pm 0.8 \sec.
\end{eqnarray}
Other hadrons have very short lifetimes and 
decay before scattering off
the background nuclei.~\footnote{
For example, lifetimes of $\pi^0$, $K_S^0$, and $\Lambda^0$ are
$\tau_{\pi^0} = 8.4\times 10^{-17} \sec$, $\tau_{K_S^0} = 0.89 \times
10^{-10}$, and $\tau_{\Lambda^0} = 2.63 \times 10^{-10}$ sec,
respectively \cite{Hagiwara:fs}.}

Since the lifetime of the neutron is relatively long, it is necessary
to see if the non-thermally produced neutrons may decay before causing
the hadrodissociation processes.  As will be discussed in the next
subsection, at low temperature $T\ll m_e$, energy-loss processes via
the electromagnetic interaction becomes ineffective for
non-relativistic neutrons.  Then, such neutron effectively scatters
off the background nuclei if the mean-free time is shorter than the
lifetime.  On the contrary, if the mean-free time is longer than the
lifetime, most of the energetic neutrons decay and become protons.
Since the proton can be stopped in the thermal bath more easily than
the neutrons, the hadrodissociation processes are suppressed 
in this case.

Once the energy-loss processes via the electromagnetic interactions
become ineffective, the effective lifetime of the neutron with energy
$E_n$ is given by $\gamma_{n}\tau_{n}$ (where $\gamma_{n}=E_n/m_n$ is
the Lorentz factor).  Then, the condition for the mean-free time
$1/\Gamma^{n}_{A_j\rightarrow A_k}$ being shorter than the effective
lifetime is given by
\begin{eqnarray}
    \label{eq:neutron_itself}
    T \gg 4.0 \times 10^{-1} \  \kev \  
    \kakko{\frac{E_{n}}{\gev}}^{-1/3}
    \kakko{\frac{\sigma^{n}_{p\rightarrow N' }\beta_{n}}
    {40 \mb} }^{-1/3}
    \kakko{\frac{\eta_{10}}{6}}^{-1/3},
\end{eqnarray}
and, for the cosmic time, 
\begin{eqnarray}
    \label{eq:neutron_itself2}
    t \ll 0.83 \times 10^{7} \  \sec \
    \kakko{\frac{E_{n}}{\gev}}^{2/3}
    \kakko{\frac{\sigma^{n}_{p\rightarrow N' }\beta_{n}}{40 \mb} }^{2/3}
    \kakko{\frac{\eta_{10}}{6}}^{2/3}.
\end{eqnarray}

Before closing this subsection, we check that the effects of the
cosmic expansion are negligible; this can be seen if the cosmic
expansion rate is smaller than the scattering rate of the nuclei.
Indeed, in the radiation dominated universe with $T\lesssim m_e$, the
expansion rate is given by
\begin{eqnarray}
    \label{eq:Hubble}
    H \simeq (2.6 \ {\rm sec})^{-1} 
    \times \left( \frac{T}{\mev} \right)^2.
\end{eqnarray}
Thus, the condition $\Gamma^{H_i}_{N\rightarrow N'}\gg H$ is satisfied
when
\begin{eqnarray}
    \label{eq:gamma_gt_H}
    T \gg 1.7 \times 10^{-2} \ {\rm eV} 
    \times f_{N}^{-1}
    \kakko{\frac{\eta_{10}}{6}}^{-1} 
    \kakko{\frac{\sigma^{H_i}_{N\rightarrow N'}}{40 \mb}}^{-1}.
\end{eqnarray}
Since we only consider the cases where $X$ decays during the
radiation-dominated epoch, this condition is automatically satisfied.

\subsection{Energy-loss of the hadrons via the electromagnetic interactions}
\label{subsec:stopping}

As we will discuss in the following sections, energy-loss of the
energetic hadrons via the electromagnetic interactions is very
important for the calculation of the abundances of the light elements.
If energetic hadrons completely lose their energy before scattering
off the background nuclei, they do not cause the dissociations of the
background nuclei (in particular, $\alpha_{\rm BG}$).  In this case,
only exothermic reactions are allowed for the hadronic particles, and
the extra-ordinary hadronic inter-converting reactions between protons
and neutrons become important.  Such processes, which occurs at
$t\lesssim 100\ {\rm sec}$, are discussed in Section
\ref{sec:interconversion}.  On the other hand, if the electromagnetic
interactions do not stop the energetic hadrons (in particular,
nucleons), they effectively scatter off the background nuclei and
induce hadrodissociation.  For $t\gtrsim 100\ {\rm sec}$, this is the
case.  In particular, (i) $\alpha_{\rm BG}$ is destructed and various
energetic debris nuclei ($n$, $p$, ${\rm D}$, ${\rm T}$, and
$\hethree$) are produced, (ii) some of these energetic nuclei (in
particular, ${\rm T}$, $\hethree$, and $\hefour$) scatter off
$\alpha_{\rm BG}$ to produce heavier nuclei ($\lisix$, $\liseven$, and
$\beseven$), and (iii) the energetic nucleons emitted during the
evolution of hadronic shower inter-convert the background proton and
neutron through hadronic collisions.  These processes will be
discussed in detail in Section~\ref{sec:hadrodis} and
Section~\ref{sec:ntlibe}.

In this subsection, we discuss the interaction of the energetic
hadronic particles with the background (in particular, photon,
electron, and positron) via the electromagnetic interaction.  The
energetic hadron $H_i$ scatters off the background particles via the
electromagnetic interaction in particular with the following
processes: the Coulomb scattering ($H_{i} + e^{\pm} \to H_{i} +
e^{\pm}$), the Compton scattering ($H_{i} + \gamma \to H_{i} +
\gamma$), the Bethe-Heitler scattering ($H_{i} + \gamma \to H_{i} +
e^{+} + e^{-}$), and the photo-pion process ($H_{i} + \gamma \to H_i'
+ \pi$).  With these processes, energetic hadrons (gradually) lose
their energy.  The energy-loss rate of $H_i$ can be expressed in the
following form:
\begin{eqnarray}
    \label{eq:tot_dedt}
    \kakko{\frac{dE_{H_i}}{dt}} = 
    \kakko{\frac{dE_{H_i}}{dt}}_{\rm Coulomb} +
    \kakko{\frac{dE_{H_i}}{dt}}_{\rm CP}  + 
    \kakko{\frac{dE_{H_i}}{dt}}_{\rm BH} +
    \kakko{\frac{dE_{H_i}}{dt}}_{\rm photo\mbox{-}pion},
\end{eqnarray}
which is the sum of the various processes listed above.  (The concrete
expressions of the energy loss rates are given in Appendix
\ref{sec:energy_loss_rates}.)

For our purpose, it is important to estimate how the hadrons lose
their energy in the thermal bath before scattering off the background
nuclei.  If the energy-loss rate via the electromagnetic processes is
large enough, hadrons are stopped before scattering off the background
nuclei via the hadronic interactions.  To estimate whether a hadron
$H_i$ is stopped or not through the electromagnetic interaction before
scattering off the background proton or $\alpha$, we
calculate~\footnote{
$R^{H_i}$ depends also on the $\hefour$ fraction $Y$.  We
properly take into account the $Y$ dependence in our numerical
calculations.}
\begin{eqnarray}
    \label{eq:r_stop}
    R^{H_i}_{A_j\rightarrow A_k} (E^{\rm (in)}_{H_i}, E_{H_i}'; T)
    \equiv
    n_{A_j}
    \int_{E^{\rm (in)}_{H_i}}^{E_{H_i}'}
    \sigma^{H_i}_{A_j\rightarrow A_k} \beta_{H_{i}}
    \kakko{\frac{dE_{H_i}}{dt}}^{-1} dE_{H_i},
\end{eqnarray}
and
\begin{eqnarray}
    \label{eq:r_tot}
    R^{H_i} (E^{\rm (in)}_{H_i}, E_{H_i}'; T)
    \equiv
    \sum_{j,k} R^{H_i}_{A_j\rightarrow A_k} 
    (E^{\rm (in)}_{H_i}, E_{H_i}'; T),
\end{eqnarray}
where $(dE_{H_i}/dt)$ is the energy-loss rate via the electromagnetic
interactions which is given in Eq.\ (\ref{eq:tot_dedt}), and the sum
in Eq.\ (\ref{eq:r_tot}) is over all the possible hadronic processes.
Since $p$ and $\alpha$ are the most abundant in the background among
the nuclei, hadronic scattering processes are dominated by the
scatterings with $p$ and $\alpha_{\rm BG}$; in our study, we use the
approximation
\begin{eqnarray}
    \sum_{j} n_{A_j} \sigma^{H_i}_{A_j \rightarrow A_k} =
    n_{p} \sigma_{H_i+p\rightarrow A_k+\cdots}
    + n_{\alpha} \sigma_{H_i+\alpha\rightarrow A_k+\cdots}.
\end{eqnarray}

$R^{H_i}(E^{\rm (in)}_{H_i},E_{H_i}';T)$ parameterizes the number of
the hadronic scatterings before the energy of the hadron $H_i$ with
its {\sl initial} energy $E^{\rm (in)}_{H_i}$ decreases to $E_{H_i}'$
via the electromagnetic interactions.  Thus, if $R^{H_i}(E^{\rm
(in)}_{H_i}, E^{\rm (th)}_{H_{i}};T)\lesssim 1$ (with $E^{\rm
(th)}_{H_{i}}$ being the threshold energy of some hadrodissociation
process), the high energy hadron $H_i$ is (mostly) stopped and does
not cause the hadrodissociation process.  On the contrary, if
$R^{H_i}(E^{\rm (in)}_{H_i}, E^{\rm (th)}_{H_{i}};T)$ is larger than
unity, the high energy hadrons are not stopped through the
electromagnetic interaction and cause hadrodissociation processes.  In
particular, if $R^{H_i}(E^{\rm (in)}_{H_i}, E^{\rm
(th)}_{H_{i}\alpha};T)\gtrsim 1$ with $E^{\rm (th)}_{H_{i}\alpha}$
being the threshold energy for the destruction process of $\alpha_{\rm
BG}$, high energy projectile hadron with its initial energy $E^{\rm
(in)}_{H_i}$ effectively destroys $\alpha_{\rm BG}$.  The number
density of the $\alpha_{\rm BG}$ becomes abundant after the cosmic
time $t\sim 200$ sec ($T \sim$ 0.1 MeV).  Thus, if $X$ decays after
this epoch, abundances of the light elements (in particular, ${\rm D}$
and ${\rm ^3He}$) may be significantly changed by the direct
destruction of $\alpha_{\rm BG}$.

Using the quantity $R^{H_i}$ given in Eq.\ (\ref{eq:r_tot}), we
estimate the energy of $H_i$ just before scattering off the background
proton or $\alpha_{\rm BG}$.  For stable particles, such a energy for
a given initial energy $E^{\rm (in)}_{H_i}$, which is denoted as
$\tilde{E}^{(R=1)}_{H_i}(E^{\rm (in)}_{H_i})$, is estimated by solving
the following relation
\begin{eqnarray}
    R^{H_i} (E^{\rm (in)}_{H_i}, \tilde{E}^{(R=1)}_{H_i}; T) = 1.
    \label{R=1}
\end{eqnarray}
For the neutron, we should take account of the fact that the neutron
may decay before scattering off the background nuclei.  Thus, for the
neutron, we define $\tilde{E}^{(R=1)}_n(E^{\rm (in)}_n)$ as follows.
For the given initial energy of the neutron $E^{\rm (in)}_n$, we
calculate the Lorentz factor $\gamma^{\rm (in)}_n\equiv E^{\rm
(in)}_n/m_n$ as well as the total scattering rate for the hadronic
processes $\Gamma^n_{\rm had}\equiv n_{p}\sigma_{n+p\rightarrow\cdots}
+n_{\alpha}\sigma_{n+\alpha\rightarrow\cdots}$.  (Here, the cross
sections are estimated with the initial energy of the neutron.)  Then,
if $\Gamma^n_{\rm had}>(\gamma^{\rm (in)}_n\tau_n)^{-1}$, we expect
that the neutron scatters off the background nuclei before it decays,
and  we estimate $\tilde{E}^{(R=1)}_n(E^{\rm (in)}_n)$ using Eq.\ 
(\ref{R=1}).  On the contrary, if $\Gamma^n_{\rm had}<(\gamma^{\rm
(in)}_n\tau_n)^{-1}$, neutron is likely to decay before scattering off
the background nuclei.  In this case, the energetic neutron is
equivalent to the proton for the calculation of the hadrodissociation
processes.  Thus, for this case we use the relation
$\tilde{E}^{(R=1)}_n(E^{\rm (in)}_n)= \tilde{E}^{(R=1)}_p(E^{\rm
(in)}_n)$.~\footnote{
Neutron loses its energy before it decays and hence the energy of the
proton produced by the neutron is not exactly equal to $E^{\rm
(in)}_n$.  However, the energy-loss rate of the neutron via the
electromagnetic processes is much less efficient than that of the
proton and hence we can neglect the energy-loss of the neutron before
it decays.}

\begin{figure}
    \centering
    \centerline{{\vbox{\epsfxsize=8.0cm\epsfbox{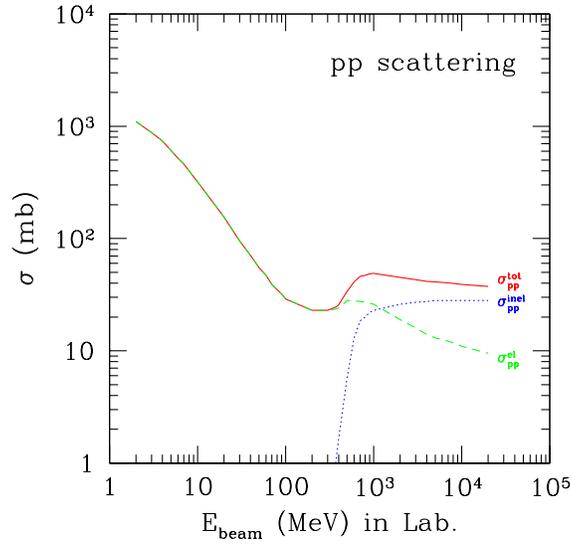}}}}
    \caption{Cross sections of the $pp$ scattering processes as
    functions of the kinetic energy of the beam (projectile) proton in
    the laboratory system. The solid line is the total cross section
    while the dashed (dotted) line is the elastic (inelastic) cross
    section.}
    \label{fig:ppsig}
\end{figure}

\begin{figure}
    \centering
    \centerline{{\vbox{\epsfxsize=8.0cm\epsfbox{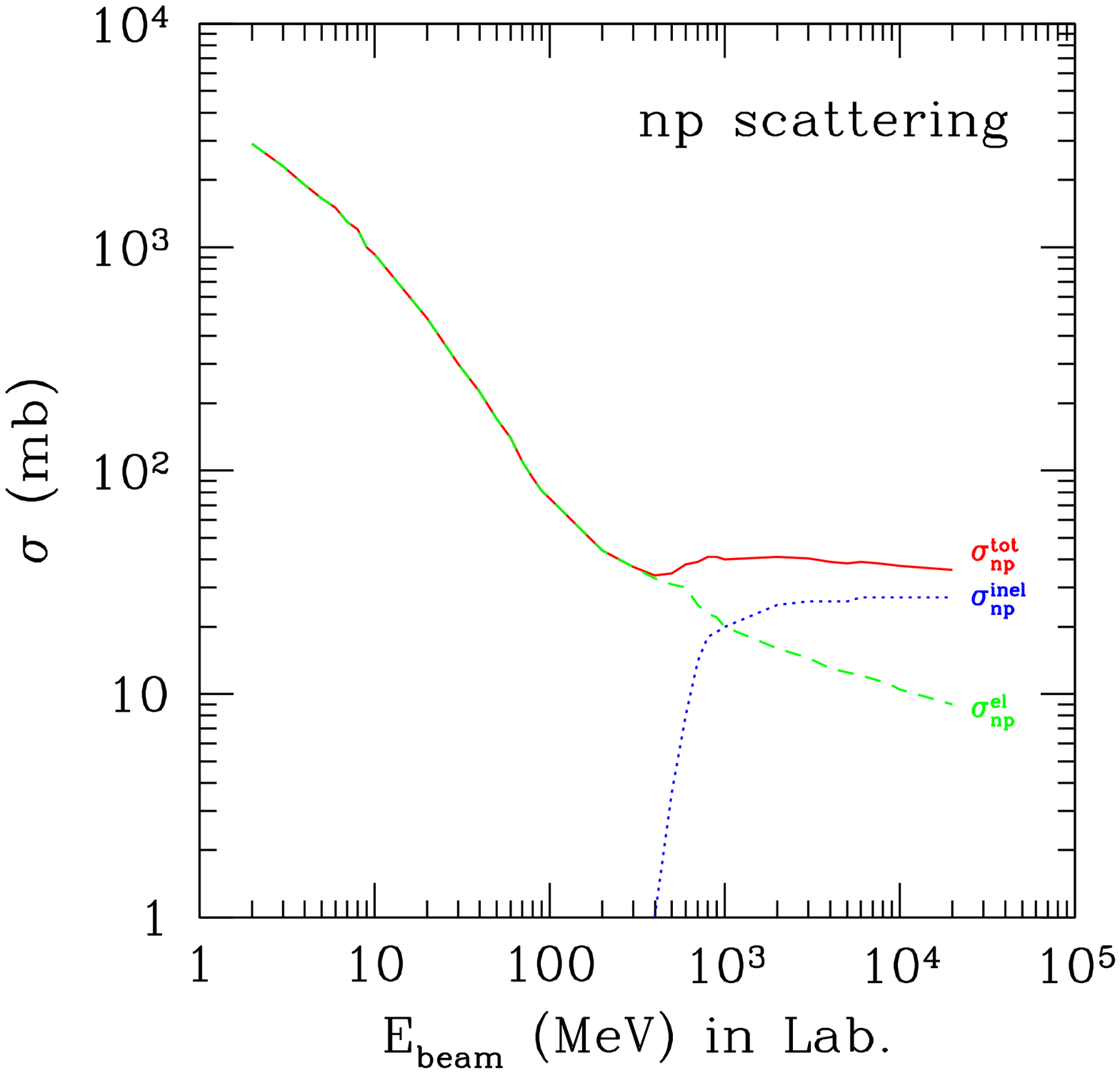}}}}
    \caption{Cross sections of the $np$ scattering processes as
    functions of the kinetic energy of the beam (projectile)
    proton in the laboratory system. The solid line is the total cross
    section while the dashed (dotted) line is the elastic (inelastic)
    cross section.}
    \label{fig:npsig}
\end{figure}

\begin{figure}
  \centering
  \centerline{{\vbox{\epsfxsize=8.0cm\epsfbox{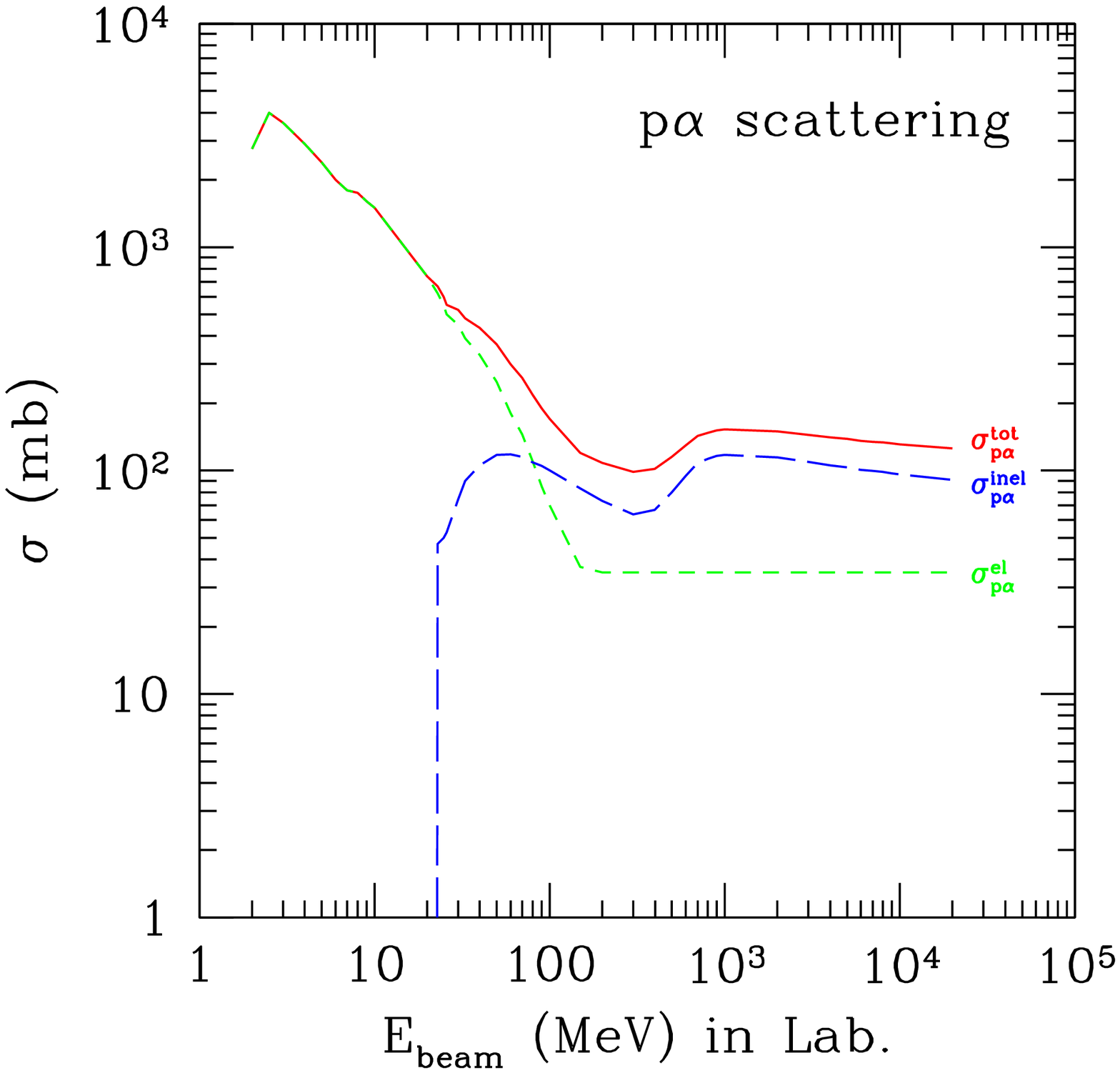}}}}
  \caption{Cross sections of the $p\alpha$ scattering processes as
    functions of the kinetic energy of the beam (projectile) proton
    in the laboratory system. The solid line is the total cross
    section while the dashed (dotted) line is the elastic (inelastic)
    cross section.}
  \label{fig:sig4}
\end{figure}

\begin{figure}
  \centering
  \centerline{{\vbox{\epsfxsize=8.0cm\epsfbox{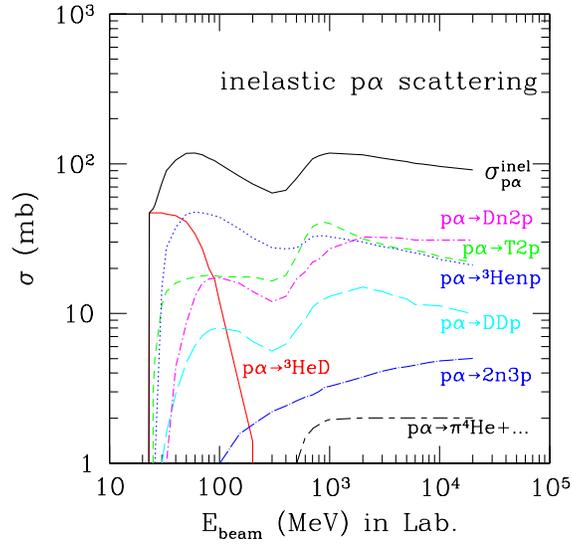}}}}
  \caption{Cross sections of the inelastic $p\alpha$ scattering
    processes.}
  \label{fig:inelsig4_2}
\end{figure}

For various nuclei, we calculate $\tilde{E}^{(R=1)}_{H_i}$ as
functions of the initial energy $E^{\rm (in)}_{H_i}$.  In order to
precisely calculate $\tilde{E}^{(R=1)}_{H_i}$, we need informations
about the cross sections for the hadronic processes.  For energetic
proton and neutron, we use detailed experimental data of cross
sections for $pp$, $np$, and $p\alpha$ collisions~\cite{meyer:1972}.
In Figs.\ \ref{fig:ppsig}, \ref{fig:npsig}, \ref{fig:sig4}, and
\ref{fig:inelsig4_2}, we plot the data of total, elastic and inelastic
cross sections for these collision processes, respectively.

\begin{figure}[t]
    \centering
    \centerline{{\vbox{\epsfxsize=8.0cm\epsfbox{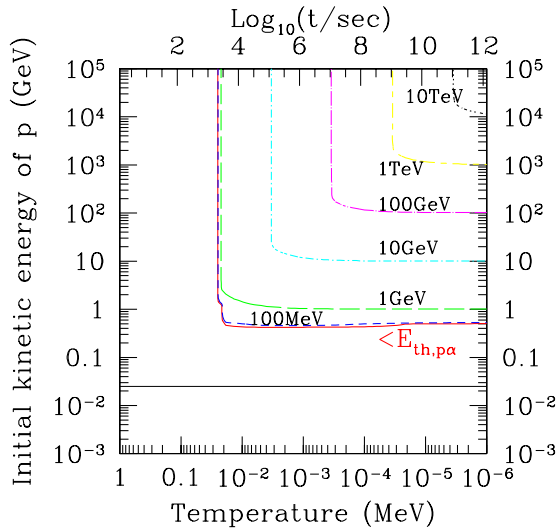}}}}
    \caption{Contours of the constant $\tilde{E}^{(R=1)}_p$,
    the energy of the proton just before it scatters off  the
    background nuclei ($p$ or $\hefour$).  The horizontal axis is the
    temperature while the vertical one is the initial energy of the
    proton $E_p^{\rm (in)}$.  The numbers in the figure are the values
    of $\tilde{E}^{(R=1)}_p$, and the solid line with ``$<E_{{\rm
    th},p\alpha}$'' shows the contour for $\tilde{E}^{(R=1)}_p$ being
    equal to the threshold energy for the destruction process of
    $\hefour$, $E^{\rm (th)}_{p\alpha}=25\ {\rm MeV}$. The horizontal
    thin-solid line denotes $E_p^{\rm (in)}=E^{\rm
    (th)}_{p\alpha}$. Here we use $\eta =6.1 \times 10^{-10}$ and
    $Y=0.25$.}
    \label{fig:Eloss_p}
\end{figure}

In Fig.~\ref{fig:Eloss_p}, we plot the contours of the constant
$\tilde{E}^{(R=1)}_p$ on the $T$ vs.\ $E_{p}^{\rm (in)}$ plane.  Here
we use $\eta =6.1 \times 10^{-10}$ and $Y=0.25$.  From this figure, we
see that the protons are completely stopped when the temperature is
sufficiently high.  This is because the number of the background
electron is not Boltzmann-suppressed at high temperature, resulting in
enhanced energy-loss rate.

For a more quantitative discussion, it is convenient to define the
typical energy-loss rate through the electromagnetic processes:
\begin{eqnarray}
    \Gamma^{H_i}_{\rm EM} \equiv 
    \frac{1}{E_{\rm kin}} \frac{dE_{H_i}}{dt}.
\end{eqnarray}
If $\Gamma^{H_i}_{\rm EM}$ is larger than the hadronic scattering
rate, energy loss is effective and the energy of the hadron is
significantly reduced before scattering off the background nuclei.  On
the contrary, if $\Gamma^{H_i}_{\rm EM}\lesssim
\Gamma^{H_i}_{A_j\rightarrow A_k}$, energy loss is not important.

At the temperature $T\sim {\cal O}(10 - 100)\ {\rm keV}$, Coulomb
scattering is the most important for the energy loss.  Using the
energy-loss rates given in Appendix \ref{sec:energy_loss_rates}, we
can estimate $\Gamma^p_{\rm EM}$ for the case where the Coulomb
scattering is the dominant process:
\begin{eqnarray}
    \left[ \Gamma^p_{\rm EM} \right]_{\rm Coulomb} \simeq
     \left(1.4 \times 10^{-14} \sec \right)^{-1}
    \times \beta_p^{-1} \Lambda
    \left( \frac{E_{\rm kin}}{m_N} \right)^{-1}
    x_e^{-3/2} e^{-x_e},
    \label{tau_p(EM)}
\end{eqnarray}
where $x_e\equiv m_e/T$.  (Here, we consider protons with $\beta_p$
larger than the thermal velocity of the background electron, in which
we are mostly interested at this temperature.)  Since the number
density of the electron becomes smaller as the temperature gets
decreased, $\Gamma^p_{\rm EM}$ in this case decreases as the
temperature becomes lower.  Comparing Eq.\ (\ref{tau_p(EM)}) with Eq.\ 
(\ref{eq:gamma^i_nn}), we can see that the proton is completely
stopped when $T\gtrsim {\cal O}(10)\ {\rm keV}$.  

For ultra-relativistic protons, in fact, the Compton energy-loss (and
the Bethe-Heitler process) becomes effective in particular at lower
temperature.  Typical behavior given in Fig.\ \ref{fig:Eloss_p} is
indeed understood by using the formula for,  e.g., the Compton
process.  Using Eq.\ (\ref{eq:dedt_compton}), we can obtain
$\Gamma^p_{\rm EM}$ for the ultra-relativistic case:
\begin{eqnarray}
    \left[ \Gamma^p_{\rm EM} \right]_{\rm Compton} \simeq
     \left(13\ {\rm sec} \right)^{-1}
    \times
    \left( \frac{E_{\rm kin}}{\rm 100\ GeV} \right)
    \left( \frac{T}{\rm 1\ keV} \right)^4.
    \label{tau_p(EM)'}
\end{eqnarray}
Thus, for the ultra-relativistic protons, we can see that the energy
loss is effective for the temperature $T\gtrsim {\cal O}(0.1)\ {\rm
keV}\times (E_{\rm kin}/{\rm 100\ GeV})^{-1}$.

\begin{figure}
  \centering
  \centerline{{\vbox{\epsfxsize=8.0cm\epsfbox{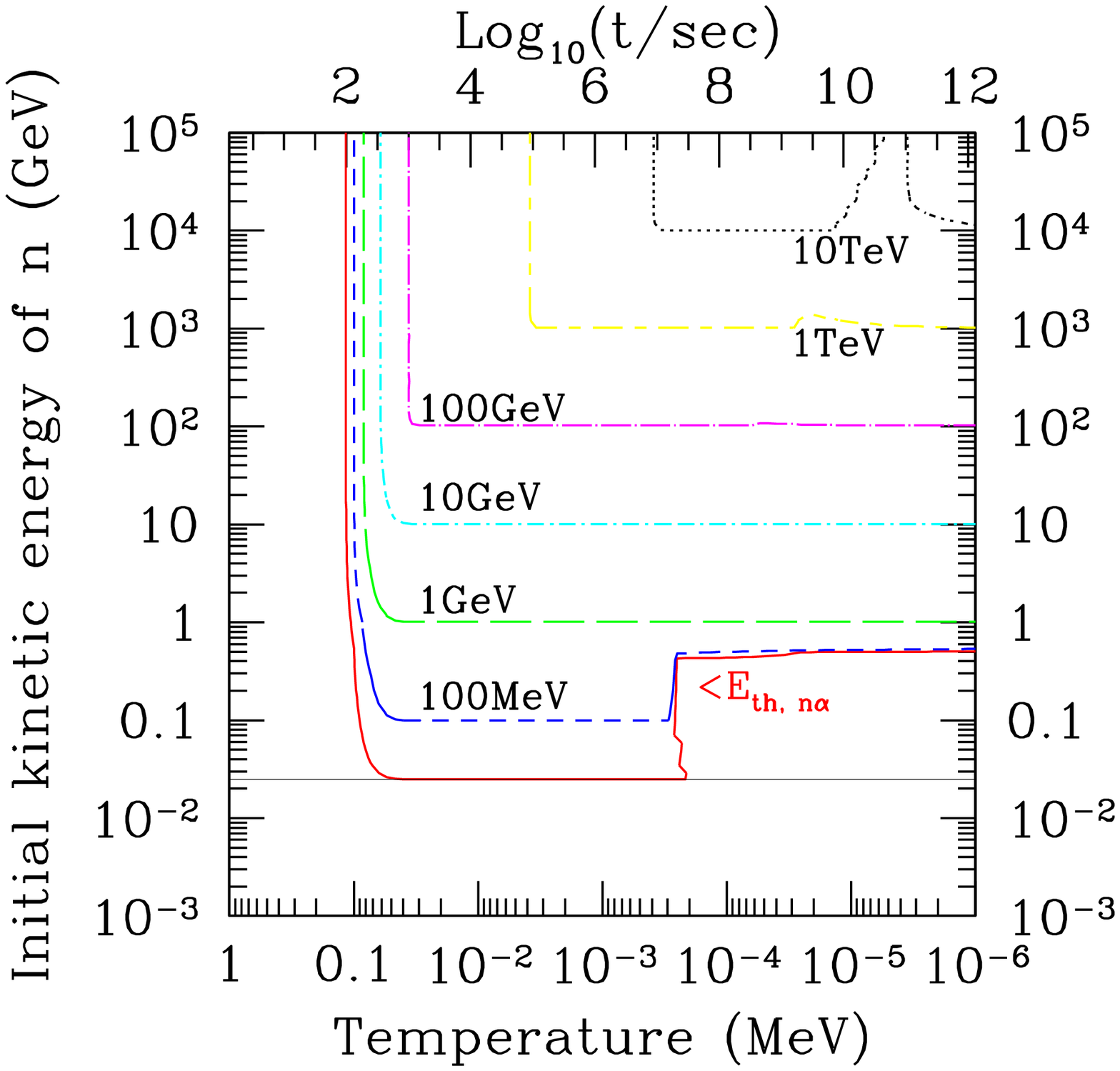}}}}
  \caption{Same as Fig.~\ref{fig:Eloss_p}, but for neutron.
  (The horizontal thin-solid line denotes $E_n^{\rm (in)}=E^{\rm
  (th)}_{n\alpha}$ accordingly.)}
  \label{fig:Eloss_n}
\end{figure}

For the neutron, we plot the contours of constant
$\tilde{E}^{(R=1)}_n$ in Fig.~\ref{fig:Eloss_n}.  For the $n\alpha$
scatterings, we use the cross section for $p\alpha$ processes assuming
the isospin symmetry as well as small number of data for $n\alpha$
scatterings.  Empirically, the size of the isospin breaking in this
case is estimated to be within 20 \% (10\%) for $E_{n}^{\rm {(in)}}$ =
25 MeV (100 MeV).  As we will discuss later, we will adopt 20\ \%
errors to all the hadronic cross sections in our Monte Carlo analysis,
which will also take account of this uncertainty related to the
isospin breaking.

As one can see, non-relativistic neutrons are effectively stopped when
the temperature is so high that the number density of the background
electron is large enough.  For the complete stopping of the neutron,
the temperature should be higher than $\sim 100\ {\rm keV}$, which is
slightly higher than the stopping temperature of the proton; using
Eq.\ (\ref{de/dt(n)}), one can calculate $\Gamma_{\rm EM}^n$ and see
that $\Gamma_{\rm EM}^n$ for non-relativistic neutron becomes smaller
than the hadronic scattering rate when $T\lesssim 100\ {\rm keV}$.
Thus, at lower temperature, energy-loss of the neutron becomes
inefficient.  When the temperature becomes low enough, however, time
scale for the hadronic scattering becomes longer than the lifetime of
the neutron.  This is the reason of the drastic change of
$\tilde{E}^{(R=1)}_n$ at the temperature $T\sim 0.3\ {\rm keV}$ for
non-relativistic neutron.

\begin{figure}
    \centering
    \centerline{\vbox{\epsfxsize=8.0cm\epsfbox{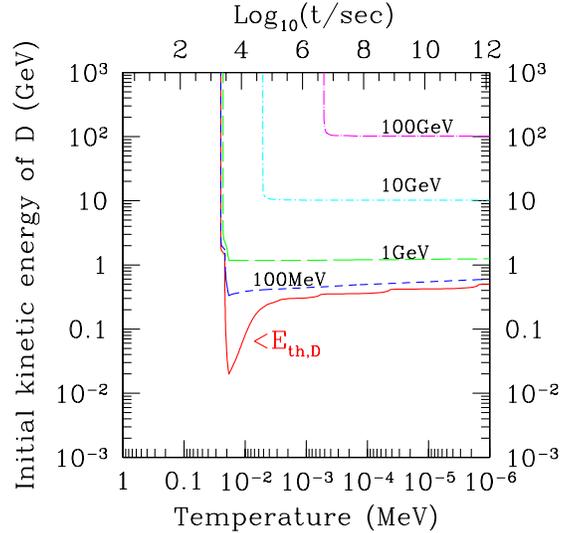}}}
    \caption{Same as Fig.~\ref{fig:Eloss_p}, but for D. 
    The solid line with ``$<E_{{\rm th},\D}$'' shows the contour for
    $\tilde{E}^{(R=1)}_{\D}$ being equal to the threshold energy for
    the destruction process ${\rm D}+p_{\rm BG}\rightarrow 2p+n$.}
    \label{fig:Eloss_D}
\end{figure}

\begin{figure}
    \centering
    \centerline{{\vbox{\epsfxsize=8.0cm\epsfbox{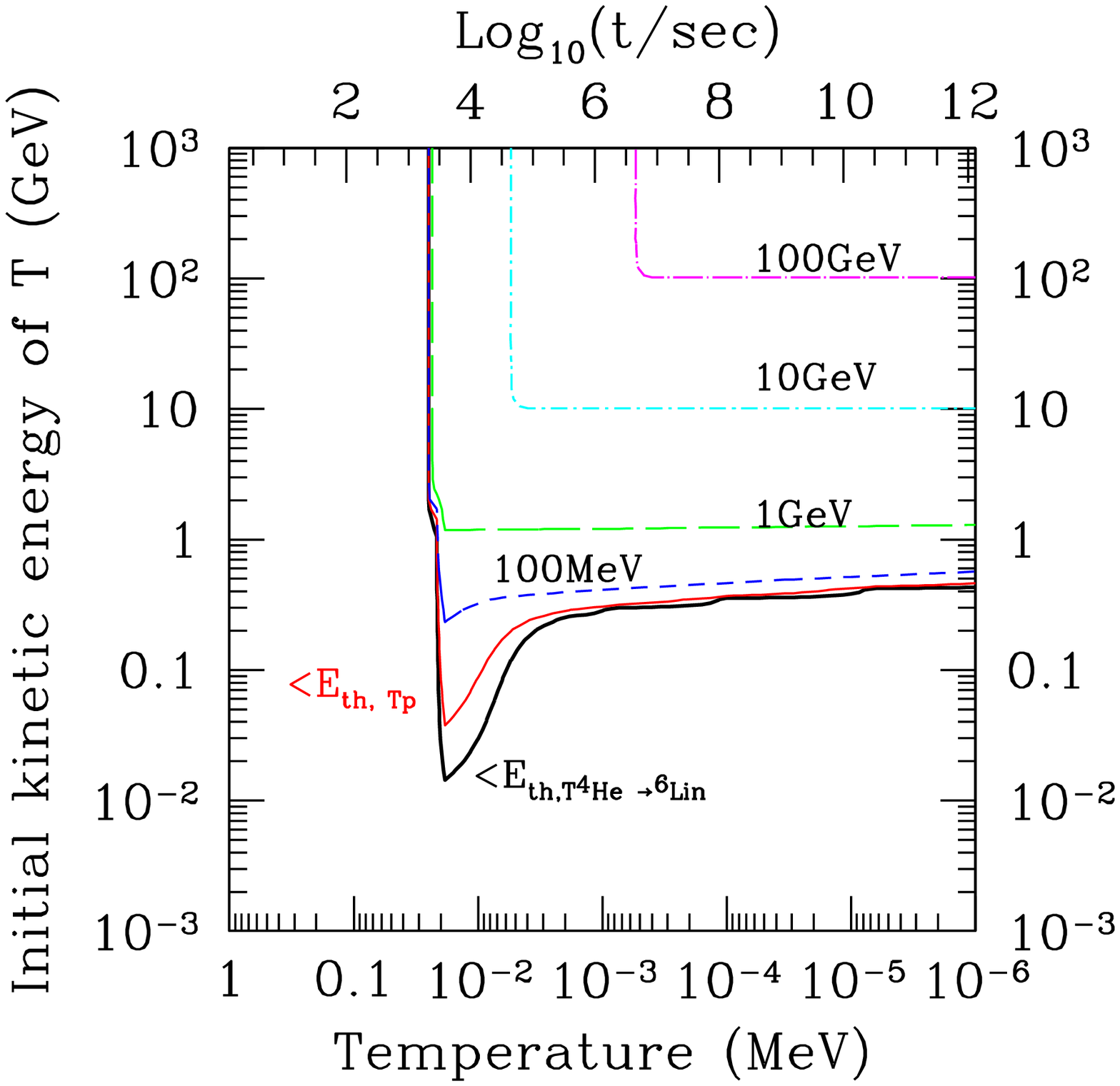}}}}
    \caption{Same as Fig.~\ref{fig:Eloss_p}, but for T. 
    The solid line with ``$<E_{{\rm th},\T p}$'' shows the contour for
    $\tilde{E}^{(R=1)}_{\T}$ being equal to the threshold energy for
    the destruction process of the projectile T through ${\rm
    T}+p_{\BG}$, while the thick solid line is the contour for
    $\tilde{E}^{(R=1)}_{\T}$ being equal to the threshold energy for
    the process $\T+\alpha_{\rm BG}\to\lisix +n$.}
    \label{fig:Eloss_T}
\end{figure}

\begin{figure}
    \centering
    \centerline{{\vbox{\epsfxsize=8.0cm\epsfbox{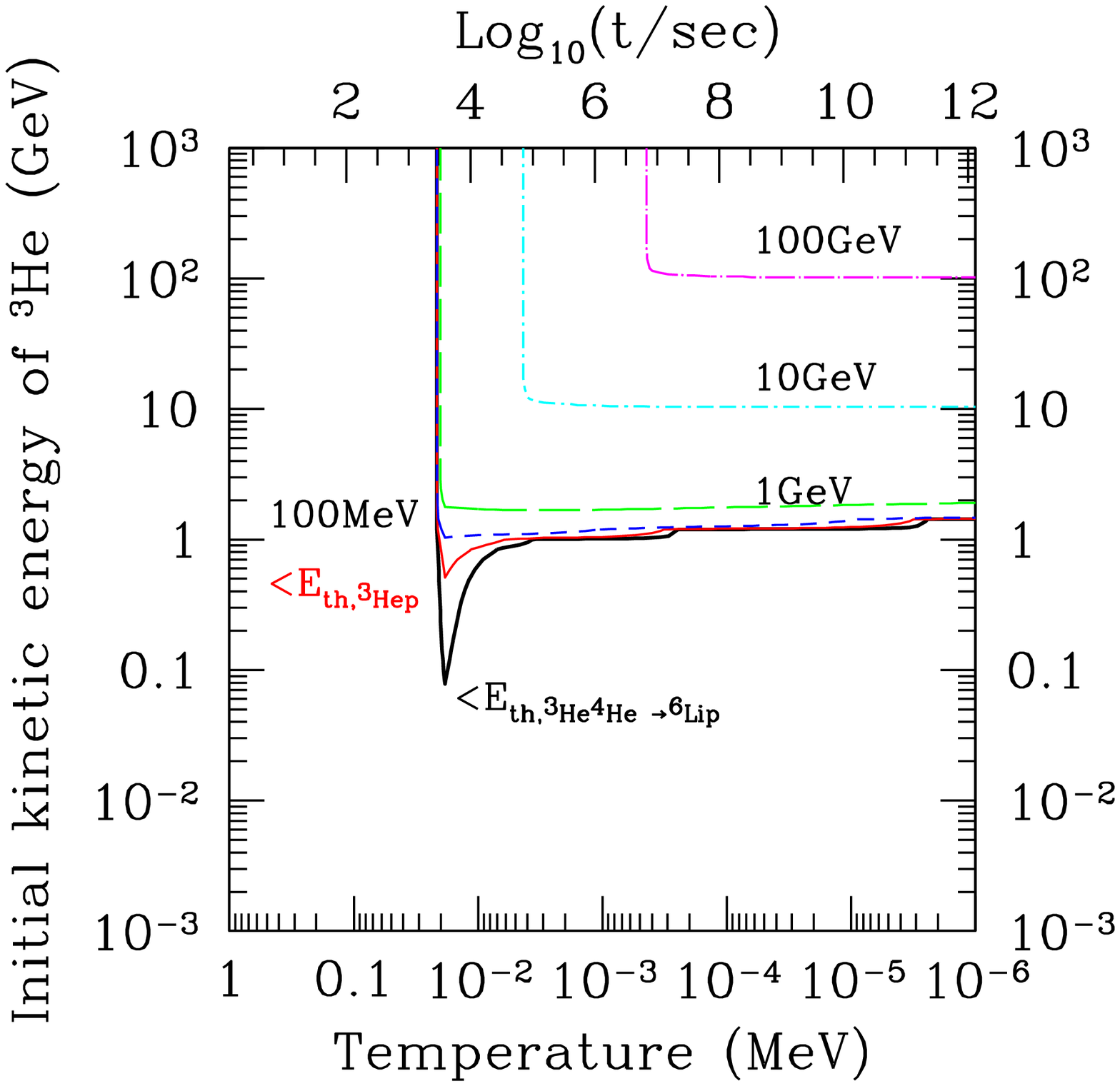}}}}
    \caption{Same as Fig.~\ref{fig:Eloss_p}, but for $\hethree$. 
    The solid line with ``$<E_{{\rm th},\hethree p}$'' shows the
    contour for $\tilde{E}^{(R=1)}_{\hethree}$ being equal to the
    threshold energy for the destruction process of the projectile
    $\hethree$ through $\hethree +p_{\BG}$, while the thick solid line
    is the contour for $\tilde{E}^{(R=1)}_{\hethree}$ being equal to
    the threshold energy for the process $\hethree + \alpha_{\rm BG}
    \to \lisix + p$.}
    \label{fig:Eloss_He3}
\end{figure}

\begin{figure}
    \centering
    \centerline{{\vbox{\epsfxsize=8.0cm\epsfbox{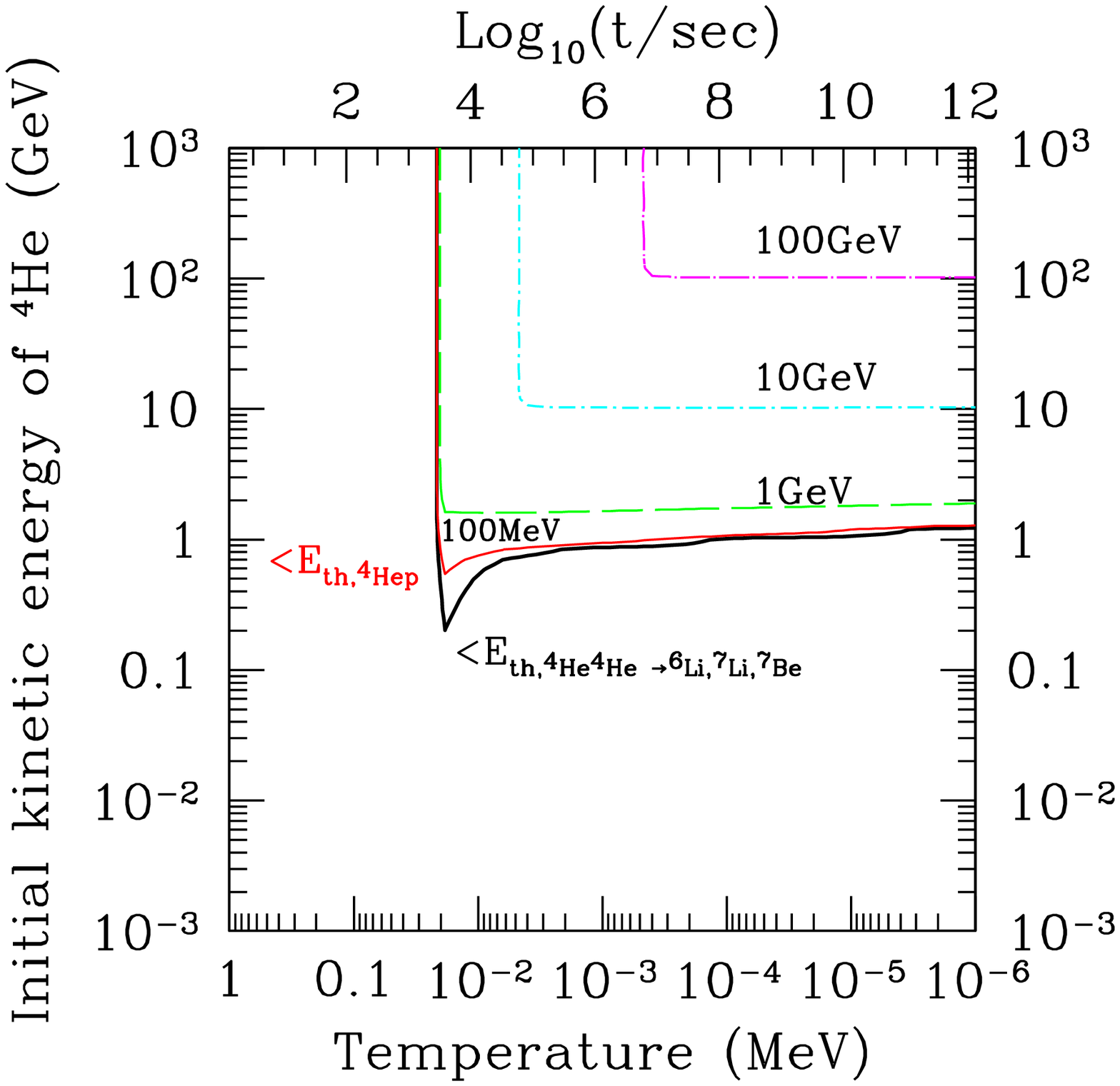}}}}
    \caption{Same as Fig.~\ref{fig:Eloss_p}, but for $\hefour$. 
    The solid line with ``$<E_{{\rm th},\hefour p}$'' shows the
    contour for $\tilde{E}^{(R=1)}_{\hefour}$ being equal to the
    threshold energy for the destruction process of the projectile
    $\hefour$ through $\hefour +p_{\BG}$, while the thick solid line
    denotes the contour for $\tilde{E}^{(R=1)}_{\hefour}$ being equal
    to the threshold energy for the process $\hefour + \alpha_{\rm BG}
    \to \liseven + \tenten$.}
    \label{fig:Eloss_He4}
\end{figure}

\begin{figure}
  \centering
  \centerline{{\vbox{\epsfxsize=8.0cm\epsfbox{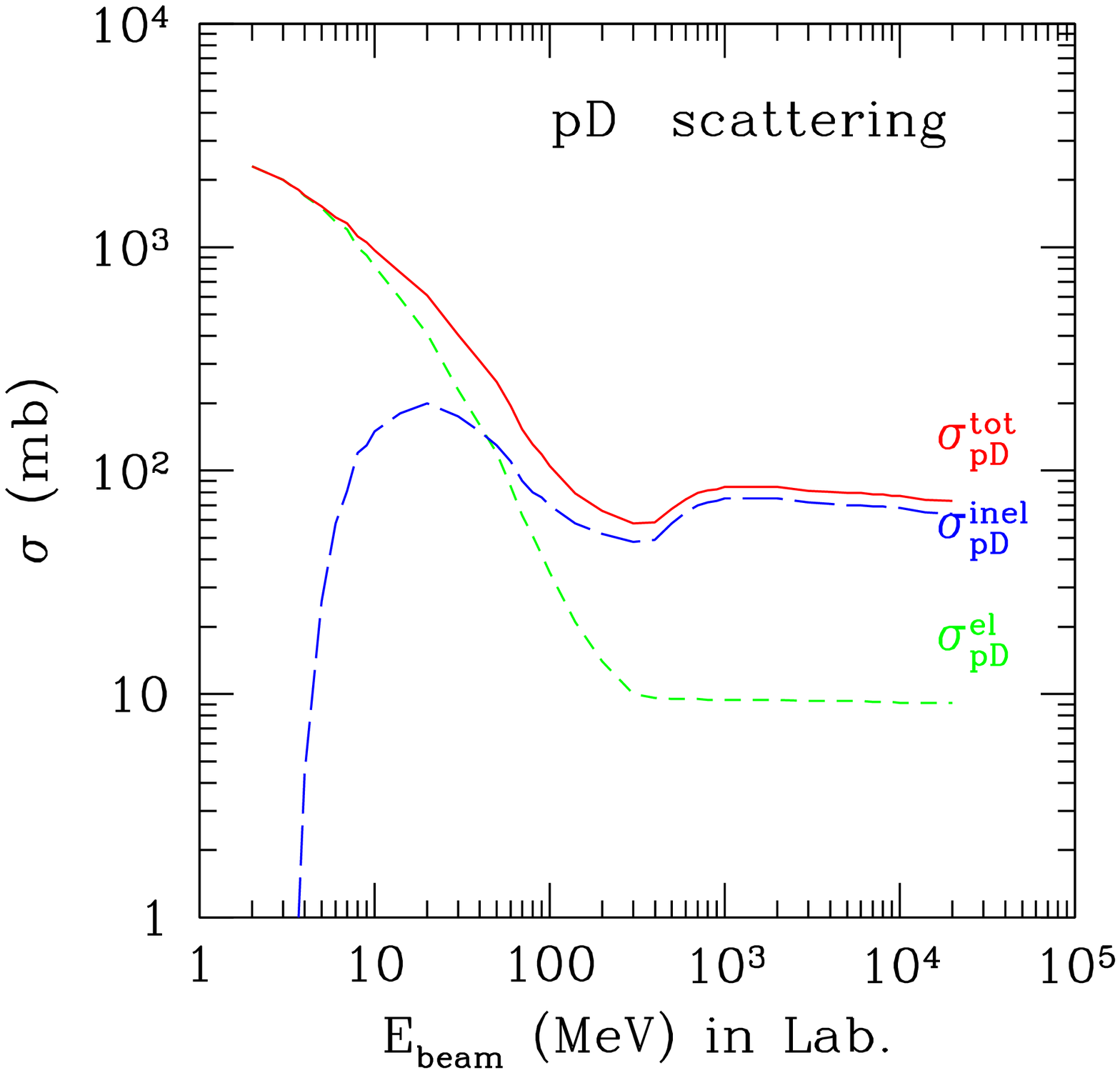}}}}
  \caption{Cross sections of the $p{\rm D}$ scattering processes.}
  \label{fig:sigpd}
\end{figure}

\begin{figure}
  \centering
  \centerline{{\vbox{\epsfxsize=8.0cm\epsfbox{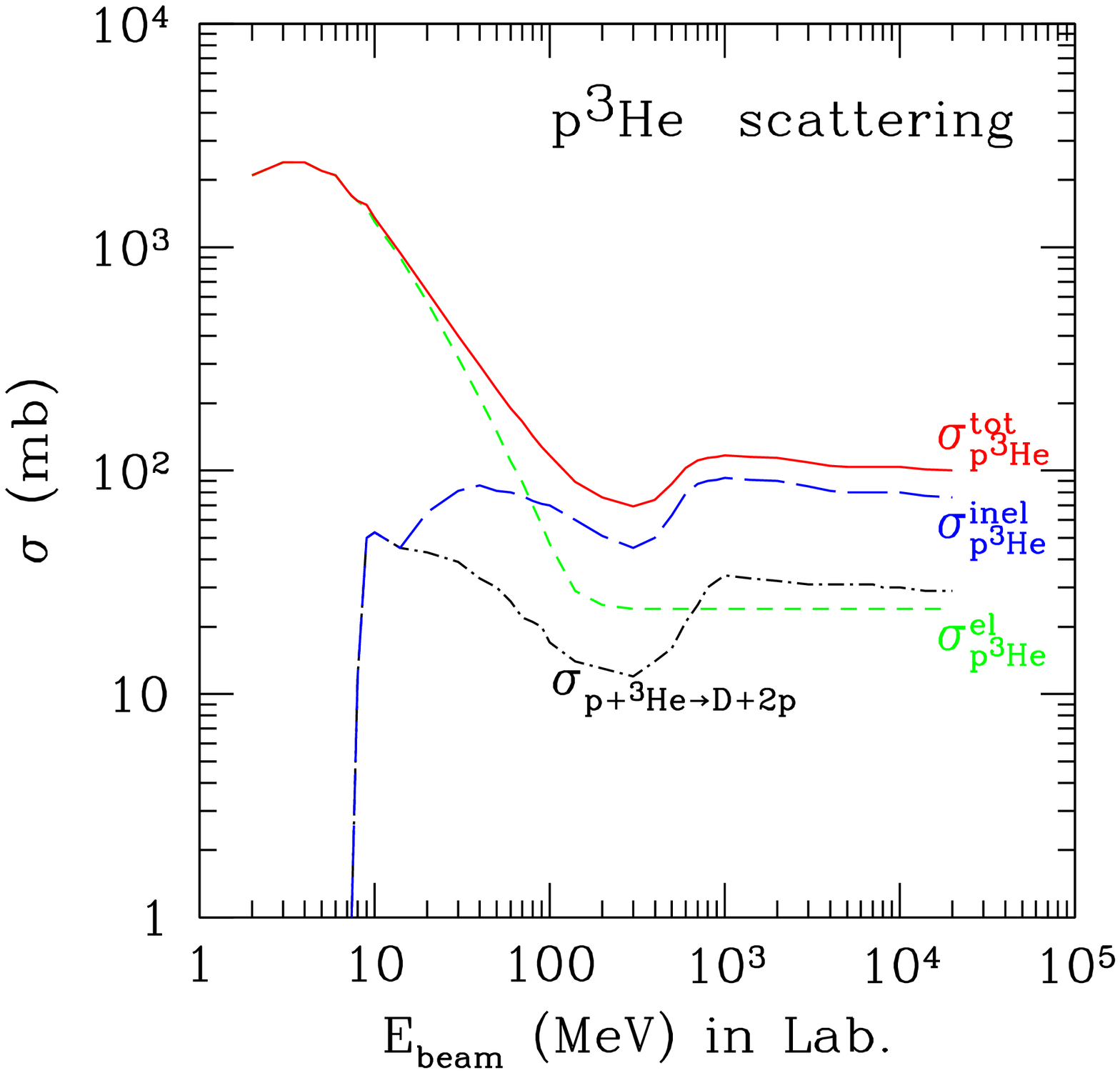}}}}
  \caption{Cross sections of the $p{\rm ^3He}$ scattering processes.}
  \label{fig:sigphe3}
\end{figure}

We also calculate the typical energy of ${\rm D}$, ${\rm T}$, ${\rm
^3He}$ and ${\rm ^4He}$ just before scattering off the background
nuclei.  We plot the contours of $\tilde{E}^{(R=1)}_{\rm D}$,
$\tilde{E}^{(R=1)}_{\rm T}$, $\tilde{E}^{(R=1)}_{\rm ^3He}$ and
$\tilde{E}^{(R=1)}_{\rm ^4He}$ in Figs.\ \ref{fig:Eloss_D},
\ref{fig:Eloss_T}, \ref{fig:Eloss_He3}, and \ref{fig:Eloss_He4},
respectively.  (For readers' convenience, we also plot the cross
sections for the $p{\rm D}$ and $p{\rm ^3He}$ processes in
Figs.~\ref{fig:sigpd} and \ref{fig:sigphe3}, respectively.)  For these
charged particles $A_i$, behavior of $\tilde{E}^{(R=1)}_{A_i}$ is
similar to the case of the proton.

In this paper, we approximate that all the hadrons $H_i$ have the
energy $\tilde{E}^{(R=1)}_{H_i}$ just before they scatter off the
background proton or $\alpha_{\rm BG}$.  

\section{Inter-Conversion between Neutron and Proton}
\label{sec:interconversion}
\setcounter{equation}{0}

So far, we have discussed the propagation of the hadronic particles in
the expanding universe, paying particular attention to their
energy-loss.  Those particles cause various hadronic processes before
and after being stopped by the electromagnetic scatterings and affect
the abundances of the light elements.

The first effect we discuss is the inter-converting $p\leftrightarrow
n$ reaction caused by hadrons emitted from
decaying $X$.~\footnote{
Of course, pions generated from the high-energy photons via the
process $\gamma+\gamma_{\rm BG}\rightarrow \pi^++\pi^-$ are also
included in the present work.}
Implications of such effect was first considered in
Ref.~\cite{Reno:1987qw}.  In our study, we basically 
follow the strategy of Ref.~\cite{Reno:1987qw} with 
several improvements, which are discussed in
this section.

Especially, for relatively early epoch $T \gtrsim 0.1$ MeV (i.e.,
$t\lesssim 100\ {\rm sec}$), the emitted hadrons extraordinarily
inter-convert the ambient protons and neutrons by the strong
interaction even after the conventional freeze-out time of the neutron
in the SBBN.  Thus, for massive particle $X$ with relatively short
lifetime ($\tau_{X}\sim (10^{-2}-100)\ {\rm sec}$), the
inter-conversion effect may induce a significant change of the
light-element abundances.  In particular, since the proton is more
abundant than the neutron, $n/p$ ratio is enhanced if such
inter-conversion is effective, resulting in
overproduction of $\hefour$.~\footnote{
D/H and $^{7}$Li/H are also sensitive to $n/p$ even  at $t \gtrsim
100$~sec. This effect is studied together  with hadrodissociation
effect in the next section.
} 
Thus, in this section, we discuss the inter-conversion processes.  In
particular, we present the thermally averaged cross sections for the
relevant processes.  In our analysis, for all the inter-conversion
cross sections given in this section, we adopt 50\ \% uncertainties
when we perform the Monte Carlo analysis.

\subsection{Cross sections of hadron-nucleon scattering}
\label{subsec:cross-section}

First, let us summarize the cross sections for the relevant reactions.
As we have discussed in the previous sections, once a high energy
hadron is injected into the thermal bath at the early stage of the BBN
(more precisely, when $T \gtrsim 0.1$ MeV), energy-loss processes via
the electromagnetic interactions are very effective (except for the
neutral kaon).  Thus, even if the non-thermally produced hadrons (in
particular, $\pi^{\pm}, K^{\pm}$, $p$, and $n$) are quite energetic
when they are produced, they are quickly thermalized.  In this case,
the typical kinetic energy of these hadrons becomes $\sim T$.  Thus,
for the inter-conversion processes, only the exothermic reactions are
relevant since the kinetic energy of the thermalized hadrons are
expected to be too small to induce endothermic reactions.  In
addition, the inter-conversion processes occur mostly with very small
velocity.

Concerning the exothermic reactions, it is well-known that the cross
section $\sigma$ is approximately inversely proportional to the
velocity $\beta$ when the velocity is very small; namely $\sigma\beta$
becomes almost constant. Thus, we use the threshold cross section.
The cross sections given in the following are from Ref.\ 
\cite{Reno:1987qw}.

The thermally-averaged cross sections for the processes
$\pi^++n\rightarrow p+\pi^0$ and $\pi^-+p\rightarrow n+\pi^0$ are
given by
\begin{eqnarray}
    \label{eq:sigma_pi}
    \langle\sigma \beta\rangle^{\pi^+}_{n \rightarrow p} &=& 
    1.7 \ \mb, \\ 
    \langle\sigma \beta\rangle^{\pi^-}_{p \rightarrow n} &=&
    1.5C^{2}_{\pi}(T) \ \mb,
\end{eqnarray}
respectively.  Here, $C^{2}_{H_i}(T)$ is the Coulomb correction factor
which takes account of the modification of the wave-function of the
charged particle due to the Coulomb field.  For the opposite-sign
charged particles, Coulomb correction factor is given by \cite{Schiff}
\begin{equation}
    \label{eq:coulomb}
    C^{2}_{H_i}(T) = 
    \frac{2\pi\alpha_{\rm em}\sqrt{\mu_{H_i}/2T}}
         {1 - e^{-2\pi \alpha_{\rm em}\sqrt{\mu_{H_i}/2T}}},
\end{equation}
where $\alpha_{\rm em}$ is the fine structure constant and $\mu_{H_i}$
is the reduced mass of the hadron $H_i$ and the target nucleon.

Next, we consider the effects of the nucleons and anti-nucleons
directly produced by the decay of $X$.  In our study, we only consider
the case where the numbers of such $\bar{p}$ and $\bar{n}$ produced by
the hadronic decay of $X$ are individually the same as those of $p$
and $n$.

When the temperature is high enough, nucleons are stopped in the
thermal bath by the electromagnetic interactions.  (Such temperature
is given by $T\gtrsim 20\ {\rm keV}$ and $100\ {\rm keV}$ for (anti-)
proton and (anti-) neutron, respectively.)  In such a case, we treat
nucleon-anti-nucleon pair $N_i\bar{N}_i$ like a meson, following
Ref.~\cite{Reno:1987qw}.  Then the $N_i\bar{N}_i$ meson induces the
inter-conversion processes: $N_i\bar{N}_i+N_j\rightarrow N'_j+\cdots$.
If the nucleons are thermalized, we can use the threshold cross
sections:
\begin{eqnarray}
    \label{eq:sigma_nuc}
    \langle\sigma \beta\rangle^{n\bar{n}}_{n \rightarrow n} 
    &=&  37 \ \mb, \\ 
    \langle\sigma \beta\rangle^{n\bar{n}}_{p \rightarrow n} 
    &=&  28 \ \mb, \\ 
    \langle\sigma \beta\rangle^{p\bar{p}}_{n \rightarrow p} 
    &=&    28 \mb, \\  
    \langle\sigma \beta\rangle^{p\bar{p}}_{p \rightarrow p}
    &=&  37 C^{2}_{p}(T)\ \mb. 
\end{eqnarray}

When the temperature becomes lower, on the contrary, nucleons cannot
be easily stopped.  Then, they also induce the hadrodissociation
processes and hence effects of the anti-nucleons cannot be studied
just by taking account of the conversion effects.  Importantly,
however, once the hadrodissociation processes become effective, a
large number of protons and neutrons are produced while the
anti-nucleons are not produced in the hadronic shower.  In this case,
anti-nucleons directly produced by the decay of $X$ become irrelevant
since their numbers are much smaller than those of non-thermally
produced nucleons.  Thus, when most of the energetic nucleons scatter
off the background nuclei before being stopped, inter-conversion by
the anti-nucleons becomes unimportant.  When the anti-nucleons are not
stopped, it is difficult to estimate their energies with which the
inter-conversion cross sections should be evaluated.  Thus, we do not
include the inter-conversion process due to the anti-nucleon in such a
case.  Notice that we have numerically checked that the resultant
constraints on the properties of $X$ does not change even if we
include the inter-conversion by the anti-nucleons with the threshold
cross sections at such a later epoch.

Finally, we comment on our treatment of the kaons.  Since the kaons
have relatively long lifetimes, they may also contribute to the
inter-conversion processes.  Unfortunately, however, cross sections of
many of the conversion processes related to the kaons are not
available.  In addition, the neutral kaon $K_L^0$ is hardly stopped in
the thermal bath and hence it is difficult to estimate its effects on
the inter-conversion without making some assumptions.  In fact, in
Refs.~\cite{Reno:1987qw,Kohri:2001jx}, effects of the inter-conversion
by the kaons are studied with some assumptions and approximations.  We
have followed the method given in those references and estimated the
effects of the kaons.  Then, we have found that, with the procedure
given in Refs.~\cite{Reno:1987qw,Kohri:2001jx}, our resultant
constraint (i.e., upper bound on the parameter $m_XY_X$) from the over
production of ${\rm ^4He}$ becomes slightly severer.  In our analysis,
however, we neglect the inter-conversion effects of the kaons since
the inclusion of the kaons introduce some uncontrollable
uncertainties.  Effects of the kaons are expected to make the
constraints severer and hence, in order to derive conservative
constraints, our treatment of the kaons is justified.

\subsection{Formulation}
\label{subsec:formulation}

In this subsection, we formulate the time evolution equations with the
late-time ``meson'' injection.  As we have mentioned, the hadron
injection at the beginning of the BBN epoch enhances the
inter-converting reactions between neutron and proton, and the
freeze-out value of $n/p$ may be increased. The inter-conversion term
in the Boltzmann equations for the number density of the nucleon
$N(=p,n$) is written as
\begin{equation}
    \label{eq:difeqN}
    \left[\frac{dn_N}{dt}\right]_{\rm IC} 
    = 
    - B_h \Gamma_X n_X \sum_{N'}
    \left(
    K_{N \rightarrow N'} - K_{N' \rightarrow N} \right),
\end{equation}
where $K_{N \rightarrow N'}$ denotes the averaged number of the
transition $N \rightarrow N'$ per one hadronic decay of $X$.

The averaged number of the transition $N\rightarrow N'$ for one
hadronic decay of $X$ is expressed as
\begin{equation}
    \label{eq:Knn}
    K_{N \rightarrow N'} = 
    \sum_{H_i} N^{H_i}
    P^{H_i}_{N \rightarrow N'},
\end{equation}
where $H_i$ runs over the hadronic species which are relevant to the
nucleon inter-converting reactions (i.e., $H_{i}=\pi^{+}$, $\pi^{-}$,
$n\bar{n}$ and $p\bar{p}$).  In addition, $N^{H_i}$ is the averaged
number of $H_i$ produced by one hadronic decay of $X$.  Note that
we assume that two hadronic jets are produced in one decay of $X$ in
this section.  $N^{H_i}$ is shown in Fig.~\ref{fig:ee_nch2002_large}
as a function of $2 E_{\rm jet}$, where $E_{\rm jet}$ is the energy of
one hadronic jet.~\footnote{
Here, we only consider the effects of the ``mesons'' ($\pi^{+}$,
$\pi^{-}$, $n\bar{n}$ and $p\bar{p}$) directly produced by the decay
of $X$.  Notice that, for the period when the inter-conversion becomes
important, the background temperature is so high that the hadrons we
are interested in are stopped before scattering off the background
nuclei.  Thus, the hadronic shower does not occur in this case, and we
do not have to consider the secondary production of the mesons and
baryons.}
Furthermore, $P^{H_i}_{N \rightarrow N'}$ denotes the probability that
a hadron species $H_i$ induces the nucleon transition $N \rightarrow
N'$ and is represented by
\begin{equation}
    \label{eq:trans_prob}
    P^{H_i}_{N \rightarrow N'} =
    \frac{\Gamma^{H_i}_{N \rightarrow N'}}{\Gamma^{H_i}_{\rm dec} +
    \Gamma^{H_i}_{\rm abs}},
\end{equation}
where $\Gamma^{H_i}_{\rm dec}$ is the decay rate of $H_i$, and
$\Gamma^{H_i}_{\rm abs}\equiv\Gamma^{H_i}_{p\rightarrow
p}+\Gamma^{H_i}_{p\rightarrow n}+\Gamma^{H_i}_{n\rightarrow
p}+\Gamma^{H_i}_{n\rightarrow n}$ is the total absorption rate.

\section{Hadrodissociation of Background $\alpha$}
\label{sec:hadrodis}
\setcounter{equation}{0}

\subsection{Basic equations}

As was discussed in Section~\ref{sec:hadonicdecay}, when $R^{H_i}$ is
more than unity, the hadronic scattering processes between the emitted
high-energy nucleons and the background proton or $\hefour$ (called
$\alpha_{\rm BG}$) become effective.  In particular, $\alpha_{\rm BG}$
can be destroyed and energetic nuclei, like D, T, $\hethree$ are
produced.  We call this type of the hadronic destruction
``hadrodissociation.''

\begin{figure}[t]
    \centering
    \centerline{{\vbox{\epsfxsize=0.8\textwidth
    \epsfbox{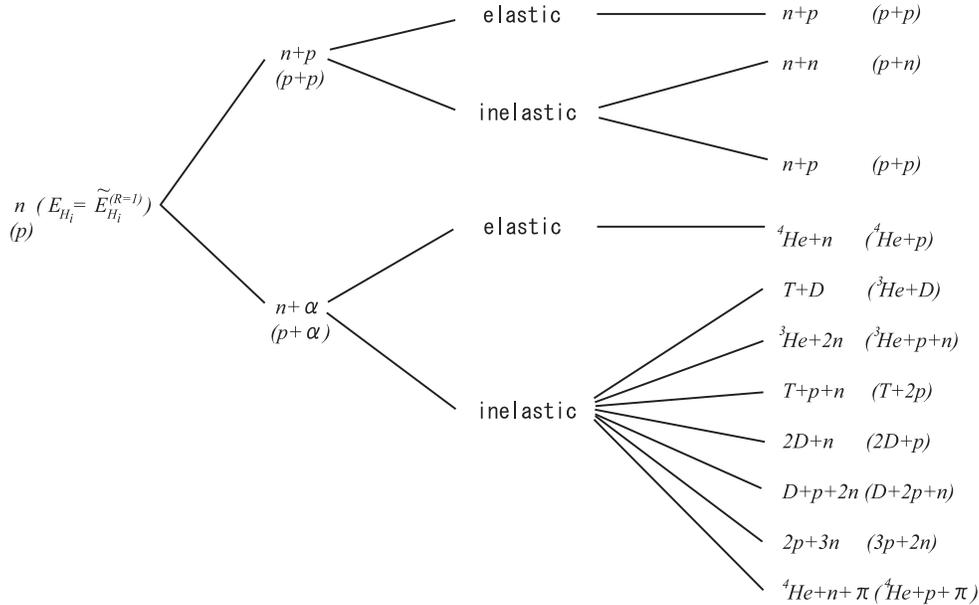}}}}
    \vspace{1cm}
    \caption{Schematic picture of hadron shower induced by 
    a high energy neutron (proton) which scatters off the background
    proton or the background $\alpha_{\rm BG}$.}
    \label{fig:hadron-showers}
\end{figure}

In order to study the effects of the hadrodissociation, we follow the
approach of Ref.~\cite{DimEsmHalSta} with several improvements.  In
Fig.~\ref{fig:hadron-showers}, we show the schematic picture of the
hadronic shower induced by a high-energy neutron and proton.
Hereafter, we discuss how we study the processes shown in
Fig.~\ref{fig:hadron-showers}.

Before going to the main discussion, however, let us comment on the
treatment of our high-energy anti-nucleons.  In our analysis, we
neglect hadrodissociation processes induced by high-energy
anti-protons and anti-neutrons since we do not have sufficient
experimental data for the scattering processes between an energetic
anti-nucleon and a nucleus.  Such anti-nucleons may change the
abundances of the light elements by dissociating background nuclei (as
well as by the inter-conversion effect which was discussed in Section\ 
\ref{sec:interconversion}).  We expect, however, that the resultant
constraints do not change much by this approximation since the
anti-nucleons are not produced secondarily in the hadronic shower.  Of
course, such energetic anti-nucleon directly produced by the decay of
$X$ may produce energetic hadrons by scattering off the background
nuclei, which may evolve into a hadronic shower.  Number of such a
process is at most the same as that of the hadronic shower induced by
the energetic $p$ and $n$ and hence the numbers of the
hadrodissociation processes may be underestimated at most by a factor
of two.  

Although the hadronic shower contains multiple scatterings of the
hadrons in the thermal bath, the evolution of the shower can be
followed by taking account of two types of elementally processes,
which are both discussed in the previous sections: one is the
electromagnetic processes through which hadrons gradually lose their
kinetic energy, and the other is the hadronic scatterings which change
the number of the hadronic particles.

Let us see what happens to a hadron injected into the thermal bath
with the initial energy $E_{H_i}^{\rm (in)}$ in more detail.  Such a
high energy hadron may be a direct decay product of $X$ or may be a
daughter particle produced in the hadronic showers.  As mentioned in
the previous section, once a high energy hadron is injected into the
thermal bath, it loses its energy down to
$\sim\tilde{E}^{(R=1)}_{H_i}$ defined in Eq.\ (\ref{R=1}).  In our
study, we approximate that the hadrons injected into the thermal bath
with the initial energy $E_{H_i}^{\rm (in)}$ scatters off the
background nuclei (i.e., $p_{\rm BG}$ and $\alpha_{\rm BG}$) with the
energy $\tilde{E}^{(R=1)}_{H_i}$ given above.

Since there are various hadronic processes, many possible final states
exist even if we fix the initial-state particles.  We specify the
individual processes by denoting $(i,j;k)$; here, $i$ and $j$ specify
the projectile and target nuclei, respectively, while $k$ is for the
final state.  Then, we approximate the probability at which the
projectile hadron $H_i$ scatters off a background nuclei via the
process $(i,j;k)$ as
\begin{eqnarray}
    \label{eq:p_or_alpha}
    P^{H_i}_{(i,j;k)} (E_{H_i}^{\rm (in)}; T) \equiv 
    \left[ \frac{n_{j}(T) \sigma_{(i,j;k)}}
    {\displaystyle{
    \sum_{m=p_{\rm BG},\alpha_{\rm BG}} \sum_l n_{m}(T) 
    \sigma_{(i,m;l)}} }
    \right]_{E_{H_i}=\tilde{E}^{(R=1)}_{H_i}},
\end{eqnarray}
where $n_{j}$ is the number density of the background nucleus $j$, and
$\sigma_{(i,j;k)}$ is the cross section for the process $(i,j;k)$ as a
function of $E_{H_i}^{\rm (in)}$.~\footnote{
Notice that the probability $P^{H_i}_{(i,j;k)}$ depends not only on
$E_{H_i}^{\rm (in)}$ and $T$ but also on the $^4$He mass fraction $Y$.
Such $Y$-dependence of $P^{H_i}_{(i,j;k)}$ is taken into account in
the numerical calculations.}

For each processes, we need to calculate the energy distribution of
the final-state nuclei.  The relevant final-state nuclei are $n$, $p$,
$\D$, $\T$, $\hethree$, and $\hefour$ in this
case.~\footnote{
In the hadronic scattering processes discussed so far, ${\rm Li}$ and
${\rm Be}$ are not produced.  Treatment of those nuclei will be
discussed in the next section.}
If we consider the scattering process $(i,j;k)$ in which projectile
hadron $H_i$ hits the target nucleon and produces the final-state
hadron $H_k$, the energy distribution of $H_k$ is given by
\begin{eqnarray}
    \label{eq:dist_nuc}
    f_{H_k}^{(i,j;k)}
    (E_{H_i}, E_{H_k})
    = 
    g_{H_k}^{(i,j;k)}
    \frac{1}{\sigma_{(i,j;k)}}
    \frac{d\sigma_{(i,j;k)}}{dE_{H_k}},
\end{eqnarray}
where $E_{H_i}$ here is the energy of $H_i$ at the time of the
scattering while $E_{H_k}$ is the energy of $H_k$ just after being
produced by the scattering process.  In addition, $g_{H_k}^{(i,j;k)}$
is the number of $H_k$ produced in the process $(i,j;k)$.  (For more
details, see Appendix \ref{sec:transfer_energy}.)  Using
$f_{H_k}^{(i,j;k)}$, we can also obtain the total energy distribution
of $H_k$ after the hadronic scattering of $H_i$, which is given by
\begin{eqnarray}
    \label{eq:summed_dist_nuc1}
    G_{H_i\rightarrow H_k} (E_{H_i}^{\rm (in)}, E_{H_k}; T)
    =
    \sum_{j=p,\alpha} \sum_k 
    P^{H_i}_{(i,j;k)}(E_{H_i}^{\rm (in)}; T) 
    f_{H_k}^{(i,j;k)} (\tilde{E}_{H_i}^{(R=1)}, E_{H_k}).
\end{eqnarray}
Notice that, if the initial-state particle has some energy
distribution $F_{H_i}(E_{H_i}; T)$ (where $E_{H_i}^{\rm (in)}$ is the
energy of $H_i$ just after being produced), then the distribution of
the final-state particle $H_k$ is obtained as
\begin{eqnarray}
    F'_{H_k} (E_{H_k}; T) = 
    \sum_{i} \int dE_{H_i}^{\rm (in)}
    F_{H_i}(E_{H_i}^{\rm (in)}; T)
    G_{H_i\rightarrow H_k} (E_{H_i}^{\rm (in)}, E_{H_k}; T).
\end{eqnarray}

With the relations given above, we can recursively follow the
evolution of the hadronic shower.  To make some image, let us consider
the hadronic shower induced by a primary energetic nucleon $H_i^{(0)}$
generated by the decay of $X$.  (To the primary nucleon, we assign the
generation number ``$0$.'')  We denote the initial energy of the
primary nucleon as $E_{H_i^{(0)}}^{\rm (in)}$.  As mentioned, emitted
nucleon loses its energy via the electromagnetic processes typically
down to $\tilde{E}^{(R=1)}_{H_i^{(0)}}$, which satisfies
$R^{H_i}(E_{H_i^{(0)}}^{\rm (in)},\tilde{E}^{(R=1)}_{H_i^{(0)}};T)=1$.
Then, $H_i^{(0)}$ scatters of the background nuclei ($\pbg$ or
$\alpha_{\rm BG}$) via the hadronic interactions.  We call this
``primary scattering.''  In our analysis, the primary scattering
occurs with the energy $\tilde{E}^{(R=1)}_{H_i^{(0)}}$, and we obtain
the energy distribution of the ``1st generation'' hadrons $H_k^{(1)}$
as $G_{H_i^{(0)}\rightarrow H_k^{(1)}}(E_{H_i^{(0)}}^{\rm
(in)},E_{H_k^{(1)}};T)$.  (Here and hereafter, the superscript for the
hadronic species are to identify their generation number in the
evolution of the hadronic shower.)  Thus, once the energy distribution
of the primary hadrons $F^{(0)}_{H_i}$ are known, we can calculate the
distribution of the first-generation hadrons by using the relation
\begin{eqnarray}
    F^{(1)}_{H_j} (E_{H_j}; T) = 
    \sum_{i} \int d E_{H_i}^{\rm (in)}
    F^{(0)}_{H_i} (E_{H_i}^{\rm (in)}; T)
    G_{H_i\rightarrow H_k} (E_{H_i}^{\rm (in)}, E_{H_j}; T).
\end{eqnarray}
In our study, the energy distributions of the primary (i.e., ``0-th
generation'') hadrons are calculated by using the JETSET 7.4 Monte
Carlo event generator.  Similarly, the distribution functions for the
$l$-th generation nuclei are recursively calculated by using the
following relation:
\begin{eqnarray}
    F^{(l)}_{H_k} (E_{H_k}; T) = 
    \sum_{i} \int d E_{H_i}
    F^{(l-1)}_{H_i} (E_{H_i}; T)
    G_{H_i\rightarrow H_k} (E_{H_i}, E_{H_k}; T),
    \label{eq:eq:summed_dist_1}
\end{eqnarray}
where $F^{(l)}_{H_j}$ is the distribution function of $H_k$ in $l$-th
generation.  After a large number of multiple scatterings, energy
distribution function of any hadrons for energy region above the
threshold energies of the hadrodissociation processes becomes
negligibly small and the hadrodissociation processes stop.

\subsection{Approximations}

Evolution of the hadronic shower can be in principle understood by the
recursive procedure discussed in the previous subsection.  In the
actual calculation, however, it is difficult to obtain the resultant
distributions of the shower particles without any simplification and
approximation.  One reason is that the number of hadrons contributing
to the hadronic shower is enormous so it is difficult to track all the
energy-loss processes of those hadrons.  In addition, for some of the
hadronic processes, experimental data for the cross sections are not
available.  Thus, in this subsection, we explain how we solve the
basic equations given in the previous subsection with some
simplifications and approximations.

Our primary purpose is to obtain {\sl conservative} constraints on the
properties of $X$.  By adopting reasonable experimental and
theoretical errors, the SBBN scenario predicts abundances of the light
elements consistent with the observations, as discussed in Section\ 
\ref{sec:obs_status}.  Thus, the non-standard processes usually make
the theoretical constraints inconsistent with the observations.  In
particular, if hadronic scattering processes with the
background $\alpha_{\rm BG}$ occurs too much, ${\rm D}$ and/or ${\rm
^3He}$ are overproduced.  In addition, non-thermal processes, which
will be discussed in the next section, may overproduce ${\rm ^6Li}$.

Importantly, for some processes, we do not have enough experimental
informations.  For such cases, we adopt some approximation or
assumption so that the numbers of ${\rm D}$, ${\rm ^3He}$, and ${\rm
^6Li}$ produced by the non-standard processes are minimized, resulting
in conservative constraints.  Thus, we should note that, for some
case, the resultant abundances of ${\rm D}$, ${\rm ^3He}$, and ${\rm
^6Li}$ obtained from our calculations are their lower bounds.

First of all, as we mentioned before, we simplify the treatment of the
target (background) nuclei by approximating that the energetic hadrons
scatter off only the background proton or $\alpha_{\rm BG}$.  This is
because most of the background nuclei are in the form of the proton or
${\rm ^4He}$.  Of course, some of the energetic hadrons may scatter
off other light elements in the background which may be destructed by
those processes.  If such processes are effective, however, production
of those light elements from the dissociation of the background
$\alpha_{\rm BG}$ is far more important since $\alpha_{\rm BG}$ is
more abundant than other light elements (except $p_{\rm BG}$).  Thus,
our constraints is not affected by our approximation on the target
particles of the hadronic processes.

Second simplification is that, among various hadrons generated by
hadronic scattering processes, only $p$ and $n$ are used as projectile
nuclei in the next-step hadronic process (except for the non-thermal
production processes of ${\rm Li}$ and ${\rm Be}$, which will be
discussed in the next section).  This is because most of the nuclei
produced in the shower processes are $p$ or $n$, and also because we
could not find sufficient experimental data for the cross sections for
other nuclei.  One might worry about the effects of the energetic
${\rm ^4He}$.  However, energetic ${\rm ^4He}$ is much rarer than $p$
or $n$ in the hadronic shower since the energy transfer to ${\rm
^4He}$ in the elastic $p\alpha_{\rm BG}$ scattering process is fairly
small.  Notice also that the cross sections for the inelastic
$p+\alpha_{\rm BG} \to p + \alpha + \cdots$ reactions are relatively
small.  Thus, the energetic ${\rm ^4He}$ has small effect on the
evolution of the hadronic shower.

\begin{table}[t]
    \begin{center}
        \begin{tabular}{lccc} 
            \hline\hline
            Process & $i = n$ & $i = p$ & reaction type 
            \\
            \hline
            $(i,p_{\rm BG};1)$ & $n + p_{\rm BG} \to n + p$ 
            &  $p + p_{\rm BG} \to p + p$ & elastic 
            \\
            $(i,p_{\rm BG};2)$ & $n + p_{\rm BG} \to n + p + \pi$ 
            &  $p + p_{\rm BG} \to p + p + \pi$ 
            & inelastic 
            \\ 
            $(i,p_{\rm BG};3)$ & $n + p_{\rm BG} \to n + n + \pi$ 
            &  $p + p_{\rm BG} \to p + n + \pi$ 
            & inelastic 
            \\
            \hline\hline
        \end{tabular}
        \caption{Hadronic processes with background proton $p_{\rm BG}$}
        \label{tab:bran_process1}
    \end{center}
\end{table}

\begin{table}[t]
    \begin{center}
        \begin{tabular}{lccc} 
            \hline\hline
            Process & $i = n$ & $i = p$ & reaction type \\
            \hline
            $(i,\alpha;1)$ & $n + \alpha_{\rm BG} \to n + \alpha$ 
            &  $p + \alpha_{\rm BG} \to p + \alpha$  & elastic  \\
            $(i,\alpha;2)$ & $n + \alpha_{\rm BG} \to \D + \T$ 
            &  $p + \alpha_{\rm BG} \to \D + \hethree$ & inelastic \\
            $(i,\alpha;3)$ & $n + \alpha_{\rm BG} \to 2n  + \hethree$ 
            &  $p + \alpha_{\rm BG} \to p  + n  + \hethree$ & inelastic \\
            $(i,\alpha;4)$ & $n + \alpha \to  p +  n + \T$ &  $p +
            \alpha \to  2p  + \T$ 
            & inelastic \\ 
            $(i,\alpha;5)$ & $n + \alpha_{\rm BG} \to  n  + 2 \D$ 
            &  $p + \alpha_{\rm BG} \to p + 2 \D$ & inelastic  \\
            $(i,\alpha;6)$ & $n + \alpha_{\rm BG} \to   p + 2 n  + \D $ 
            &  $p + \alpha_{\rm BG} \to 2p + n + \D$ & inelastic \\ 
            $(i,\alpha;7)$ & $n + \alpha_{\rm BG} \to 2p +  3n$ 
            &  $p + \alpha_{\rm BG} \to 3p + 2n$ & inelastic \\
            $(i,\alpha;8)$ & $n + \alpha_{\rm BG} \to  n  + \alpha + \pi$ 
            &  $p + \alpha_{\rm BG} \to p  + \alpha  + \pi$ & inelastic 
            \\
            \hline\hline
        \end{tabular}
        \caption{Hadronic processes with background $\alpha_{\rm BG}$}
        \label{tab:bran_process2}
    \end{center}
\end{table}

The hadronic scattering processes considered in our study are listed in
Tables \ref{tab:bran_process1} and \ref{tab:bran_process2}.  The
experimental data of the cross sections are summarized in
Refs.~\cite{Hagiwara:fs,meyer:1972}.  In addition, in our Monte Carlo
analysis, we adopt 20\ \% errors to all the hadronic cross sections.

In some case, we could not find sufficient experimental data and adopt
some reasonable approximations or assumptions.  In particular, the
hadronic cross sections for the energy of the projectile higher than
$\sim 20$ GeV cannot be found except for some $pp$ and $np$ reactions.
Fortunately, according to the existing data, however, the cross
sections for the $pp$ and $np$ reactions are known to become
approximately constant at high-energy region \cite{Hagiwara:fs}.
Thus, we assume that the inelastic cross section for the $p\alpha$
process is constant above $E>20\ {\rm GeV}$.  Our results are
insensitive to this assumption since the evolution of the hadronic
shower is mostly determined by the hadrons with energy less than $\sim
\order(1)$ GeV.  In addition, sufficient experimental data are not
available for the $n\alpha$ reactions.  For these processes, we use
the SU(2) isospin symmetry and use the cross sections of $p\alpha$
reactions for the $n\alpha$ reactions.  Those cross section differ due
to the Coulomb corrections.  Using the familiar formula of the Coulomb
correction factor \cite{Schiff}, however, the Coulomb correction is
estimated to be less than a few percent for the projectile energy
larger than the threshold energy for the inelastic $n\alpha$
scattering process ($\sim$ 25 MeV).  Thus, we neglect such a Coulomb
correction.

In addition, the experimental data of the hadronic scattering
processes for other processes are also insufficient.  Thus, we make
the following approximations for the daughter nuclei ${\rm D}$, ${\rm
T}$, ${\rm ^3He}$, and ${\rm ^4He}$.
\begin{itemize}
\item[(i)] In considering the hadronic process, the energetic daughter
    nuclei scatter off only the background proton and $\alpha_{\rm
    BG}$.
\item[(ii)] The daughter nucleus $A_k$ is assumed to survive only if
    (a) its typical energy just before the hadronic scattering (i.e.,
    $\tilde{E}^{(R=1)}_{A_k}$) is smaller than the threshold energy
    for the dissociation of $A_k$ by scattering off the background
    proton, and (b) typical energy of the background photon in the
    center-of-mass frame is smaller than the threshold energy for the
    photodissociation process of $A_k$.  (In fact, the second
    condition is not crucial; the resultant constraints on $X$ do not
    change much even if we do not include the condition (b).)  That
    is, the surviving probability of the daughter particle $A_k (={\rm
    D}, {\rm T}, {\rm ^3He}\ {\rm or}\ {\rm ^4He})$ is simply given by
    \begin{eqnarray}
        P_{A_k\rightarrow A_k} (E_{A_k}; T) = 
        \left\{ \begin{array}{ll}
                0 & ~~:~~
                \tilde{E}^{(R=1)}_{A_k} > E^{({\rm th},p)}_{A_k}
                {\rm \ or\ }
                \sqrt{3T E_{A_k}} > Q_{A_k}
                \\
                1 & ~~:~~
                {\rm otherwise}
            \end{array} \right. ,
    \end{eqnarray}
    where $E^{({\rm th},p)}_{A_k}$ is the threshold energy for the
    hadrodissociation process, while $Q_{A_k}$ is the binding energy
    of $A_k$.  
\item[(iii)] The daughter nucleus $A_k$ is completely destroyed into
    energetic nucleons if $P_{A_k\rightarrow A_k}=0$.  
   (For e.g., $\hefour+\pbg\to 3p+2n$.)
\end{itemize}
The approximation (i) is quite reasonable since almost all the baryons
in the universe at the epoch we are interested are in the form of the
proton or $\alpha_{\rm BG}$.  The assumptions (ii) and (iii) are
justified since our purpose is to obtain conservative constraints;
indeed, the numbers of the non-thermally produced $\D$, $\T$,
$\hethree$ and $\hefour$ are underestimated with these approximations
since the energetic nucleons produced by the dissociation of the light
elements rarely dissociate $\alpha_{\rm BG}$ to produce other light
elements.

Since the target particle is $p_{\rm BG}$ or $\alpha_{\rm BG}$, we can
rewrite $P^{N_i}_{(i,j;k)}$ given in Eq.\ (\ref{eq:p_or_alpha}) as
\begin{eqnarray}
    \label{eq:p_or_alpha2}
    P^{N_i}_{(i,j;k)} (E_{N_i}^{\rm (in)}; T) =
    \frac{n_{A_j}(T) \sigma_{(i,j;k)}(\tilde{E}^{(R=1)}_{N_i})}
    {n_{p}(T)\sigma_{N_ip}^{\rm (tot)}(\tilde{E}^{(R=1)}_{N_i}) +
    n_{\alpha}(T)
    \sigma_{N_i\alpha}^{\rm (tot)}(\tilde{E}^{(R=1)}_{N_i})},
\end{eqnarray}
where $\sigma_{N_ip}^{\rm (tot)}$ and $\sigma_{N_i\alpha}^{\rm (tot)}$
are total cross sections for the $N_ip_{\rm BG}$ and $N_i\alpha_{\rm
BG}$ scattering processes, respectively.  In our numerical
calculations, distributions of the final-state hadrons (in particular,
those of the light-elements) are calculated with this formula.

Since most of the final-state energetic hadrons are proton and neutron
(as well as light mesons), we adopt an approximation such that only
$p$ and $n$ are used as the initial-state energetic particles which
cause hadrodissociations of $\alpha_{\rm BG}$. With such
approximation, it is rather convenient to assign the generation
numbers only to $p$ and $n$ by ``integrating out'' effects of other
nuclei; using the distribution given in Eq.\ 
(\ref{eq:summed_dist_nuc1}), we define the distribution of the $p$ and
$n$ after taking account of the effects of other light elements as
\begin{eqnarray}
    \tilde{G}_{N\rightarrow N'} (E_N^{\rm (in)}, E_{N'}; T)
    &=&
    G_{N\rightarrow N'} (E_N^{\rm (in)}, E_{N'}; T)
    \nonumber \\ &&
    + \sum_{A_k \ne p,n}
    \int d E_{A_k}
    G_{N\rightarrow A_k} (E_N^{\rm (in)}, E_{A_k}; T)
    P^{A_k}_{(A_k+\cdots\rightarrow N'+\cdots)} (E_{A_k}; T)
    \nonumber \\ &&
    g_{N'}^{(A_k+\cdots\rightarrow N'+\cdots)} 
    (\tilde{E}^{(R=1)}_{A_k}, E_{N'}),
    \label{tilde-G}
\end{eqnarray}
where $N$ and $N'$ correspond to $p$ and $n$, and
$g_{N'}^{(A_k+\cdots\rightarrow N'+\cdots)}(\tilde{E}^{(R=1)}_{A_k},
E_{N'})$ is the energy distribution of $N'$ in the reaction
$A_k+\cdots\rightarrow N'+\cdots$ with the energy of $A_k$ being
$\tilde{E}^{(R=1)}_{A_k}$.  Although we have included the second term
in Eq.\ (\ref{tilde-G}), which take account of the effects of the
nucleons generated from the secondary destruction of the non-thermally
produced light elements, it is much smaller than the first term.
Indeed, for one hadronic decay of $X$, numbers of $p$ and $n$ produced
in the shower processes are of ${\cal O}(10-100)$ while the numbers of
the destructed light elements are a few or less.  Thus, our result is
in fact insensitive to the approximation (iii) mentioned above.
Furthermore, for other light elements $A_j={\rm D}$, ${\rm T}$, ${\rm
^3He}$, and ${\rm ^4He}$, we define
\begin{eqnarray}
    \label{eq:summed_dist_nuc1.5}
    \tilde{G}_{N\rightarrow A_j} (E_N^{\rm (in)}, E_{A_j}; T)
    = G_{N\rightarrow A_j} (E_N^{\rm (in)}, E_{A_j}; T)
    P_{A_j\rightarrow A_j} (E_{A_j}; T).
\end{eqnarray}

Then, we define the distribution function of the light element $A_j$
after the hadronic scattering of $l$-th generation nucleons, which we
denote $\tilde{F}_{A_j}$.  Notice that $\tilde{F}_{A_j}$ obeys the
recursion relation similar to Eq.\ (\ref{eq:eq:summed_dist_1}):
\begin{eqnarray}
    \tilde{F}^{(l)}_{A_j} (E_{A_j}; T) = 
    \sum_{N=p,n} \int d E_N
    \tilde{F}^{(l-1)}_N (E_N; T)
    \tilde{G}_{N\rightarrow {A_j}} (E_N, E_{A_j}; T),
\end{eqnarray}
where, in the above expression, $A_j$ denotes all the possible light
elements.

In our numerical analysis, scattering and energy-loss processes of the
energetic nuclei are studied by using $\tilde{G}_{N\rightarrow N'}$
and $\tilde{G}_{A_i\rightarrow A_j}$ defined above.  (For details, see
Appendix\ \ref{app:numerical}.)

With the distribution function given above, we calculate the number of
produced or destructed nuclei by the decay of $X$.  In our
calculation, target of the energetic hadrons is $p_{\rm BG}$ or
$\alpha_{\rm BG}$.  In this case, numbers of the $h=\D$, $\T$ and
$\hethree$ are always increased by the decay of $X$ (as far as we
neglect the subsequent thermal processes).  Then, with the
distribution function, we calculate the total number of the nuclei
$A_j=\D$, $\T$ and $\hethree$ produced by the hadronic decay of one
$X$:
\begin{eqnarray}
    \label{eq:def_xi}
    \xi_{A_j} (T) &=& 
    \sum_{l=1}^{\infty} \int d E_{A_j}
    \tilde{F}^{(l)}_{A_j} (E_{A_j}; T)
    \nonumber \\ &=&
    \int d E_{A_j} \int d E_N \sum_{N=n,p}
    \tilde{S}_N (E_N; T)
    \tilde{G}_{N\rightarrow {A_j}} (E_N, A_j; T).
\end{eqnarray}
where 
\begin{eqnarray}
    \tilde{S}_N (E_N; T) \equiv 
    \sum_{l=0}^{\infty}\tilde{F}^{(l)}_N (E_N; T).
\end{eqnarray}
We also calculate the total number of destroyed $\hefour$ as
\begin{eqnarray}
    \label{eq:xi-alpha}
    \xi_{\alpha} (T) &=&
    \sum_{N=n,p}
    \int d E_N
    \tilde{S}_N (E_N; T)
    \nonumber \\ &&
    \left[
        \sum_{k=2}^{7} P^N_{(N,\alpha;k)} (E_N; T)
        + \int dE_{\rm ^4He}
        \tilde{G}_{N\rightarrow {\rm ^4He}} (E_N, E_{\rm ^4He}; T)
        \left\{ 1-
            P_{\rm ^4He \rightarrow ^4He} (E_{\rm ^4He}; T)
        \right\}
    \right].
    \nonumber \\
\end{eqnarray}
We found that the hadrodissociation of the $\hefour$ is dominated by
the direct destruction in the hadronic process.  Thus, although we
have included the effects of the secondary destruction of the
$\hefour$ which is made energetic via the first hadronic scattering
processes (i.e., the second term in Eq.\ (\ref{eq:xi-alpha})), such
effect is subdominant and does not significantly change the
constraints.

We should also consider the effects of the low-energy neutrons
produced in the hadronic showers.  Such neutrons may be energetic when
they are produced, but they lose their energy as they propagate in the
thermal bath.  (The energetic neutrons mainly scatter off the
background $p$ and $\alpha_{\rm BG}$.)  Once the energy of the neutron
becomes lower than the threshold energy of the destruction processes
of the background $\alpha_{\rm BG}$ ($E_{n\alpha}^{\rm (th)}\sim
25~{\rm MeV}$), it no longer destruct the background $\alpha_{\rm
BG}$.  However, even after being thermalized, such extra-produced $n$
may affect the abundances of the light elements.  In particular, $p$
may capture such low-energy $n$ and $D$ may be produced.  Furthermore,
${\rm ^7Be}$ may be dissociated by the thermal neutron via the process
${\rm ^7Be}(n,{\rm ^3He}) {\rm ^4He}$, which reduces the resultant
abundance of ${\rm ^7Li}$.  These processes are included in our BBN
code.  Importantly, at high enough temperature, the neutron with
energy lower than $E_{n\alpha}^{\rm (th)}$ does not decay before being
thermalized since its lifetime is much longer than the thermalization
time.  Thus, effects of the low-energy neutron is taken into account
by injecting thermal neutron into the thermal bath.  The number of the
neutrons produced by the hadronic decay of $X$ as
\begin{eqnarray}
    \label{eq:xi_neutron}
    \xi_{n} (T) = 
    \int
    d E_n \tilde{F}^{(\infty)}_n (E_n; T).
\end{eqnarray}

\begin{figure}
    \centering
    \centerline{{\vbox{\epsfxsize=8.0cm
    \epsfbox{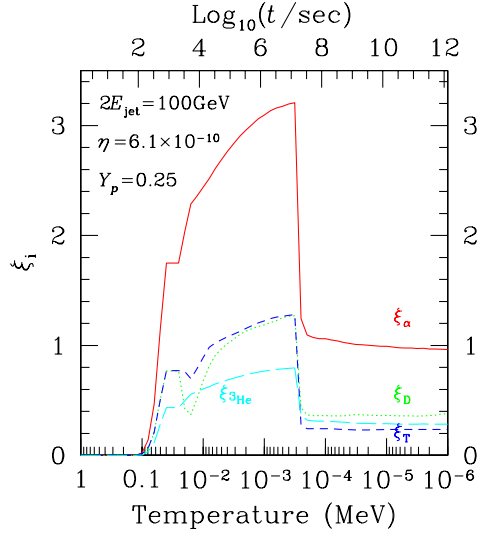}}}}
    \caption{Number of produced or destructed hadrons per one 
    hadronic decay of $X$ with $m_X=100\ {\rm GeV}$.  $\xi_{\alpha}$
    (solid line) is the number of the destructed ${\rm ^4He}$ while
    $\xi_{\D}$ (dotted line), $\xi_{\T}$ (dashed line), and
    $\xi_{\hethree}$ (long dashed line) are the number of $\D$, $\T$
    and $\hethree$ produced by $X$-decay, respectively.  We take
    $Y_{p}=0.25$ $\eta=6.1\times 10^{-10}$, and $2E_{\rm jet}=m_X$.}
    \label{fig:prod_m100gev}
\end{figure}

\begin{figure}
    \centering
    \centerline{{\vbox{\epsfxsize=8.0cm
    \epsfbox{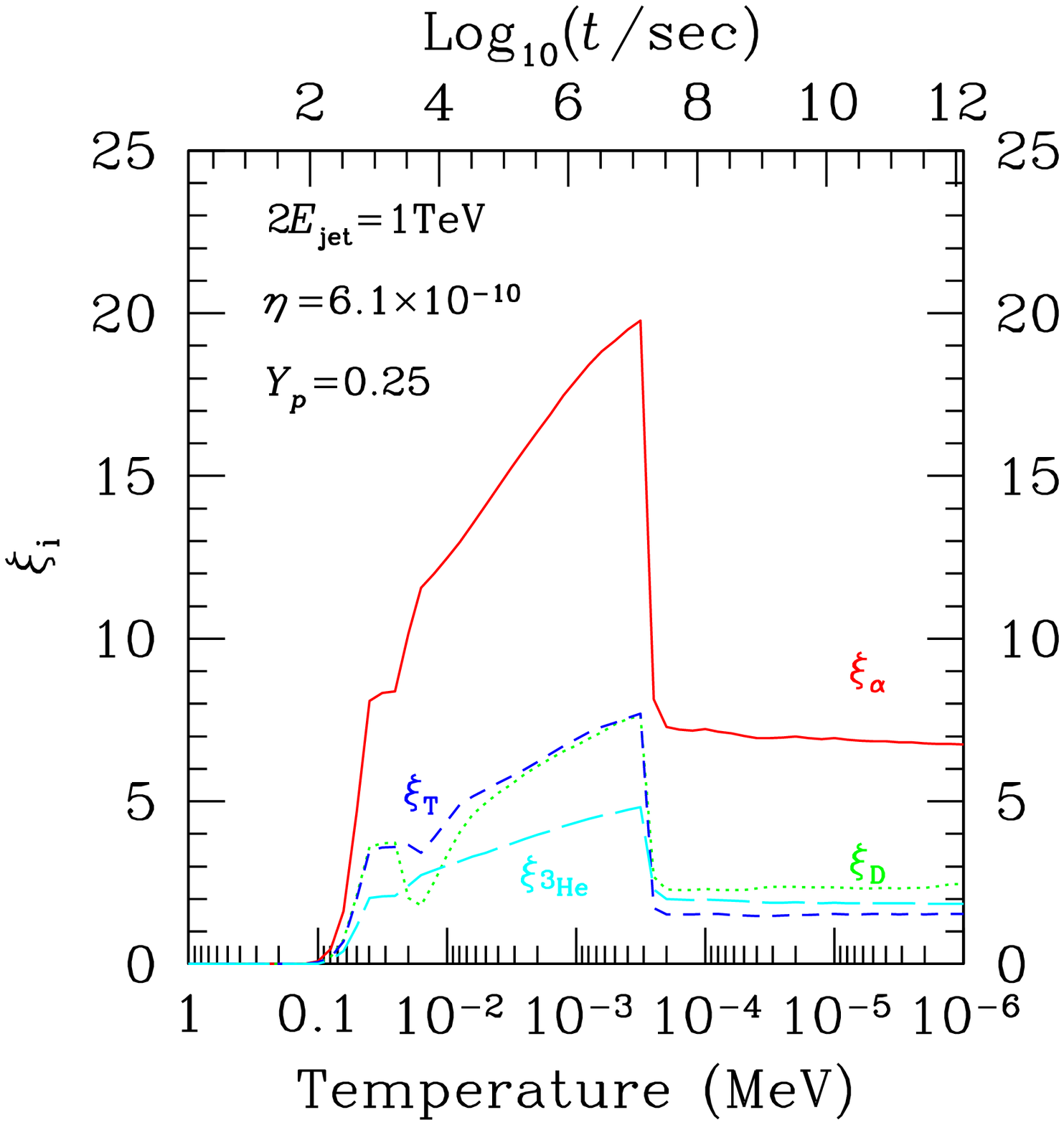}}}}
    \caption{Same as Fig.\ \ref{fig:prod_m100gev}, except for
    $m_X=1\ {\rm TeV}$.}
    \label{fig:prod_m1tev}
\end{figure}

\begin{figure}
    \centering
    \centerline{{\vbox{\epsfxsize=8.0cm
    \epsfbox{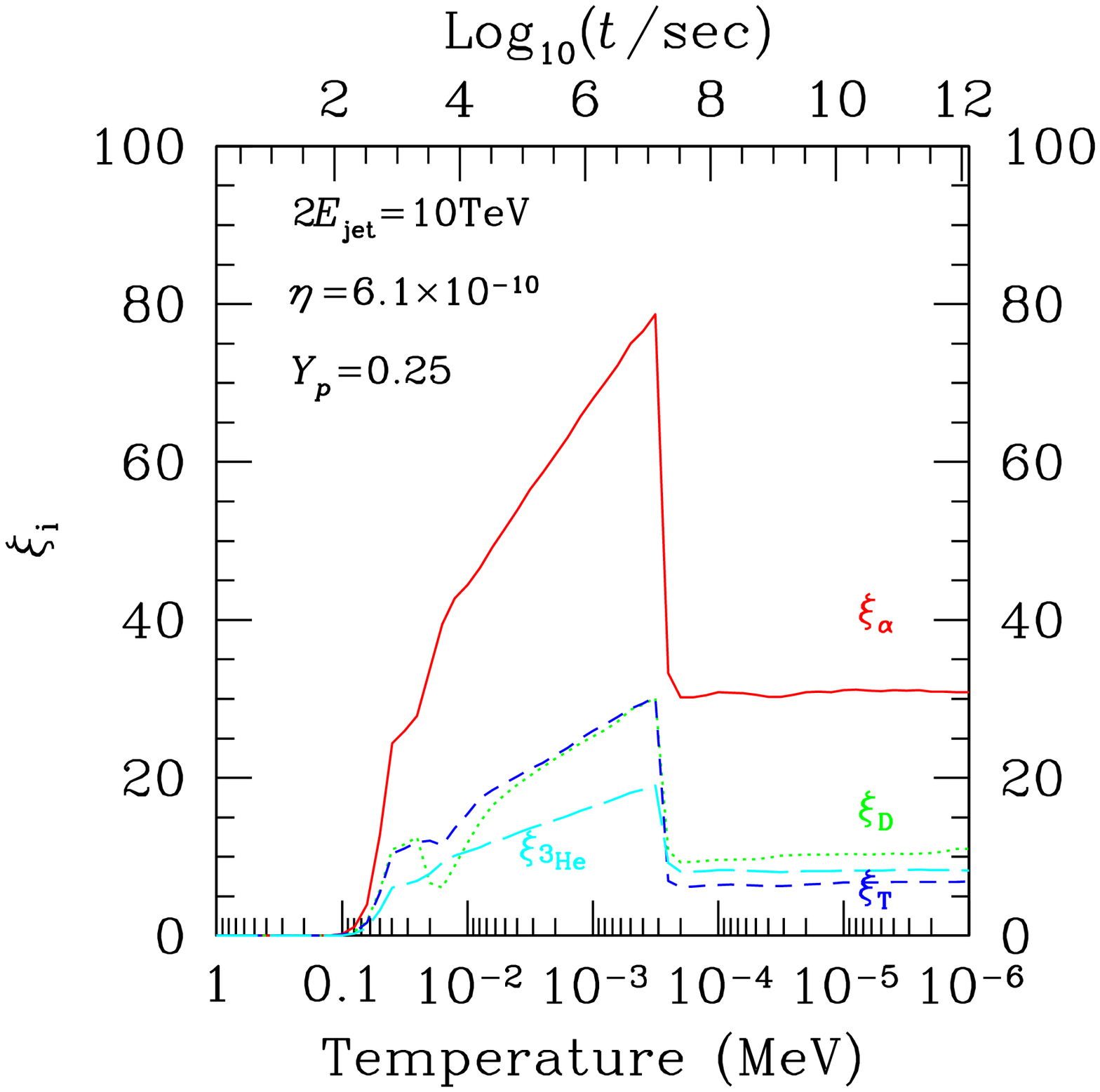}}}}
    \caption{Same as Fig.\ \ref{fig:prod_m100gev}, except for
    $m_X=10\ {\rm TeV}$.}
    \label{fig:prod_m10tev}
\end{figure}

In Figs.\ \ref{fig:prod_m100gev}, \ref{fig:prod_m1tev} and
\ref{fig:prod_m10tev}, we plot $\xi_{\D}$, $\xi_{\T}$,
$\xi_{\hethree}$ and $\xi_{\alpha}$ as functions of the temperature
$T$ for $m_X=100\ {\rm GeV}$, 1 TeV and 10 TeV.  As one can see, the
$\xi$-parameters almost vanishes at $T\gtrsim 0.1\ {\rm MeV}$.  This
is because, for such high temperature, energetic hadrons are stopped
by the electromagnetic processes before scattering off the background
nuclei.
As the temperature becomes lower, the $\xi$-parameters increases until
$T\sim 0.3 {\rm keV}$.  In this period, the hadrodissociation is
dominated by the energetic neutron since the mean-free-path of the
neutron is much longer than that of proton.  Energy-loss of the
neutron becomes less efficient as the temperature becomes lower, so
the effects of the hadrodissociations become more effective as the
temperature becomes lower.  Once the cosmic temperature becomes lower
than $T\sim 0.3 {\rm keV}$, however, the neutron decays before
scattering off $\alpha_{\rm BG}$.  Since the stopping process of the
proton is more efficient than that of the neutron, hadrodissociation
is suppressed at the low enough temperature.  Thus, we see sharp
drop-off of the $\xi$-parameters at $T\sim 0.3 {\rm keV}$.

\begin{figure}
    \centering
    \centerline{{\vbox{\epsfxsize=8.0cm
    \epsfbox{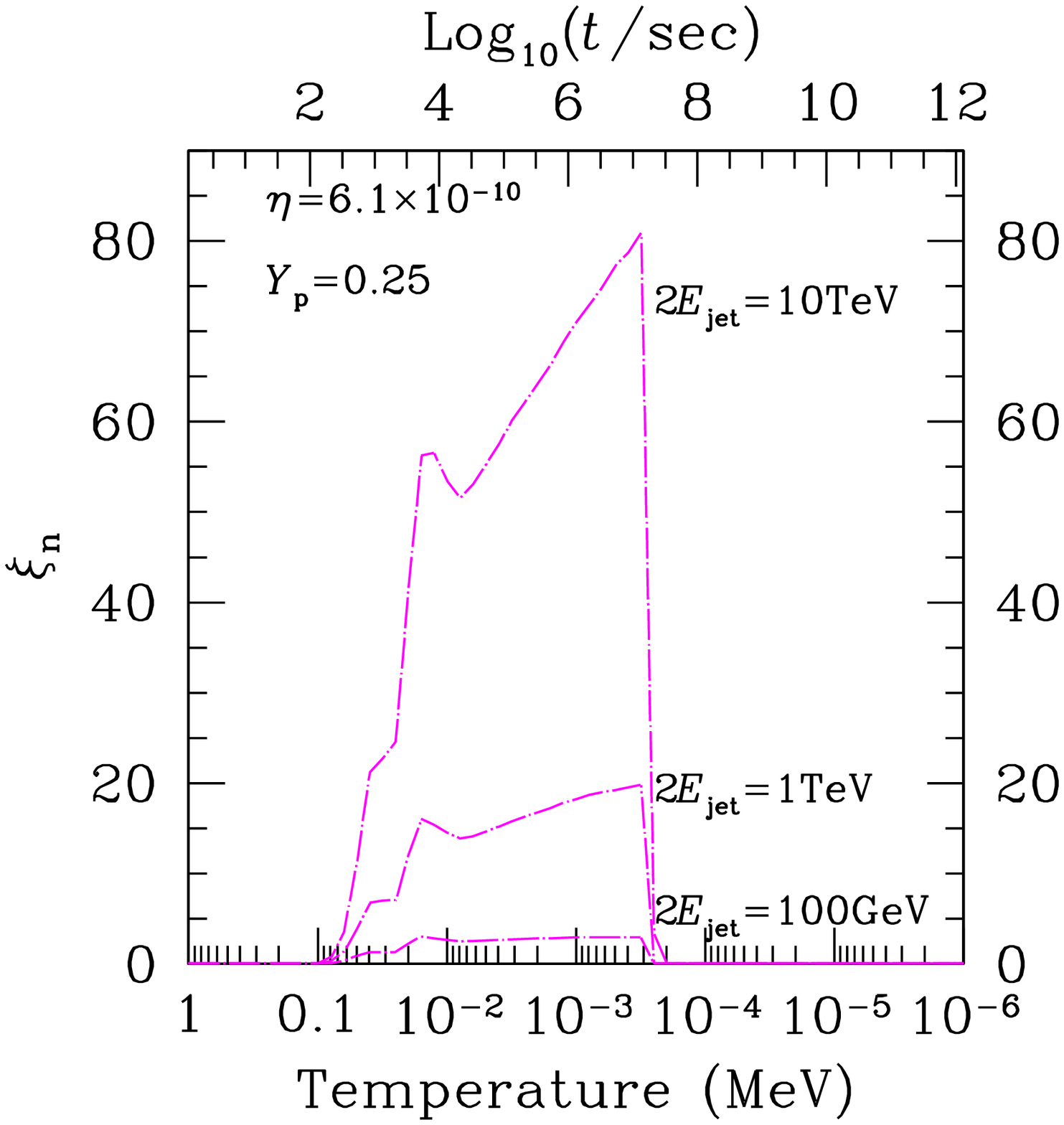}}}}
    \caption{$\xi_n$ as a function of the temperature.  Here, we take
    $Y_{p}=0.25$ and $\eta=6.1\times 10^{-10}$,  the total
    energy of the two hadronic jets is $2E_{\rm jet}=100\ {\rm GeV}$,
    $1\ {\rm TeV}$ and $10\ {\rm TeV}$.}
    \label{fig:prodneutron}
\end{figure}

We also plot $\xi_{n}$ in Fig.~\ref{fig:prodneutron}.  Note that, in
Fig.~\ref{fig:prodneutron}, we subtract the number of neutrons which
are contained in the initial spectrum of neutrons $\tilde{F}^{(0)}_n$
from $\xi_{n}$ in order to show the number of the secondary produced
neutrons.  The drastic decrease at $T\sim 0.3 {\rm keV}$
is, again, due to the neutron decay.

\section{Non-thermal Production of Lithium and Beryllium}
\label{sec:ntlibe}
\setcounter{equation}{0}

In this section, we discuss the non-thermal production processes of
${\rm Li}$ and ${\rm Be}$.  As we have discussed in the previous
sections, energetic ${\rm T}$, ${\rm ^3He}$ and ${\rm ^4He}$ can be
produced by the hadronic or photodissociation processes with the
background $\alpha_{\rm BG}$.  Such energetic nuclei may scatter off
the background $\alpha_{\rm BG}$ again and produce other nuclei, in
particular, $\lisix$, $\liseven$, and $\beseven$.  Although these
collisions are not so frequent, they are important since the
observations severely constrain the primordial abundances of $\lisix$
and $\liseven$.

First, we consider the non-thermal production of ${\rm ^6Li}$ by the
energetic ${\rm T}$ and ${\rm ^3He}$.  In this case, energetic ${\rm
T}$ and ${\rm ^3He}$ are produced by
\begin{eqnarray}
    \label{eq:energeticT}
    p (n) + \alpha_{\rm BG} \rightarrow
        \left\{ 
        \begin{array}{l}
            {\rm T} + \cdots \\
            {^3{\rm He}} + \cdots
        \end{array}
    \right. ,
\end{eqnarray}
and these $\T$ and $\hethree$ scatter off the $\alpha_{\rm BG}$ to
produce ${\rm ^6Li}$:
\begin{eqnarray}
    \label{eq:T_alpha}
    \T + \alpha_{\rm BG} &\to& \lisix + n, \\
    \label{eq:he3_alpha}
    \hethree + \alpha_{\rm BG} &\to& \lisix + p.
\end{eqnarray}

\begin{figure}
    \centering
    \centerline{{\vbox{\epsfxsize=8.0cm\epsfbox{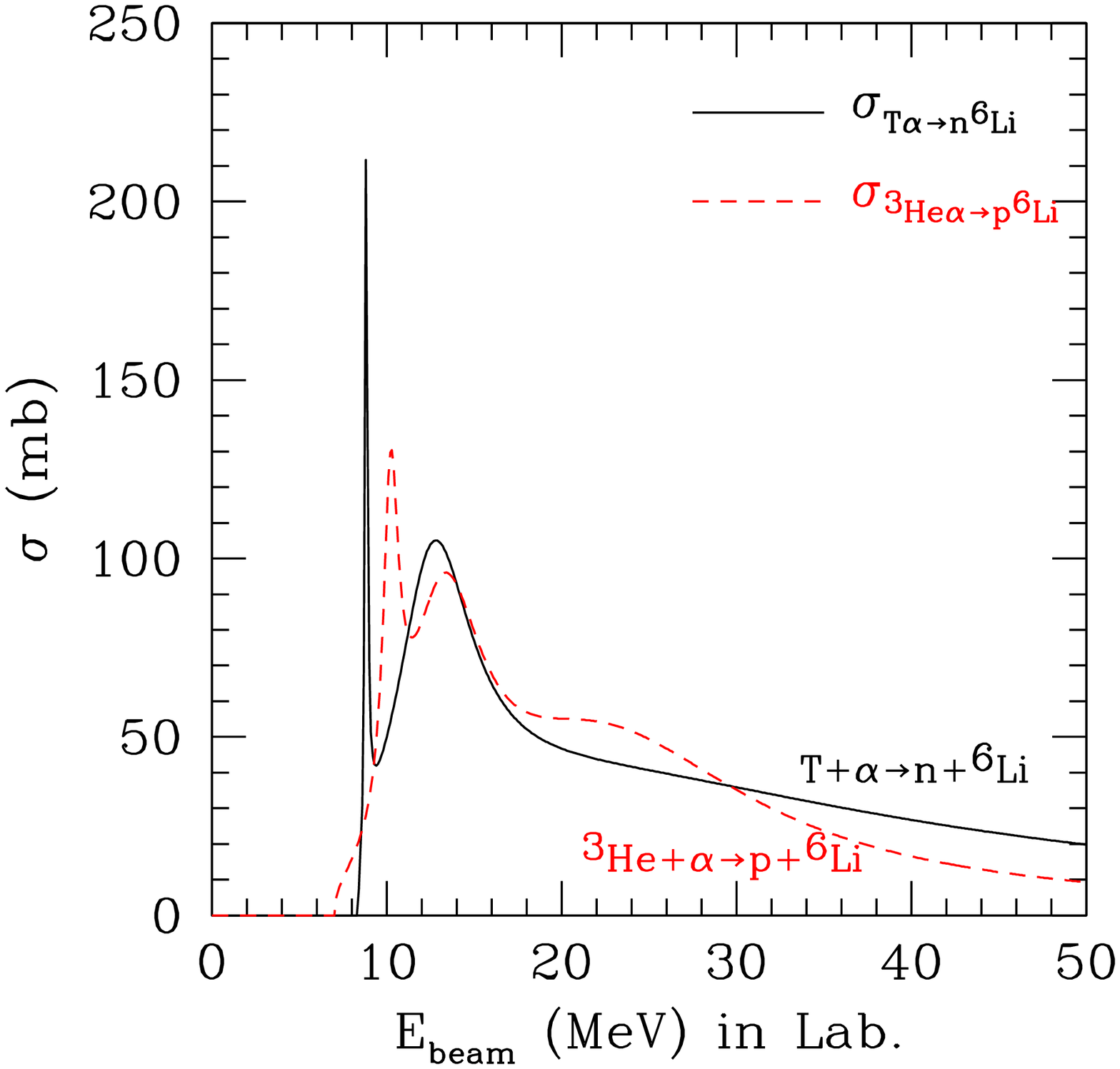}}}}
    \caption{The cross sections of the $\lisix$ production
    processes. The solid (dashed) line is for $\T+\hefour\to\lisix +n$
    ($\hethree+\hefour\to\lisix + p$).}
    \label{fig:sigli6}
\end{figure}

Once the energetic nucleus $A_i={\rm T}$ or ${\rm ^3He}$ is injected
into the thermal bath, it loses its energy via the electromagnetic
interactions by scattering off the background electron and photon
while it also scatters off the background $\alpha_{\rm BG}$. With the
energy-loss rate $(dE_{A_i}/dt)$ given in the previous section, number
of ${\rm ^6Li}$ produced by one $A_i$ is
\begin{eqnarray}
    \label{eq:rate_of_li6prod}
    \int_{\tilde{E}^{(R=1)}_{A_i}}^{E_{A_i}^{\rm (in)}} 
    d E_{A_i} \left( \frac{dE_{A_i}}{dt} \right)^{-1}
    n_{\alpha} \sigma_{A_i+\alpha_{\rm BG}\to\lisix+\cdots} 
    (E_{A_i}) \beta_{A_i},
\end{eqnarray}
where $E_{A_i}^{\rm (in)}$ is the initial energy of $A_i$,
$\beta_{A_i}$ is the velocity of $A_i$, and $\tilde{E}^{(R=1)}_{A_i}$
is the typical energy of $A_i$ just before its collision with
background proton or $\alpha_{\rm BG}$ (see
Section~\ref{sec:hadonicdecay}).  In addition,
$\sigma_{A_i+\alpha_{\rm BG}\to\lisix+\cdots}$ is the cross section of
the process (\ref{eq:T_alpha}) or (\ref{eq:he3_alpha}). (We plot the
experimental data of them in Fig.\ \ref{fig:sigli6}
\cite{koepke_brown,EXFOR,Cyburt:2002uv}.)  Summing up the
contributions of ${\rm T}$ and ${\rm ^3He}$, the number of ${\rm
^6Li}$ produced by the above process per one hadronic decay of $X$ is
given by
\begin{eqnarray}
    \label{eq:xi_Li6}
    \xi_{\lisix}^{\rm (T, ^3He)} &=& \sum_{A_i=\T,\hethree}
    \int_{0}^{\infty} d E_{A_i}^{\rm (in)}
    f_{A_i} (E_{A_i}^{\rm (in)})
    \nonumber \\ && 
    \int_{\tilde{E}^{(R=1)}_{A_i}}^{E_{A_i}^{\rm (in)}} 
    d E_{A_i} \left( \frac{dE_{A_i}}{dt} \right)^{-1}
    n_{\alpha} \sigma_{A_i+\alpha_{\rm BG}\to\lisix+\cdots} (E_{A_i})
    \beta_{A_i} P_{\rm ^6Li\rightarrow ^6Li},
\end{eqnarray}
where $P_{\rm ^6Li\rightarrow ^6Li}$ is the surviving rate of the
non-thermally produced ${\rm ^6Li}$,~\footnote{
In fact, non-thermally produced ${\rm ^6Li}$ can be also destroyed by
the process ${\rm ^6Li}(p_{\rm BG},{\rm ^4He}){\rm ^3He}$ after
being thermalized.  Such an effect is taken in account in the
standard code of the BBN calculation we used \cite{kawano}, and hence
is not included in $P_{\rm ^6Li\rightarrow ^6Li}$.}
$f_{A_i}$ is the cumulative energy-distribution function of energetic
T and $\hethree$ during whole period of the evolution of the hadronic
shower.  For the energy distribution of ${\rm T}$ and ${\rm ^3He}$
produced by the hadrodissociation processes, we use the experimental
data \cite{meyer:1972} (see Appendix \ref{sec:transfer_energy}).
Importantly, experimental result suggests that the energy distribution
of the final-state ${\rm T}$ (and ${\rm ^3He}$) is almost independent
of the energy of the initial state energetic neutron.  Thus, we use
the following formula for $f_{A_i}$ (with $A_i={\rm T}$ and ${\rm
^3He}$)
\begin{eqnarray}
    f_{A_i} (E_{A_i}) = 
    \frac{\xi_{A_i}}{\sigma_{N+\alpha\rightarrow A_i+\cdots}}
    \frac{d\sigma_{N+\alpha\rightarrow A_i+\cdots}}
    {dE_{A_i}}.
\end{eqnarray}
Fitting formula for the differential cross section obtained from the
experimental data, which is used in our analysis, is given in Eq.\ 
(\ref{dET_exp}).

The non-thermally produced ${\rm ^6Li}$ is energetic with their
kinetic energies of ${\cal O}(1-10)\ {\rm MeV}$ and might be destroyed
by scattering off the background nuclei (in particular, protons)
before it is thermalized.  To estimate the number of ${\rm ^6Li}$
destroyed after the non-thermal production, we calculate the surviving
probability $P_{\rm ^6Li\rightarrow ^6Li}$.  (For details, see
Appendix \ref{sec:lisurviving}.)  Then, for the cosmic temperature we
are interested in (i.e., $T\lesssim 100\ {\rm keV}$) we have found
that the surviving rate is very close to 1.  Thus, almost all the
non-thermally produced ${\rm ^6Li}$ survive until being thermalized.
(Same is true for ${\rm ^6Li}$, ${\rm ^7Li}$, ${\rm ^7Be}$ produced by
the non-thermal processes with energetic ${\rm ^4He}$, which will be
discussed below.)

\begin{figure}
    \centering
    \centerline{{\vbox{\epsfxsize=8.0cm
    \epsfbox{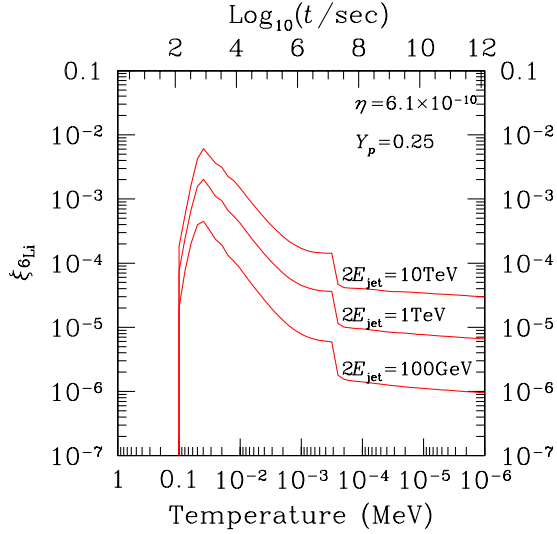}}}}
    \caption{$\xi_{\rm ^6Li}^{\rm (T, ^3He)}$ as a function of the 
    temperature for $2E_{\rm jet}=100\ {\rm GeV}$, $1\ {\rm TeV}$ and
    $10\ {\rm TeV}$.  Here, we take $Y_{p}=0.25$ and $\eta=6.1\times
    10^{-10}$.}
    \label{fig:prodli6}
\end{figure}

Using the cumulative energy-distribution function obtained by
following the evolution of the hadronic shower, we calculate the
$\xi_{\lisix}^{\rm (T, ^3He)}$ parameters for various background
temperatures (as well as other cosmological and model parameters).
The numerical result is shown in Fig.\ \ref{fig:prodli6}.  As one can
see, $\xi_{\rm ^6Li}^{\rm (T, ^3He)}$ is suppressed when $T\gtrsim
100\ {\rm keV}$.  This cut off is from the fact that, at such high
temperature, all the hadronic particles are stopped by the
electromagnetic processes before scattering off $\alpha_{\rm BG}$.
(In addition, at high temperature, surviving rate of ${\rm ^6Li}$ is
also suppressed.)  In addition, we see a sharp drop-off of $\xi_{\rm
^6Li}^{\rm (T, ^3He)}$ at $T\sim 0.3\ {\rm keV}$, which is due to the
decay of the neutron during the propagation in the universe.

\begin{figure}
    \centering
    \centerline{{\vbox{\epsfxsize=8.0cm\epsfbox{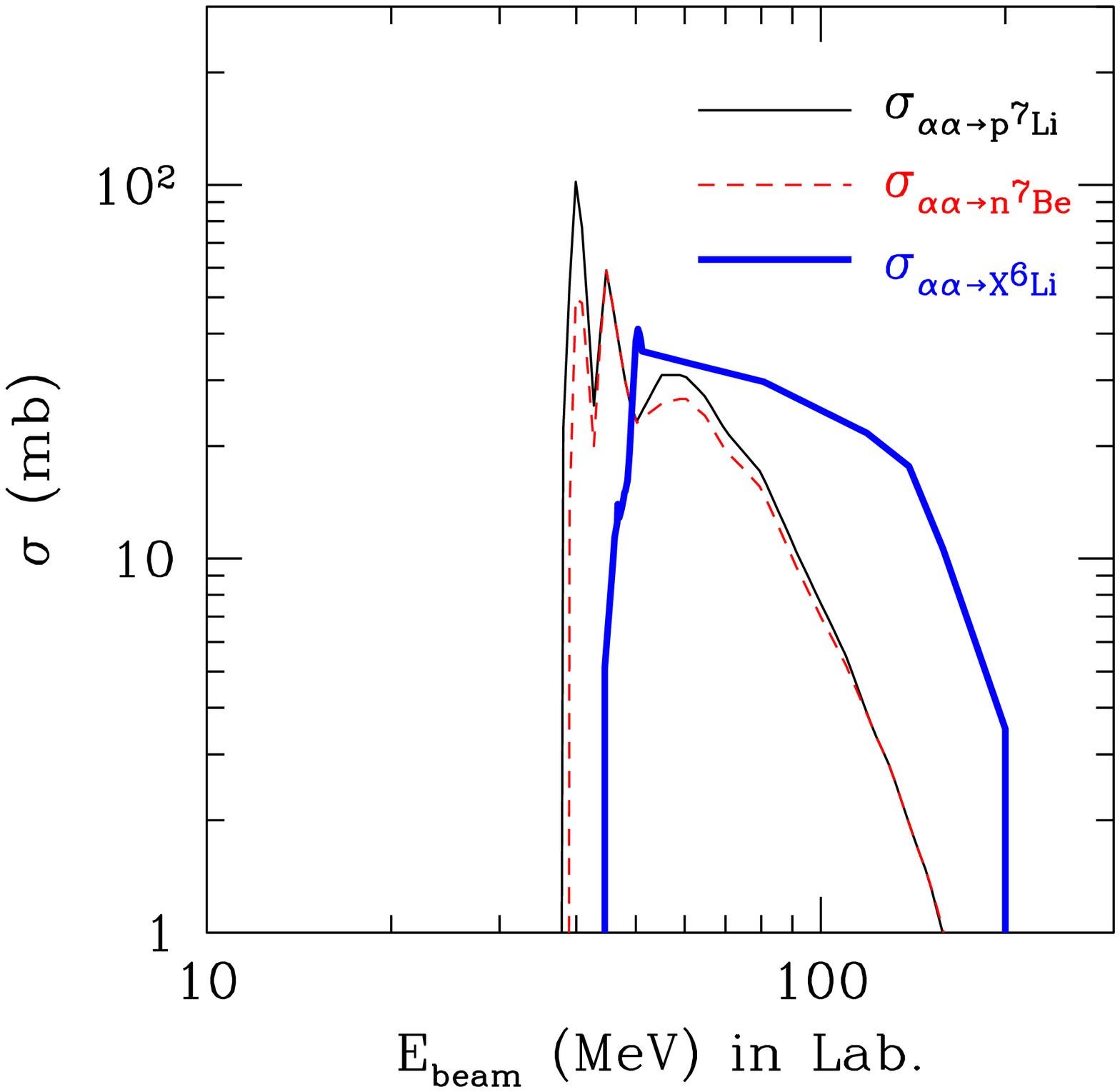}}}}
    \caption{The cross sections of the $\lisix$, $\liseven$ and
    $\beseven$ production process. The solid (dashed) line is for
    $\hefour+\hefour \to \liseven + p$ ($\hethree+\hefour \to \beseven
    + n$), while thick solid line is the total cross section for the
    process $\hefour+\hefour\to\lisix +\cdots$.}
    \label{fig:libe03}
\end{figure}

Next we consider the non-thermal production of $\lisix$, $\liseven$,
and $\beseven$ through the collision of energetic $\hefour$ with
background $\alpha_{\rm BG}$. Such energetic $\hefour$ is produced by
the elastic and inelastic scatterings between high-energy nucleons and
background $\alpha$.  The number of the non-thermally produced nuclei
per one decaying $X$ is expressed as
\begin{eqnarray}
    \label{eq:xi_LiBe}
    \xi_{A_k}^{\rm (^4He)} &=& 
    \int_{0}^{\infty} d E_{\rm ^4He}^{\rm (in)}
    f_{\rm ^4He} (E_{\rm ^4He}^{\rm (in)})
    \nonumber \\ && \times 
    \int_{\tilde{E}^{(R=1)}_{\rm ^4He}}^{E_{\rm ^4He}^{\rm (in)}} 
    d E_{\rm ^4He} \left( \frac{dE_{\rm ^4He}}{dt} \right)^{-1}
    n_{\alpha} \sigma_{{\rm ^4He}+\alpha_{\rm BG}\to A_k+\cdots} 
    (E_{\rm ^4He})
    \beta_{\rm ^4He} P_{A_k\rightarrow A_k},
\end{eqnarray}
where, here, $A_k$ = $\lisix$, $\liseven$, and $\beseven$.  In
Fig.~\ref{fig:libe03} the experimental data of the differential cross
sections for ${\rm ^4He}+{\rm ^4He}\to\liseven+p$, $\beseven+n$, and
$\lisix+X$ are plotted~\cite{glagola:1978,Read:1984,pagel:1997,EXFOR}.
As will be shown in Appendix~\ref{sec:transfer_energy}, the elastic
scattering of high-energy nucleons with $\alpha_{\rm BG}$ ($N +
\alpha_{\rm BG} \to N + \alpha$) is not important for these production
processes of $\lisix$, $\liseven$, and $\beseven$. That is because the
transfered energy to $\hefour$ in the elastic scattering is much
smaller than the case of the inelastic scattering ($N + \alpha_{\rm
BG} \to N + \alpha +\pi's$), although the cross section is fairly
large.

Although the expressions (\ref{eq:xi_Li6}) and (\ref{eq:xi_LiBe}) have
the same structure, estimation of $\xi_{A_k}^{\rm (^4He)}$ is rather
difficult.  This is because we only have insufficient data for the
transfered energies to $\hefour$ in the non-elastic scattering
processes to produce energetic ${\rm ^4He}$.  In particular, for the
process $p/n+\hefour\to p/n+\hefour +\pi +\cdots$, energy distribution
of $\hefour$ in the final state is quite uncertain at around the
threshold energy of this scattering process.  The number of the
non-thermally produced Li and Be, however, depend on the energy
distribution of $\hefour$.  At the present stage, we have to conclude
that the reliable estimation of the number of non-thermally produced
Li and Be from the process ${\rm ^4He}+\alpha_{\rm BG}\rightarrow {\rm
Li}/{\rm Be}+\cdots$ is difficult.  Thus, we will not include the
non-thermally produced Li and Be from this class of processes when we
derive the constraint on $X$.

\begin{figure}
    \centering
    \centerline{{\vbox{\epsfxsize=8.0cm
    \epsfbox{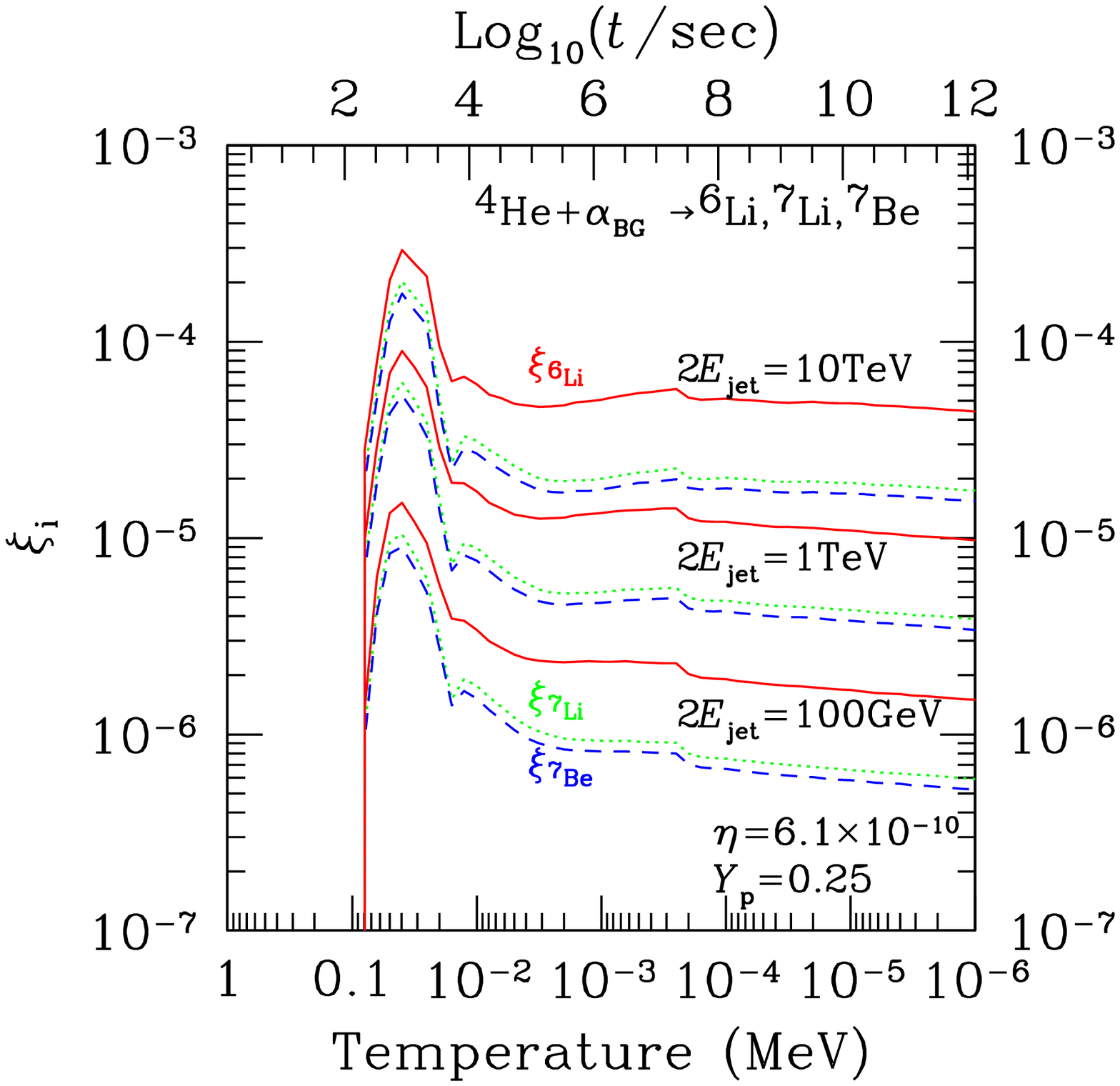}}}}
    \caption{$\xi_{\rm ^6Li}^{\rm (^4He)}$, 
    $\xi_{\rm ^7Li}^{\rm (^4He)}$, and $\xi_{\rm ^7Be}^{\rm (^4He)}$
    as functions of the temperature for $2E_{\rm jet}=100\ {\rm GeV}$,
    $1\ {\rm TeV}$ and $10\ {\rm TeV}$.  Here, we take $Y_{p}=0.25$
    and $\eta=6.1\times 10^{-10}$.}
    \label{fig:prodlibe}
\end{figure}

We can, however, estimate the number of Li and Be produced from the
process ${\rm ^4He}+\alpha_{\rm BG}\rightarrow {\rm Li}/{\rm
Be}+\cdots$ by adopting the energy distribution of ${\rm ^4He}$
generated from our shower algorithm (see
Appendix~\ref{sec:transfer_energy}):
\begin{eqnarray}
    \label{eq:f_N}
    f_{\hefour} (E_{\hefour}) =
    \sum_{N=p,n} \sum_{l=0}^{\infty} 
    \int dE_N F^{(l)} (E_N) 
    \tilde{G}_{N\rightarrow \hefour} (E_N, E_{\hefour}).
\end{eqnarray}
Using this relation, we estimate $\xi_{\rm ^6Li}^{\rm (^4He)}$,
$\xi_{\rm ^7Li}^{\rm (^4He)}$, and $\xi_{\rm ^7Be}^{\rm (^4He)}$.  The
results are shown in Fig.\ \ref{fig:prodlibe}.  The results indicate
that, by comparing with Fig.\ \ref{fig:prodli6}, the effects of ${\rm
^4He}$-$\alpha_{\rm BG}$ collision are less significant than those of
the ${\rm T}$-$\alpha_{\rm BG}$ and ${\rm ^3He}$-$\alpha_{\rm BG}$
collisions.  In addition, we have checked that, even if we adopt
$\xi_{\rm ^6Li}^{\rm (^4He)}$, $\xi_{\rm ^7Li}^{\rm (^4He)}$, and
$\xi_{\rm ^7Be}^{\rm (^4He)}$ obtained above, no significant change of
the resultant constraints on $Y_X$ is seen.

So far, we have discussed the case where the parent nucleus inducing
the non-thermal production of ${\rm Li}$ and ${\rm Be}$ is produced by
the hadronic scattering processes.  Energetic nuclei are, in fact,
also produced by the photodissociation processes of $\alpha_{\rm BG}$.
Using energetic photons produced in the electromagnetic shower
processes, energetic ${\rm T}$ and ${\rm ^3He}$ can be produced by the
processes
\begin{eqnarray}
    \gamma + \alpha_{\rm BG} \rightarrow
    \left\{ 
        \begin{array}{l}
            {\rm T} + p\\
            {^3{\rm He}}+n
        \end{array}
    \right. .
    \nonumber
\end{eqnarray}
Then, using ${\rm T}$ and ${\rm ^3He}$ produced by the above process,
non-thermal production of ${\rm ^6Li}$ is possible with the process
(\ref{eq:T_alpha}) and (\ref{eq:he3_alpha}).

We take account of these effects including the following term in the
Boltzmann equation:
\begin{eqnarray}
    \left[ \frac{dn_{^6{\rm Li}}}{dt} 
    \right]_{\gamma+\alpha_{\rm BG}\rightarrow {\rm T/^3He} +\cdots}
    &=&
    n_{^4{\rm He}} \int^{\infty}_{E_4^{\rm (th,T)}+4E_6^{\rm (th,T)}} 
    dE_{\gamma}
    \sigma_{^4{\rm He}(\gamma,p){\rm T}}(E_{\gamma}) 
    f_{\gamma}(E_{\gamma})
    \nonumber \\ &&
    \int_{E_6^{\rm (th,T)}}^{(E_{\gamma}-E_4^{\rm (th,T)})/4}
    dE_{\rm T} \left(\frac{dE_{\rm T}}{dt}\right)^{-1}
    n_{^4{\rm He}}
    \sigma_{{\rm T}(^4{\rm He},n)^6{\rm Li}}(E_{\rm T}) \beta_{\rm T}
    \nonumber \\ && + 
    n_{^4{\rm He}} 
    \int^{\infty}_{E_4^{\rm (th,^3He)}+4E_6^{\rm (th,^3He)}} 
    dE_{\gamma}
    \sigma_{^4{\rm He}(\gamma,p){\rm ^3He}}(E_{\gamma}) 
    f_{\gamma}(E_{\gamma})
    \nonumber \\ &&
    \int_{E_6^{\rm (th,^3He)}}^{(E_{\gamma}-E_4^{\rm (th,^3He)})/4}
    dE_{\rm ^3He} \left(\frac{dE_{\rm ^3He}}{dt}\right)^{-1}
    n_{^4{\rm He}}
    \sigma_{{\rm ^3He}(^4{\rm He},p)^6{\rm Li}}(E_{\rm ^3He}) 
    \beta_{\rm ^3He},
    \nonumber \\
    \label{dnli6/dt(EM)}
\end{eqnarray}
where $E_4^{\rm (th,T)}$ and $E_6^{\rm (th,T)}$ ($E_4^{\rm (th,^3He)}$
and $E_6^{\rm (th,^3He)}$) are threshold energies for the processes
$\gamma+{\rm ^4He}\rightarrow {\rm T}+p$ and ${\rm T}+{\rm
^4He}\rightarrow {\rm ^6Li}+n$ ($\gamma +{\rm ^4He}\rightarrow {\rm
^3He}+p$ and ${\rm ^3He}+{\rm ^4He}\rightarrow{\rm ^6Li}+p$,
respectively.~\footnote{
Effects of the non-thermal production of ${\rm ^6Li}$ by ${\rm T}$ and
${\rm ^3He}$ produced by the photodissociations become important at
relatively low temperature where $\tilde{E}^{(R=1)}_{\rm ^6Li}$
becomes smaller than the threshold energies.  Thus, in Eq.\
(\ref{dnli6/dt(EM)}), the threshold energies of the ${\rm ^6Li}$
productions are used for the lower bound of the integration.}
In addition, $\sigma_{{\rm T}({\rm ^4He},n){\rm
^6Li}}(E_{\rm T})$ ($\sigma_{{\rm ^3He}({\rm ^4He},p){\rm
^6Li}}(E_{{\rm ^3He}})$) is the cross section for the process ${\rm
T}+{{\rm ^4He}}\rightarrow {\rm ^6Li}+n$ (${\rm ^3He}+{\rm
^4He}\rightarrow{\rm ^6Li}+p$) with $E_{\rm T}$ ($E_{{\rm
^3He}}$)being the energy of the injected ${\rm T}$ (${\rm ^3He}$).
For these cross sections, we use the formula given in Ref.\ 
\cite{Cyburt:2002uv}; $E_{\rm T}$ dependence of $\sigma_{{\rm T}({\rm
^4He},n){\rm ^6Li}}(E_{\rm T})$ is shown in Fig.~\ref{fig:sigli6}.

\section{General Results}
\label{sec:results}
\setcounter{equation}{0}

\subsection{Outline}

In this section, we present our numerical results.  In particular, we
compare the theoretical predictions on the abundances of the light
elements with the observations and derive constraints on the
properties of $X$.  

In our analysis, we first calculate the evolution of the number
density of $X$ using
\begin{eqnarray}
    \frac{dn_X}{dt} = -3H n_X - \Gamma_X n_X.
\end{eqnarray}
At each temperature, photodissociation rates are calculated by
numerically integrating the photon spectrum and the relevant cross
sections.  In addition, we also calculate the $\xi$-parameters defined
in the previous sections.

Then, we obtain the Boltzmann equations for the light elements.
Evolution of the nucleons $N(=p,n)$ is governed by
\begin{eqnarray}
    \frac{dn_N}{dt} = 
    \left[ \frac{dn_N}{dt} \right]_{\rm SBBN}
    + \left[ \frac{dn_N}{dt} \right]_{\rm photodis}
    + B_h n_X \Gamma_X \xi_N
    + \left[\frac{dn_N}{dt}\right]_{\rm IC}.
\end{eqnarray}
For other light elements with atomic number 2 or 3 (i.e., $A_i={\rm
D}$, ${\rm T}$, ${\rm ^3He}$), we obtain
\begin{eqnarray}
    \frac{dn_{A_i}}{dt} = 
    \left[ \frac{dn_{A_i}}{dt} \right]_{\rm SBBN}
    + \left[ \frac{dn_{A_i}}{dt} \right]_{\rm photodis}
    + B_h n_X \Gamma_X \xi_{A_i},
\end{eqnarray}
while for ${\rm ^4He}$,
\begin{eqnarray}
    \frac{dn_{\rm ^4He}}{dt} = 
    \left[ \frac{dn_{\rm ^4He}}{dt} \right]_{\rm SBBN}
    + \left[ \frac{dn_{\rm ^4He}}{dt} \right]_{\rm photodis}
    - B_h n_X \Gamma_X \xi_\alpha.
\end{eqnarray}
For ${\rm ^6Li}$, we include the non-thermal secondary production
process discussed in the previous section and hence we obtain
\begin{eqnarray}
    \frac{dn_{\rm ^6Li}}{dt} =
    \left[ \frac{dn_{{\rm ^6Li}}}{dt} \right]_{\rm SBBN}
    + \left[ \frac{dn_{\rm ^6Li}}{dt} \right]_{\rm photodis}
    + \left[ \frac{dn_{{\rm ^6Li}}}{dt} 
    \right]_{\gamma+\alpha_{\rm BG}\rightarrow {\rm T/^3He} +\cdots}
    + B_h n_X \Gamma_X  \xi_{\lisix}^{\rm (T,^3He )}.
    \nonumber \\ 
    \label{dnli6/dt}
\end{eqnarray}
Here, the terms with the subscript ``SBBN'' represent the SBBN
contributions to the Boltzmann equations (including the effect of the
cosmic expansion).

In order to solve these equations, as we mentioned, we have
modified the Kawano Code (Version 4.1, with the nuclear cross sections
being updated), including the new subroutines which take account of
photodissociation, inter-conversion, and hadrodissociation processes.
The photodissociation and hadrodissociation processes included in our
analysis are summarized in Tables \ref{table:photodis},
\ref{tab:bran_process1}, and \ref{tab:bran_process2}.  In addition,
our treatments of the inter-conversion and non-thermal production of
${\rm Li}$ are discussed in Sections \ref{sec:interconversion} and
\ref{sec:ntlibe}, respectively.  

\subsection{Predicted light-element abundances}

\begin{figure}
    \centering
    \centerline{{\vbox{\epsfxsize=8.0cm\epsfbox{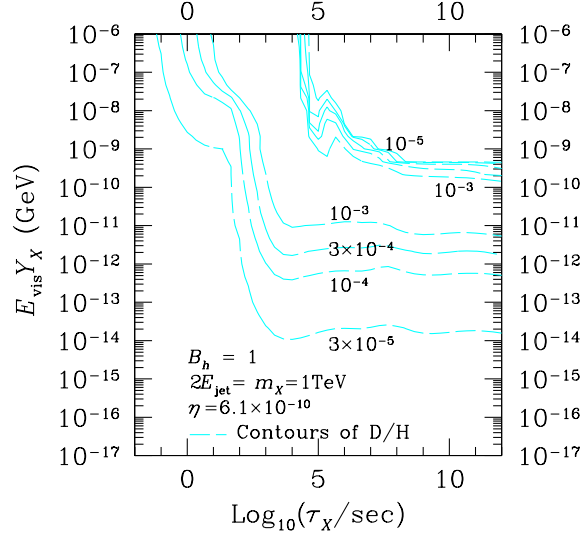}}}}
    \caption{Contours of constant ${\rm D}/{\rm H}$ 
    on the $\tau_{X}$ vs.\ $E_{\rm vis}Y_X$ plane for $m_{X}=1\ {\rm
    TeV}$.  Here we take $B_h=1$, $E_{\rm vis}=m_X$, and $X$ is assume
    to decay into two hadronic jets with $2E_{\rm jet}=m_{X}$.  Here,
    we take $\eta=6.1\times 10^{-10}$.  In the SBBN, the theoreical
    predication is $\left({\rm D}/{\rm H} \right)_{\rm SBBN} =
    2.78\times 10^{-5}$.}
    \label{fig:myxhdisy2}
\end{figure}

\begin{figure}
    \centering
    \centerline{{\vbox{\epsfxsize=8.0cm\epsfbox{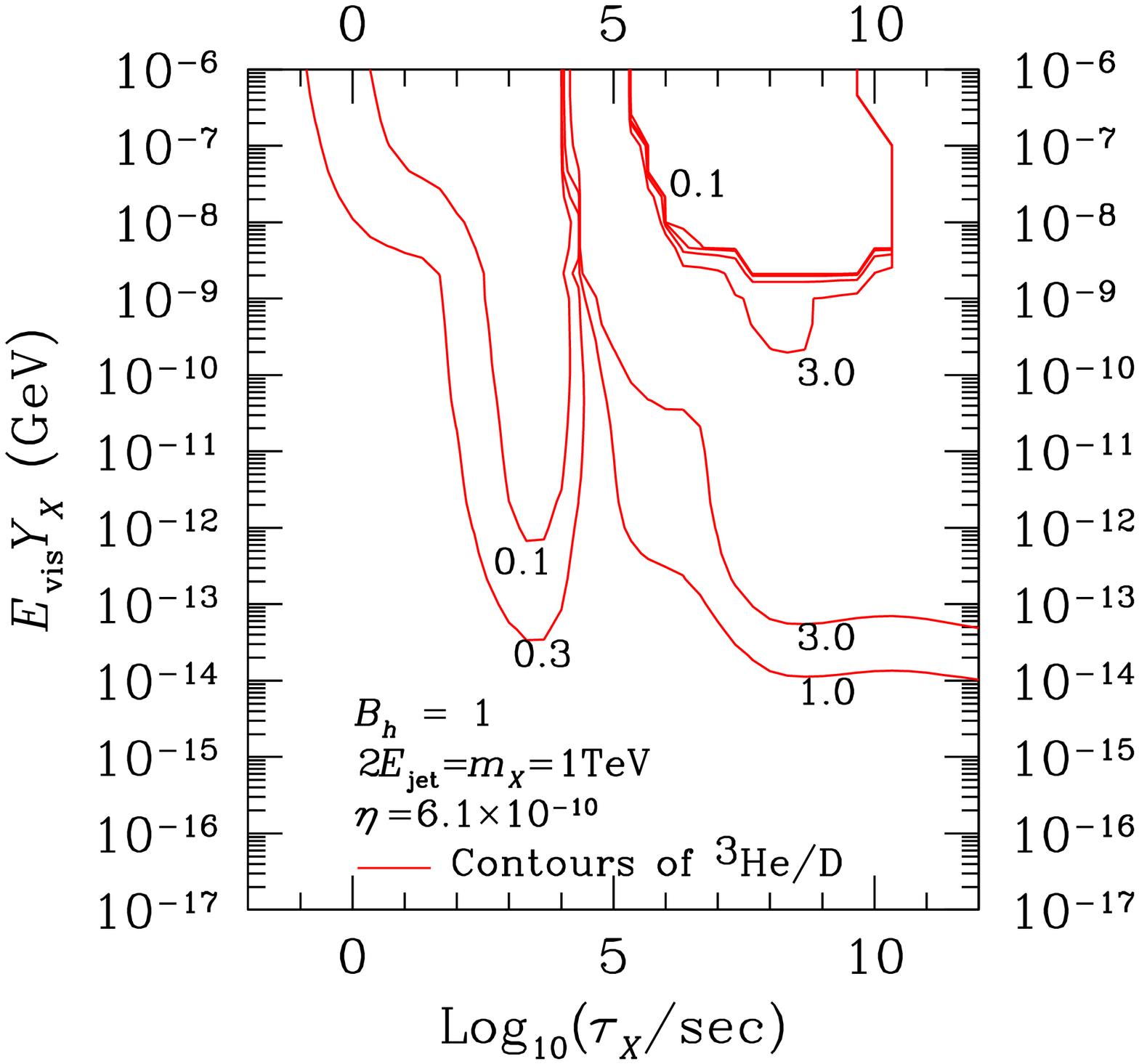}}}}
    \caption{Contours of constant $\hethree$/D.  Cosmological and
    model parameters are the same as Fig.~\ref{fig:myxhdisy2}. 
    In the SBBN, the theoreical predication is
    $\left(\hethree/\D \right)_{\rm SBBN} = 0.335$.}
    \label{fig:myxhdisy3}
\end{figure}

\begin{figure}
    \centering
    \centerline{{\vbox{\epsfxsize=8.0cm\epsfbox{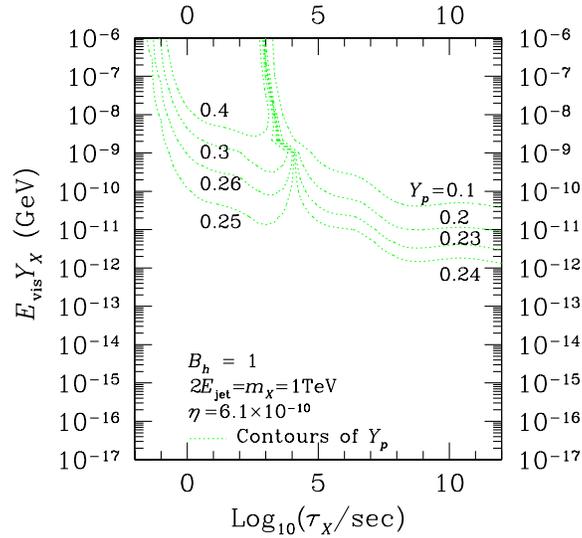}}}}
    \caption{Contours of constant $Y_{p}$.  Cosmological and
    model parameters are the same as Fig.~\ref{fig:myxhdisy2}.  In the
    SBBN, the theoreical predication is $\left(Y_{p} \right)_{\rm
    SBBN} = 0.249$.}
    \label{fig:myxhdisy4}
\end{figure}

\begin{figure}
    \centering
    \centerline{{\vbox{\epsfxsize=8.0cm\epsfbox{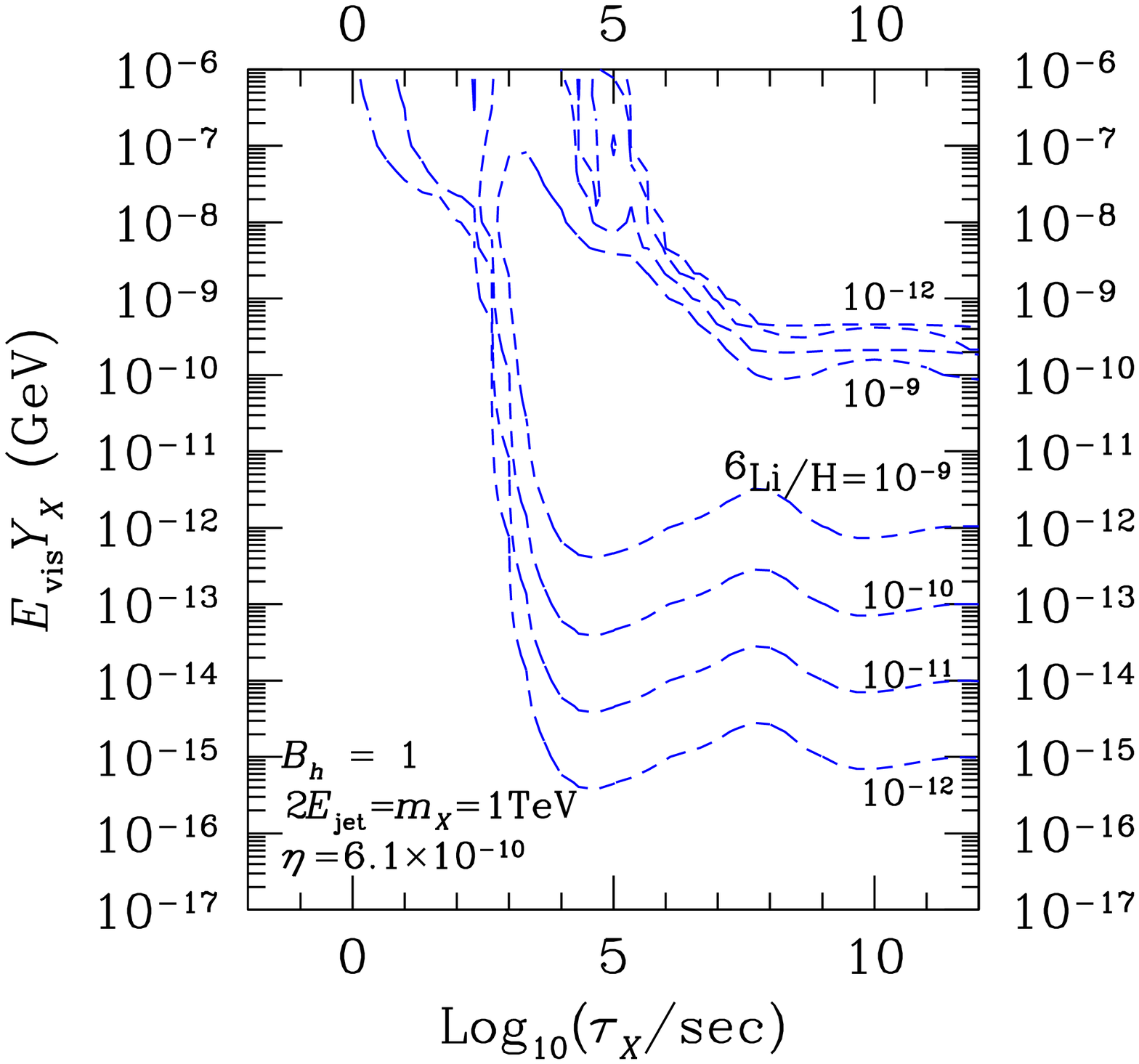}}}}
    \caption{Contours of constant  $\lisix$/H.  Cosmological and
    model parameters are the same as Fig.~\ref{fig:myxhdisy2}.  In the
    SBBN, the theoreical predication is $\left({\rm \lisix}/{\rm H}
    \right)_{\rm SBBN} = 1.30\times 10^{-14}$.}
    \label{fig:myxhdisy6}
\end{figure}

\begin{figure}
    \centering
    \centerline{{\vbox{\epsfxsize=8.0cm\epsfbox{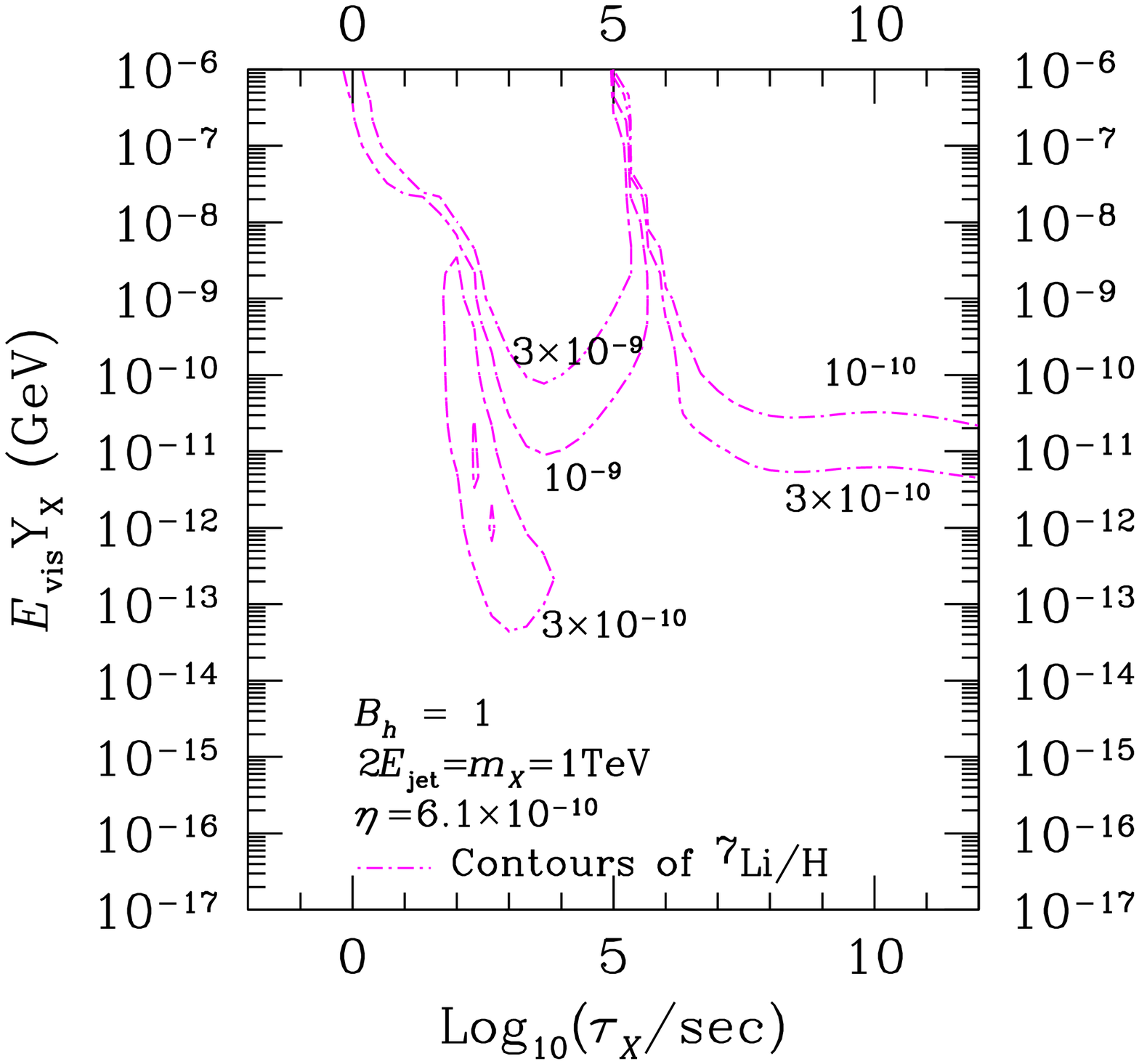}}}}
    \caption{Contours of constant $\liseven$/H.  Cosmological and
    model parameters are the same as Fig.~\ref{fig:myxhdisy2}. In the
    SBBN, the theoreical predication of the abundance is $\left({\rm
        \liseven}/{\rm H} \right)_{\rm SBBN} = 3.81\times 10^{-10}$.}
    \label{fig:myxhdisy7}
\end{figure}

To see how the abundances of the light elements behave, we estimated
the light-element abundances using the center values of the cross
sections and model parameters.  In Figs.\ \ref{fig:myxhdisy2},
\ref{fig:myxhdisy3}, \ref{fig:myxhdisy4}, \ref{fig:myxhdisy6} and
\ref{fig:myxhdisy7}, we plot contours of D/H, $\hethree$/D, $Y_{\rm
p}$, $\lisix$/H, and $\liseven$/H, in ($\tau_{X}$, $m_{X}Y_{X}$)
plane.  In the caculations, effects of photodissociation and
hadrodissociation are both included.  In addition, we take $m_{X}$ =
1 TeV and $B_h=1$, and consider the case where $X$ decays into two
hadronic jets with with the energy $2E_{\rm jet}=m_{X}$.

As one can see, for $\tau_X\gtrsim 10^{3}-10^{4}\ {\rm sec}$, ${\rm
^4He}$ abundance decreases as the primordial abundance of $X$ becomes
larger.  This is because, as $Y_X$ increases, hadrodissociation and
photodissociation processes of ${\rm ^4He}$ become more effective and
hence larger number of ${\rm ^4He}$ is destroyed.  In addition, since
the destruction processes of ${\rm ^4He}$ are followed by the creation
processes of ${\rm D}$, ${\rm ^3He}$, and ${\rm ^6Li}$, abundances of
these light elements first enhanced as $Y_X$ increases.  If $Y_X$ is
extremely large, however, all the light elements are destroyed; in
this case, abundances of ${\rm D}$, ${\rm ^3He}$, and ${\rm ^6Li}$ are
also decreased.  Contrary to ${\rm D}$, ${\rm ^3He}$ and ${\rm ^6Li}$,
${\rm ^7Li}$ is hardly produced.  Thus, for larger value of $Y_X$,
more ${\rm ^7Li}$ is destructed and hence the abundance of ${\rm
^7Li}$ decreases as the primordial abundance of $X$ increases.

On the other hand, for shorter lifetime ($\tau_X\lesssim 10^{3}\ {\rm
sec}$), inter-conversion between the proton and the neutron becomes
more effective.  In this case, $Y$ and D/H increase as $Y_X$
increases.

In Fig.~\ref{fig:myxhdisy7}, we can see a distinctive trend of
decrease of ${\rm ^7Li/H}$ at $\tau_X \sim 10^{3}$ sec and $E_{\rm
vis} Y_{X} \gtrsim 10^{-13}$ GeV. That is because free neutrons
produced by hadronic showers are captured by ${\rm ^7Be}$ through
${\rm ^7Be}(n,{\rm ^3He}) {\rm ^4He}$, which reduces the resultant
abundance of ${\rm ^7Li}$. This phenomenon was also pointed in
\cite{Jedamzik04a}.

\subsection{Constraints}
\label{subsec:MC}

Now we show the constraints on the primordial abundances of $X$,
taking into account the theoretical and observational errors.  For the
execution of Monte Carlo simulation in BBN computation, we should
understand the error of all of the reaction rates concerning both
radiative and hadronic decay processes. For radiative decay processes,
we have shown them in Table \ref{table:photodis}.

\begin{figure}
    \centering
    \centerline{{\vbox{\epsfxsize=8.0cm\epsfbox{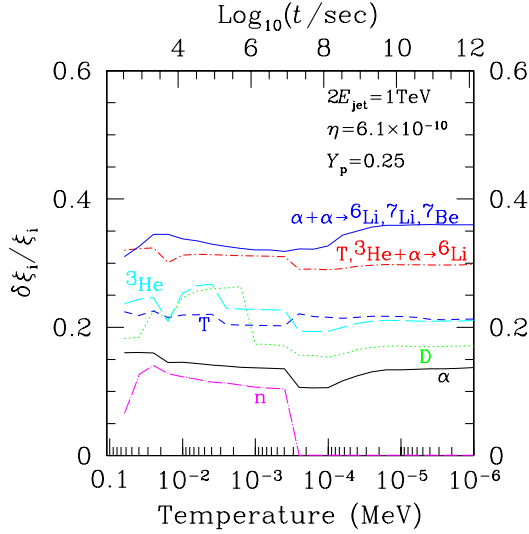}}}}
    \caption{Estimated theoretical errors of the $\xi$-parameters for
    $m_{X}=1\ {\rm TeV}$, $Y_{p}=0.25$ and $\eta=6.1\times 10^{-10}$.}
    \label{fig:dprod}
\end{figure}

As for the reaction rates related with the hadrodissociation
processes, which arediscussed in Section~\ref{sec:hadrodis} and
Section~\ref{sec:ntlibe}, we should estimate their errors in advance
of executing the Monte Carlo simulation. In this work, we assume that
the cross sections and the other model parameters of transfered
energies, which are used for computing $\xi$'s, obey the Gaussian
distribution with their 1$\sigma$ errors.  Computing $\xi$'s with such
errors sufficiently many times (in our caes, one thousand times), we
statistically evaluate the errors of $\xi$'s.  Here we adopt 20$\%$
error for all the hadronic cross sections ($\delta\sigma/\sigma =
0.2$), which is larger than typical errors of their experimental data.
For the errors of the transfered energies to nuclei in final states
after the collisions (the inverse slope parameter $K_{T}$ in inelastic
nucleon-$\alpha$ scattering, and slope parameters $B_{\rm sl}$ in
elastic nucleon-nucleon and nucleon-$\alpha$ scatterings (see,
Appendix \ref{sec:transfer_energy}), we adopt 20\ $\%$ errors.   In
addition, as we mentioned in Appendix \ref{sec:transfer_energy}, the
$\xi$ parameters do not change much with the variations of the
inelasticities ($\kappa_p$ and $\kappa_\alpha$).  Thus, we neglect
their uncertainties.  Furthermore, we have checked that $\xi$'s are
insensitive to the variation of $\eta$.  Thus, we also neglect its
uncertainty in evaluating the errors of $\xi$ parameters.

In Fig.~\ref{fig:dprod} we plot the errors of $\xi$'s as functions of
the temperature.  These are the case of $m_{X}$ = 1 TeV, $Y_{\rm
p}=0.25$ and $\eta=6.1\times 10^{-10}$. We have checked that the
errors do not change much even if we change the values of $m_{X}$ and
$Y_{\rm p}$.  Based on this result, we use the following errors in our
Monte Carlo analysis: $\delta\xi_{n}/\xi_{n}=0.15$,
$\delta\xi_{\D}/\xi_{\D}=0.2$, $\delta\xi_{\T}/\xi_{\T}=0.2$,
$\delta\xi_{\hethree}/\xi_{\hethree}=0.2$,
$\delta\xi_{\alpha}/\xi_{\alpha}=0.2$, and $\delta\xi_{\lisix}^{\rm
(T,^3He)}/\xi_{\lisix}^{\rm (T,^3He)}=0.3$.

For the hadron-nucleon inter-conversion reaction rate which was
discussed in Section~\ref{sec:interconversion}, we adopt 50$\%$ error
for each cross section because there are not any adequate experimental
data for the uncertainties of cross sections. Therefore, we take the
larger errors to get a conservative bound here, for the details, see
Ref.~\cite{Kohri:2001jx}.

To study the abundance of the light elements with $X$, we have
included the effects of the photodissociation, inter-conversion, and
the hadrodissociation processes into the BBN calculation.  In order to
estimate the theoretical uncertainties, we performed the Monte Carlo
simulation.  Here, we follow the basic procedure explained in
\cite{Holtmann:1998gd}.  In addition, for the BBN calculation, we take
account of the observational error of $\eta$ reported by the WMAP
collaborations~\cite{Spergel:2003cb} (see Eq.\ (\ref{eq:eta_WMAP})).

In our statistical analysis, with a given set of the model parameters,
we calculate the theoretical values of the light-element abundances
and calculate $\chi^2_i$, the likelihood variable for individual
statistical variable $x_i$.  For $x_i= (n_{\rm D}/n_{\rm H})$, and
$Y$, we use both the upper and lower bounds from the observation, and
hence
\begin{eqnarray}
    \chi^2_i = 
    \frac{ ( \bar{x}_i^{\rm th} - \bar{x}_i^{\rm obs} )^2}
    { (\sigma_i^{\rm th})^2 + (\sigma_i^{\rm obs})^2 }
    ~~ {\rm for}\ x_i = (n_{\rm D}/n_{\rm H})~{\rm and}~Y, 
\end{eqnarray}
where $\bar{x}_i^{\rm th}$ and $\bar{x}_i^{\rm obs}$ are the center
values of $x_i$ determined from the theoretical calculation and
observations, while $\sigma_i^{\rm th}$ and $\sigma_i^{\rm obs}$ are
their errors, respectively.  In our analysis, $(\sigma_i^{\rm th})^2$
is calculated by the Monte Carlo analysis.  Notice that the $\chi^2$
depends on the model parameters through $x_i^{\rm th}$ and
$\sigma_i^{\rm th}$.  For $x_i=r_{3,2}$ $(n_{\rm ^6Li}/n_{\rm H})$ and
$\log_{10}[(n_{\liseven}/n_{\rm H})]$ we only use the upper bound.  In
this case case, we define $\chi^2_i$ as
\begin{eqnarray}
    \chi^2_i =
    \left\{ 
        \begin{array}{ll}
            {\displaystyle{
            \frac{ ( \bar{x}_i^{\rm th} - \bar{x}_i^{\rm obs} )^2}
            { (\sigma_i^{\rm th})^2 + (\sigma_i^{\rm obs})^2 } } }
            &:\bar{x}_i^{\rm th} < \bar{x}_i^{\rm obs} 
            \\ \\
            0 &:{\rm otherwise}
        \end{array} 
    \right.
    {\rm for}\ x_i = r_{3,2},~(n_{\rm ^6Li}/n_{\rm H})~{\rm
    and}~\log_{10}[(n_{\liseven}/n_{\rm H})]. 
\end{eqnarray}
Notice that, contrary to the case of SBBN, we do not use the lower
bound on $(n_{\liseven}/n_{\rm H})$.  This is because we do
not include the non-thermal $\liseven$ production processes through
$\alpha$-$\alpha$ collisions.  All the observational constraints on
primordial abundances of the light elements have been summarized in
Section \ref{sec:obs_status}.

\begin{figure}
    \centering
  \centerline{{\vbox{\epsfxsize=8.0cm\epsfbox{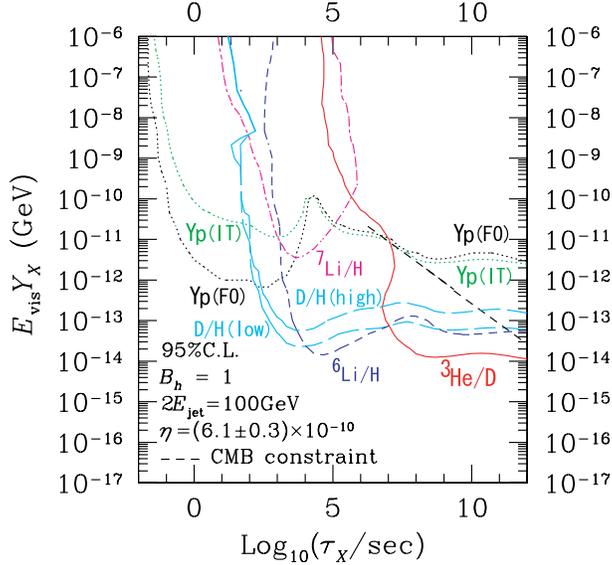}}}}
    \caption{Upper bounds on $m_{X}Y_{X}$ at 95\% C.L. for $B_h=1$ and
    $m_X=100\ {\rm GeV}$.  The horizontal axis is the lifetime of $X$.
    Here, the lines with ``D/H (low)'' and ``D/H (high)'' are for
    the constraints (\ref{lowd}) and (\ref{highd}), respectively. The
    straight dashed line is the upper bound by the deviation from the
    Planck distribution of the CMB.}
    \label{fig:myx100gev}
\end{figure}

\begin{figure}
    \centering
    \centerline{{\vbox{\epsfxsize=8.0cm\epsfbox{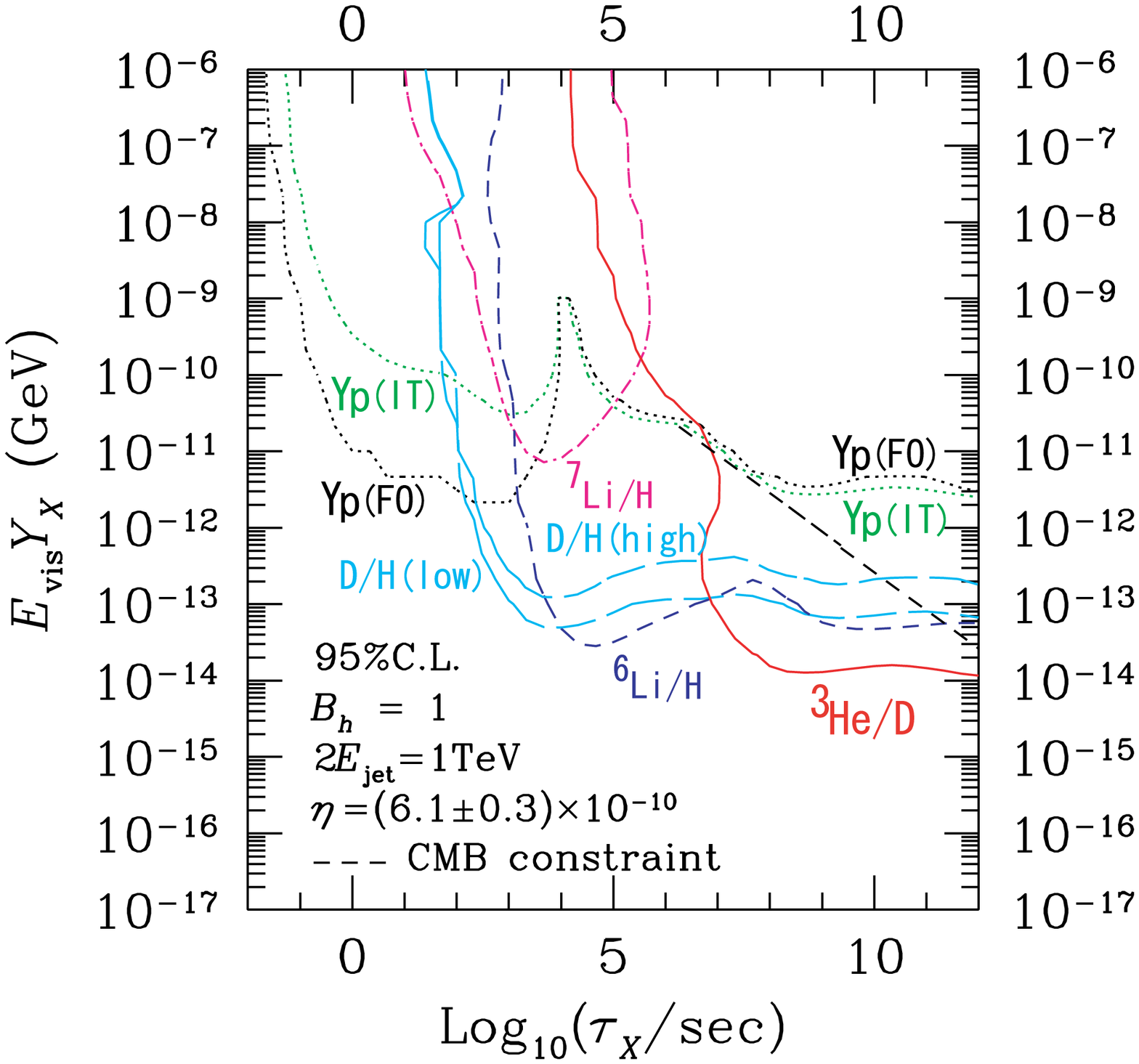}}}}
    \caption{Same as Fig.~\ref{fig:myx100gev}, but for 
    $m_{X}=1\ {\rm TeV}$.}
    \label{fig:myx1tev}
\end{figure}

\begin{figure}
    \centering
    \centerline{{\vbox{\epsfxsize=8.0cm\epsfbox{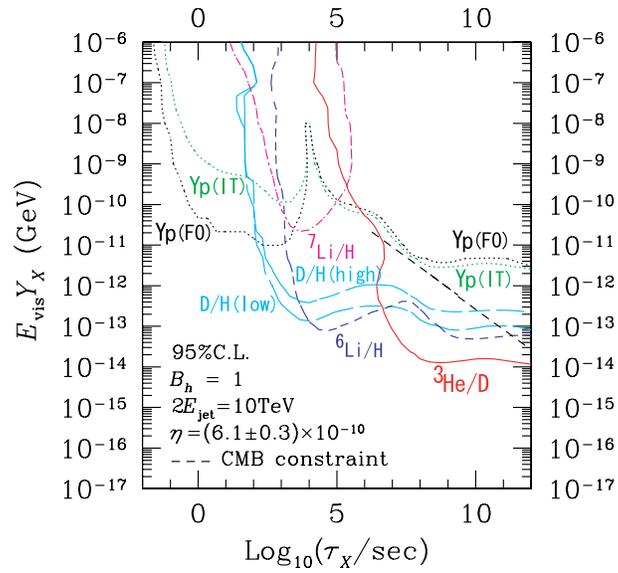}}}}
    \caption{Same as Fig.~\ref{fig:myx100gev}, but for 
    $m_{X}=10\ {\rm TeV}$.}
    \label{fig:myx10tev}
\end{figure}

In Figs.\ \ref{fig:myx100gev}, \ref{fig:myx1tev} and
\ref{fig:myx10tev}, we plot the results of the $\chi^2$ analysis at
$95\ \%$ C.L. (i.e., $\chi^2_i=3.84$ for $x_i= (n_{\rm D}/n_{\rm H})$
and $Y$; $\chi^2_i=2.71$ for $x_i=r_{3,2}$, ($n_{\rm ^6Li}/n_{\rm H}$)
and $\log_{10}[(n_{\liseven}/n_{\rm H})]$) on the $\tau_X$ vs.\ 
$E_{\rm vis}Y_{X}$ plane for $m_X=100\ {\rm GeV}$, $1\ {\rm TeV}$, and
$10\ {\rm TeV}$, respectively.  Here, the hadronic branching ratio is
unity, and $X$ decays into two hadronic jets with the energy $2 E_{\rm
jet} = m_{X}$. As mentioned in Section \ref{sec:obs_status}, the
constraint with use of the highest observed value of D/H
(Eq.~(\ref{highd})) is shown together with that obtained by taking our
standard value (Eq.~(\ref{lowd})).  One can see that the constraint
from D/H changes by a factor $2-3$ by different adoption of the
observed value.  We also plot the upper bound not to disturb the
Planck distribution of the cosmic microwave background.  As one can
see, constraints on the combination $E_{\rm vis}Y_{X}$ is quite
insensitive to the mass of $X$.  This fact implies that the
constraints for the case of $2E_{\rm jet}\neq m_{X}$ can be roughly
estimated by rescaling the bounds given in the figures.

\begin{figure}
    \centering
    \centerline{{\vbox{\epsfxsize=6.0cm
    \epsfbox{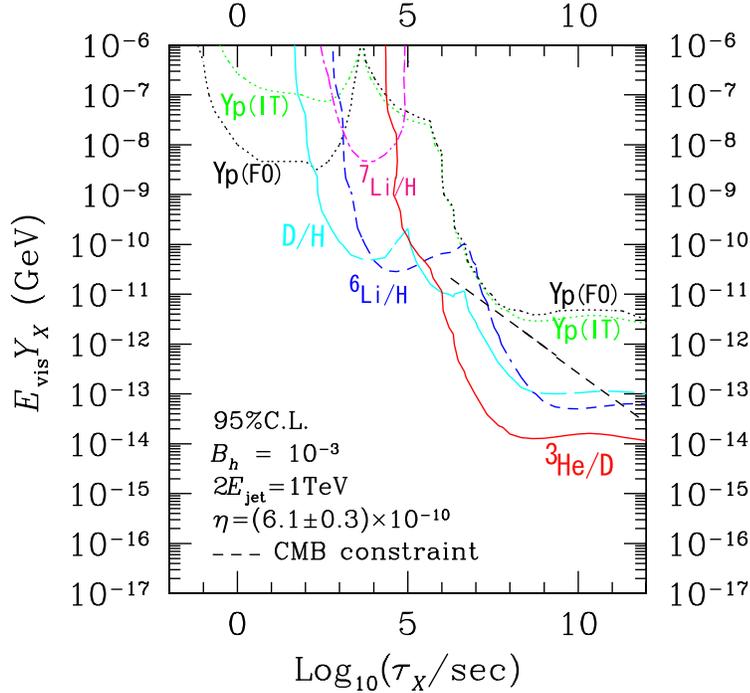}}}}
    \vspace{1.5cm}
    \caption{Same as Fig.~\ref{fig:myx100gev}, but for 
    $m_{X}=1\ {\rm TeV}$ and $B_h=10^{-3}$.}
    \label{fig:myx1tev_bm3}
\end{figure}

\begin{figure}
    \centering
    \centerline{{\vbox{\epsfxsize=5.cm
    \epsfbox{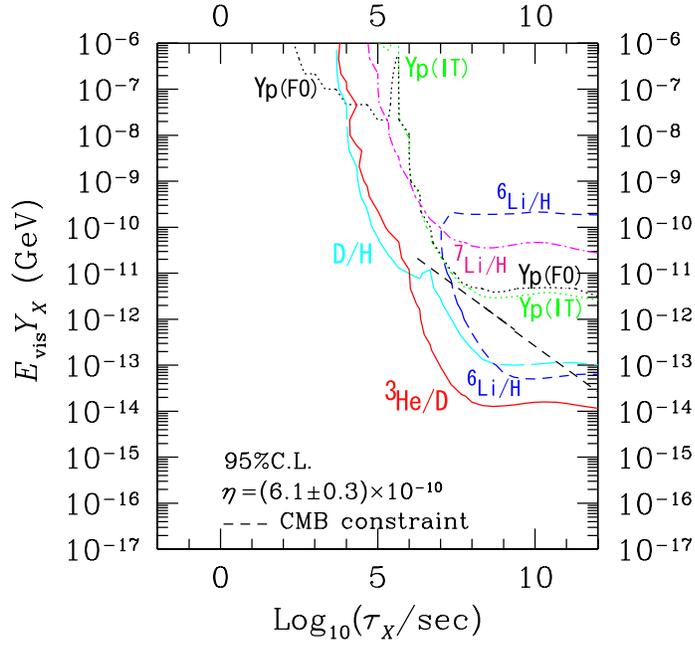}}}}
    \vspace{1.5cm}
    \caption{Same as Fig.~\ref{fig:myx100gev}, but for 
    $m_{X}=1\ {\rm TeV}$ and $B_h=0$ (no hadronic decay mode).}
    \label{fig:myxphotodis}
\end{figure}

\begin{figure}
    \centering
    \centerline{{\vbox{\epsfxsize=8.0cm
    \epsfbox{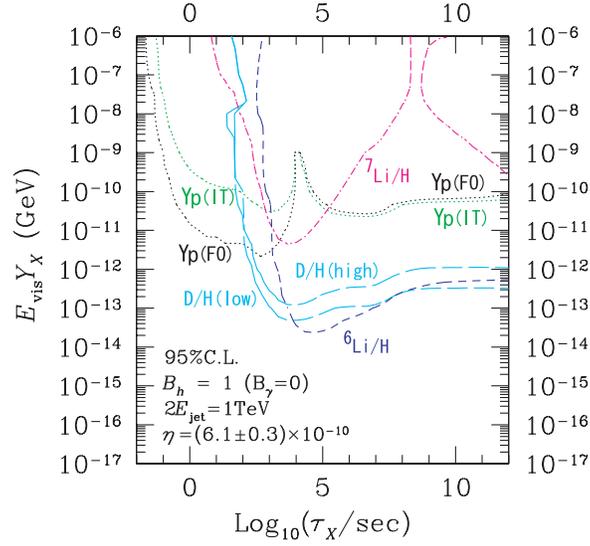}}}}
    \caption{Same as Fig.~\ref{fig:myx100gev}, but for 
    $m_{X}=1\ {\rm TeV}$ and no photo-dissociation.}
    \label{fig:myxonlyhadron}
\end{figure}

In order to see the dependence on hadronic branching ratio, we also
show the results for $B_h=10^{-3}$ and $B_h=0$ in Figs.\ 
\ref{fig:myx1tev_bm3} and \ref{fig:myxphotodis},
respectively.~\footnote
{Our constraint for the case of $B_h=0$ is in a reasonable
agreement with the results obtained by previous studies (in
particular, by Cyburt et al.\ \cite{Cyburt:2002uv}).  Cyburt et al.\
did not consider the constraint on $n_{\rm ^3He}/n_{\rm D}$ and, for
$n_{\rm D}/n_{\rm H}$, observational constraint milder than ours is
used ($1.3\times 10^{-5}<(n_{\rm D}/n_{\rm H})^{\rm obs}<5.3\times
10^{-5}$).  If we adopt $(n_{\rm D}/n_{\rm H})^{\rm obs}$ used in
\cite{Cyburt:2002uv}, the difference between the upper bound on $Y_X$
from our analysis and theirs is within the factor of $\sim 3$ or so
and is quite mild.  It may be due to the difference of the photon
spectrum used in the analysis.  (For the comparison, notice that we
have normalized the yield variable $Y_X$ by the entropy density $s$,
and also that $E_{\rm vis}\simeq\frac{1}{2}m_{X}$ is assumed in
\cite{Cyburt:2002uv}.)} 
(Here, we considered only our standard value of the observed D/H
(Eq.~(\ref{lowd})).) As one can see, with larger value of the
hadronic branching ratio, upper bounds on $E_{\rm vis}Y_{X}$ become
severer.  In addition, even with a relatively small value of the
hadronic branching ratio (i.e., $B_h=10^{-3}$), the hadronic decay
mode may significantly affect the light-element abundances.  In
addition, in order to separate out the effects of the hadronic decay,
we also calculated the light element abundance only taking account of
the inter-conversion and hadrodissociation processes.  The resultant
constraint is shown in Fig.\ \ref{fig:myxonlyhadron}.  Notice that, in
this figure, effects of the photodissociation is not included so the
situation is unrealistic; this figure is shown just for demonstration.

With the hadronic decay modes, the most significant constraint on
$E_{\rm vis}Y_{X}$ depends on the lifetime $\tau_X$:
\begin{itemize}
\item For $10^{-2}\ {\rm sec}\lesssim\tau_X\lesssim 10^{2}\ {\rm
    sec}$, the inter-conversion processes are efficient.  In this case,
    significant amount of $p$ may be converted to $n$ and,
    consequently, $Y$ may be enhanced.  In this case, the constraint from
    the overproduction of ${\rm ^4He}$ is the most significant.
\item For $10^{2}\ {\rm sec}\lesssim\tau_X\lesssim 10^{7}\ {\rm sec}$,
    energetic hadrons (in particular, neutron) is hardly stopped by
    the electromagnetic processes and hence the hadrodissociation
    processes become the most efficient.  In particular, in this case,
    non-thermal productions of ${\rm D}$ and ${\rm ^6Li}$ provide the
    most stringent constraint.
\item For $10^{7}\ {\rm sec}\lesssim\tau_X\lesssim 10^{12}\ {\rm
    sec}$, energetic neutron is likely to decay before scattering off
    the background nuclei.  In this case, effects of the hadronic
    decay modes become less significant compared to the case with
    shorter lifetime.  Furthermore, in this case, effects of the
    photodissociation becomes comparable to or more significant than
    the hadrodissociation.  Then, the strongest constraint is from the
    overproduction of ${\rm ^3He}$.
\end{itemize}

Importantly, for the case with relatively short lifetime (i.e.,
$\tau_X\lesssim 10^7\ {\rm sec}$), the hadrodissociation and the
inter-conversion processes are the most important.  Consequently, the
constraints strongly depend on the hadronic branching ratio.  In this
case, the upper bounds on $E_{\rm vis}Y_X$ are approximately
proportional to $B_{\rm h}$.  On the contrary, for longer lifetime
(i.e., $\tau_X\gtrsim 10^7\ {\rm sec}$), the most significant constraint
is from the overproduction of ${\rm ^3He}$ by the photodissociation of
$\alpha_{\rm BG}$.  Then, the upper bound on $E_{\rm vis}Y_X$ becomes
insensitive to $B_{\rm h}$.

Figs.\ \ref{fig:myx100gev} $-$ \ref{fig:myxphotodis} are our main
results and one can read off the constraints on the primordial
abundance of long-lived exotic particles.  Since our analysis does not 
assume any special properties of $X$, our results can be applied to
various classes of the long-live particles.  In the next section, we
will discuss one important application of our results, i.e., the
application to the gravitino problem.

\section{Application to Unstable Gravitino}
\label{sec:grav}
\setcounter{equation}{0}

In the previous section, we have derived constraints on the primordial
abundance of the late-decaying particle $X$.  Now, we apply our
results to one of the most important cases, i.e., the case with
unstable gravitino.  In supergravity theory, the gravitino, which is
the superpartner of the graviton, exists.  The gravitino acquires a
mass $m_{3/2}$ from the effect of the supersymmetry breaking.  In
large class of the models, the gravitino mass is comparable to or larger
than $\sim {\cal O}(100)\ {\rm GeV}$.  Importantly, the interaction of
the gravitino is suppressed by inverse powers of the (reduced) Planck
scale and hence its interaction is very weak.  Thus, if the gravitino
is unstable, its lifetime becomes very long.  This fact means that, if
the gravitino dominantly decays into visible-sector particle and its
superpartner, the decay of the gravitino in the early universe may
significantly change the predictions of the SBBN unless the primordial
abundance of the gravitino is small enough \cite{Weinberg:zq}.

With the inflation in the early stage of the universe, the primordial
gravitino is once diluted but it is produced after the reheating
starts.  Thus, even in the inflationary models, we may still have the
gravitino problem \cite{Krauss:1983ik}.  We can calculate the abundance
of the gravitino $Y_{3/2}$ as a function of the reheating temperature
$T_R$, which is defined in this paper as
\begin{eqnarray}
    T_{\rm R} \equiv 
    \left( 
        \frac{10}{g_* \pi^2} M_*^2 \Gamma_{\rm inf}^2 
    \right)^{1/4},
\end{eqnarray}
with $\Gamma_{\rm inf}$ being the decay rate of the inflaton and $g_*$
is the effective number of massless degrees of freedom.  (Here, we use
$g_*=228.75$.)  In our analysis, we reanalyzed the gravitino production
processes in the early universe.  In particular, we have used the
thermally-averaged gravitino production cross section given in Ref.\
\cite{Bolz:2000fu}, which properly takes account of the effect of the
thermal mass of the gauge bosons, and numerically solved the Boltzmann
equation for the gravitino production.  (Thus, the following fitting
formula is different from the previous ones given in
Refs.~\cite{KawMor,Bolz:2000fu}.)  The details are discussed in Appendix
\ref{app:Ygrav}.  Then, we find that the gravitino abundance after the
inflation is well-fitted by the following formula:
\begin{eqnarray}
    \label{eq:Yx-new}
    Y_{3/2} &\simeq& 
    1.9 \times 10^{-12}
    \nonumber \\ &&
    \times \left( \frac{T_{\rm R}}{10^{10}\ {\rm GeV}} \right)
    \left[ 1 
        + 0.045 \ln \left( \frac{T_{\rm R}}{10^{10}\ {\rm GeV}} 
        \right) \right]
    \left[ 1 
        - 0.028 \ln \left( \frac{T_{\rm R}}{10^{10}\ {\rm GeV}} 
        \right) \right].
    \nonumber \\
\end{eqnarray}
Importantly, the primordial abundance of the gravitino $Y_{3/2}$ is
approximately proportional to the reheating temperature $T_{\rm R}$.
Consequently, for the inflation models with high reheating temperature,
the gravitino abundance may become so large that the light-element
abundances are too much affected by the decay of the gravitino.  Thus,
we obtain a upper bound on the reheating temperature after inflation.

In order to derive the upper bound on $T_{\rm R}$, we have to specify
the decay mode of the gravitino to calculate its lifetime, $B_h$, and
so on.  In this paper, we consider two typical cases.  The first case
is that the gravitino can directly decay into the pair of colored
particles.  In particular, we consider the case where the gravitino
can decay into gluon and gluino pair: $\psi_\mu\rightarrow
g+\tilde{g}$ (see Fig.\ \ref{fig:feyngravhad}), producing one
hadronic jet with $E_{\rm jet} = \frac12 m_{3/2}$.  Assuming that this
is the dominant decay mode, the lifetime of the gravitino is given by
\begin{eqnarray}
    \label{eq:tau_gluon-gluino}
    \tau_{3/2} (\psi_\mu\rightarrow g+\tilde{g}) \simeq 
    6 \times 10^7\ {\rm sec} \times
    \left( \frac{m_{3/2}}{100\ {\rm GeV}} \right)^{-3},
\end{eqnarray}
where we have neglected other decay modes (in particular, decay into
the lightest neutralino and its superpartner).  In this case, the
hadronic branching ratio is expected to be very large, so we use
$B_h=1$.  In our analysis, we calculated the hadrodissociation
rates with the approximation that the numbers of hadrons (i.e.,
proton, neutron, pion, and so on) from the single gluon jet are the
same as those from the single quark jet.  In addition, since the
hadronization processes of the decay products of the gluino are
uncertain, we neglected the effects of the gluino.  Such a treatment
of the gluino might have made the upper bound on $T_{\rm R}$ less
stringent by the factor of $\sim 2$ or so.  For the photodissociation
rates, since the gluino is a colored particle, most of its initial
energy is expected to be converted to that of radiation.  Thus, even
though it is expected that some fraction of the initial energy of
gluino is carried away by the lightest neutralino (which is assumed to
be the LSP), we use $E_{\rm vis}=m_{3/2}$ in this case.

Direct decay of the gravitino into the colored superparticles may be,
however, kinematically blocked.  Thus, we also consider the case where
the gravitino dominantly decays into the photon and the lightest
neutralino (which we call ``photino'' in this section):
$\psi_\mu\rightarrow\gamma+\tilde{\gamma}$.  In this case, the
lifetime of the gravitino is obtained as
\begin{eqnarray}
    \label{eq:tau_photon-photino}
    \tau_{3/2}(\psi_\mu \rightarrow \gamma + \tilde{\gamma})
    \simeq 
    4\times 10^8{\rm ~sec}
    \left(\frac{m_{3/2}}{100 {\rm ~GeV}} \right)^{-3}.
\end{eqnarray}
Even if the gravitino dominantly decays into a photon and a photino,
the hadronic branching ratio is non-vanishing since the
quark-anti-quark pair can be attached at the end of the virtual photon
line.  (See Fig.\ \ref{fig:feyngravrad}.)  In this case, $B_h$ is
expected to be of order ${\cal O}(\alpha_{\rm em}/4\pi)$, and hence we
adopt $B_h=10^{-3}$.  In addition, because almost half of the
initial energy is carried away by the photino in this case, we use the
relation $E_{\rm vis}=\frac{1}{2}m_{3/2}$. Here we assume that
the decaying gravitino produces two hadronic jets with $E_{\rm jet} =
\frac13 m_{3/2}$.

\begin{figure}[t]
    \begin{center}
        \centerline{{\vbox{\epsfxsize=8.0cm
        \epsfbox{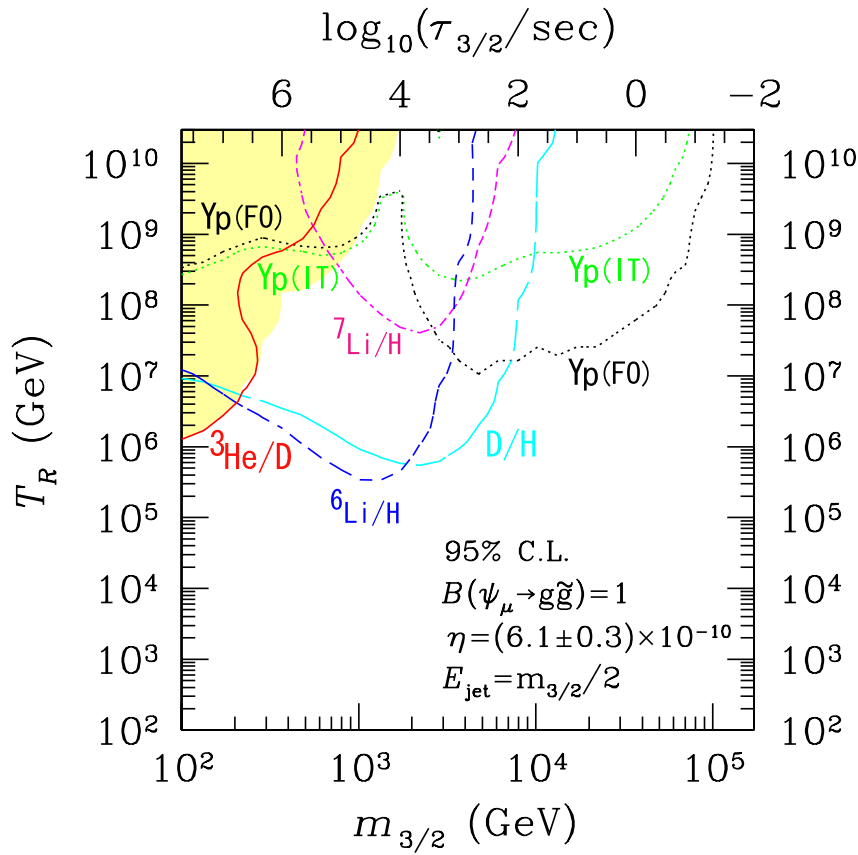}}}}
       \vspace{1.5cm}
        \caption{Upper bounds on the reheating temperature as a
        function of the gravitino mass for the case where the
        gravitino dominantly decays into gluon-gluino pair. Here, we
        take $B_h=1$, $E_{\rm vis}=m_{3/2}$, and $E_{\rm
        jet}=\frac{1}{2}m_{3/2}$.  The shaded region is the excluded 
        region for the case with $B_{\rm h}=0$.}
        \label{fig:mtr_gluino}
    \end{center}
\end{figure}

\begin{figure}[t]
    \begin{center}
        \centerline{{\vbox{\epsfxsize=10.0cm
        \epsfbox{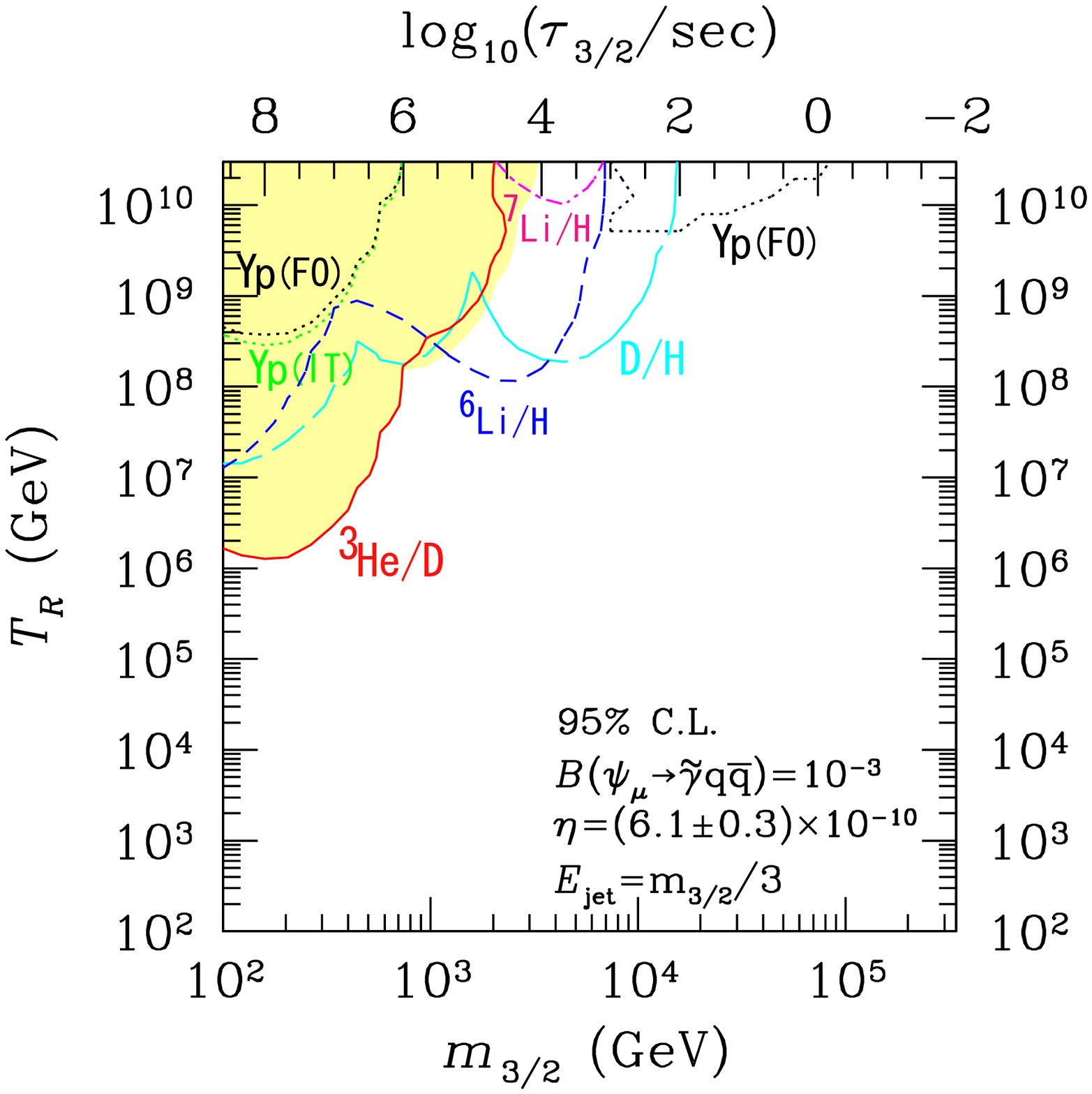}}}}
        \caption{Upper bounds on the reheating temperature as a
        function of the gravitino mass for the case where the
        gravitino dominantly decays into photon-photino pair. 
        Here, we take $B_h=10^{-3}$,  
        $E_{\rm vis}=\frac{1}{2}m_{3/2}$, and
        $E_{\rm jet}=\frac{1}{3}m_{3/2}$. 
        The shaded region is the excluded 
        region for the case with $B_{\rm h}=0$.}
        \label{fig:mtr_photino}
    \end{center}
\end{figure}

In Figs.\ \ref{fig:mtr_gluino} and \ref{fig:mtr_photino}, we plot the
upper bound on the reheating temperature for the cases where the
gravitino dominantly decays into gluon-gluino and photon-photino pairs
at 95\% C.L., respectively.  For comparison, we also calculated the
light-element abundances for the case neglecting the effects of the
hadronic decay modes (i.e., for $B_{\rm h}=0$), which only takes into
account the effects of the photodissociation.  Excluded region for
such a case is also shown in the figures by the shaded region.

If the hadronic branching ratio is large (i.e, $B_h\sim 1$), the
constraint on the reheating temperature is very stringent.  For example,
in the gravity-mediated supersymmetry breaking scenario, the gravitino
mass becomes comparable to the masses of the superpartners of the
standard-model particles, and hence $m_{3/2}\sim {\cal O}(100)\ {\rm
GeV}$ is expected.  In this case, the reheating temperature should be
lower than $10^5-10^7\ {\rm GeV}$.  In the anomaly-mediated
supersymmetry breaking scenario \cite{amsb}, however, the gravitino mass
can be two to three orders of magnitude larger than the masses of the
squarks and sleptons.  In this case, the upper bound is slightly
relaxed, and we obtain $T_{\rm R}\lesssim 10^7-10^{10}\ {\rm GeV}$.

For the case where the gravitino dominantly decays into the
photon-photino pair, the upper bound becomes higher.  However, the
constraint is still much more stringent than the case with $B_h=0$.
For $m_{3/2}\sim {\cal O}(100)\ {\rm GeV}$, the upper bound is given
by $10^6-10^8\ {\rm GeV}$.  With larger gravitino mass, the constraint
becomes less stringent.  Since we are assuming $B_h=10^{-3}$ here,
however, colored superparticles should be heavier than $m_{3/2}$.
Such a mass spectrum looks quite unnatural from the naturalness point
of view.

Note that the above upper bounds on the reheating temperature can be
alleviated if the primordial gravitinos are diluted by the decay of
particles (such as the thermal inflaton or the moduli fields).
Although again gravitino may be also produced directly by the decay or
after the reheating via the scattering processes, there are a set of
model parameters to make the bounds milder and resolve the gravitino
problem.  (For the recent study, see, for e.g.,
Ref.~\cite{Kohri:2004qu}).

The upper bound on the reheating temperature provides significant
information about the evolution of the universe, in particular, for
the successful scenario of baryogenesis.  Since the primary baryon
asymmetry is diluted by the inflation, baryogenesis should occur at
the temperature lower than $T_{\rm R}$.  For some scenario of
baryogenesis, this becomes a very stringent constraint.  In
particular, for the Fukugita-Yanagida scenario for baryogenesis
\cite{Fukugita:1986hr} where the lepton asymmetry generated by the
decay of the right-handed neutrino is converted to the baryon
asymmetry via the spharelon process, mass of the right-handed neutrino
should be heavier than $10^9-10^{10}\ {\rm GeV}$ \cite{M_nuR}.  Thus,
for successful scenario of leptogenesis, the reheating temperature is
required to be higher than $10^9-10^{10}\ {\rm GeV}$.  Based on our
analysis of the BBN, however, such high reheating temperature is
allowed in a very limited case, if the gravitino is unstable.  Thus,
in order to realize the Fukugita-Yanagida scenario, we may have to
make some assumption on the property of the gravitino, like a very
heavy gravitino, a light stable gravitino, or an invisible decay of
the gravitino (like decay into the axion-axino pair).

\section{Conclusions and Discussion}
\label{sec:conclusion}
\setcounter{equation}{0}

In this paper, we have studied the BBN scenario with exotic
late-decaying particle $X$.  If there exist an exotic particle with
the lifetime longer than $\sim 0.1\ {\rm sec}$, it decays after the
BBN starts.  Decay products of $X$ may cause various non-standard
processes, like the photodissociation and hadrodissociation of the
background nuclei, and the inter-conversion between the proton and
neutron.  As a result, abundances of the light elements are affected
by the decay of $X$.  Since the theoretical predictions of the SBBN
scenario is in a reasonable agreements with the observation, decay of
such long-lived particle during/after the BBN may spoil the success of
the SBBN scenario.

We have derived upper bound on the primordial abundance $X$, paying
particular attention to the hadronic decay mode of $X$; as we have
seen, as the hadronic branching ratio $B_{\rm h}$ becomes larger,
upper bound on the primordial abundance of $X$ becomes smaller.  In
particular, for the case where the lifetime of $X$ is $10^{-2}\ {\rm
sec}\lesssim\tau_X\lesssim 10^{2}\ {\rm sec}$, inter-conversion
induced by the emitted mesons and nucleons provides the most stringent
constraint.  For longer lifetime ($10^{2}\ {\rm
sec}\lesssim\tau_X\lesssim 10^{7}\ {\rm sec}$), hadrodissociation of
the background $\alpha$ becomes effective, resulting in non-thermal
productions of various light elements (in particular, ${\rm D}$ and
${\rm ^6Li}$).  If we consider the case with the long enough lifetime
($10^{7}\ {\rm sec}\lesssim\tau_X\lesssim 10^{12}\ {\rm sec}$), ${\rm
^3He}$ is overproduced by the photodissociation and
hadrodissociation of $\alpha_{\rm BG}$, which gives the most
stringent constraint on the primordial abundances of $X$.
(See figures in Section \ref{sec:results}.)

Since our analysis has been done without specifying the detailed
properties of $X$, our results can be applied to various exotic
long-lived particles.  Some of the examples are the gravitinos,
cosmological moduli fields, and the NLSP for the case where the
gravitino is the LSP.  As an application of our analysis, we
considered the unstable gravitino and applied our results to the
gravitino problem.  In particular, for the cases where the gravitino
dominantly decays into gluon-gluino and photon-photino pairs, we
derived the upper bound on the reheating temperature after inflation.
For the case of the gravity-mediated supersymmetry breaking, gravitino
mass is expected to be $\sim {\cal O}(100)\ {\rm GeV}$.  In this case,
even if the hadronic branching ratio is $\sim {\cal O}(10^{-3})$, the
reheating temperature is constrained to be smaller than
$10^6-10^8~{\rm GeV}$.  If the gravitino mass is much larger than
$\sim {\cal O}(100)\ {\rm GeV}$, the constraint on $T_{\rm R}$ may be
relaxed.  With such gravitino, however, the hadronic branching ratio
would be close to $1$ since, in such a case, all the superpartners of
the standard-model particles are expected to be lighter than the
gravitino from the naturalness point of view.  (Such a mass spectrum
may be realized in the anomaly-mediated supersymmetry breaking
scenario.)  For $m_{3/2}\sim {\cal O}(10-100)\ {\rm TeV}$ with $B_{\rm
h}\sim 1$, the upper bound is given by $T_{\rm R}\lesssim
10^7-10^{10}~{\rm GeV}$.  (See figures in Section \ref{sec:grav}.)

\section*{Acknowledgements}

The authors wish to thank J.~Arafune, A.~Brandenburg, R.H.~Cyburt,
T.~Hatsuda, J.E. Kim, A.~Kohama and K.~Maki for valuable discussions
and suggestions.  K.K. also thanks T.~Asaka, O.~Biebel, S.~Mihara,
M.M.~Nojiri, H.~Sakai, T.~Sj\"ostrand, F.~Takahara and J.~Yokoyama for
useful comments. The comments by the anonymous referee are also
suggestive  in making the final draft. This work was partially
supported by the Grants-in Aid of the Ministry of Education, Science,
Sports, and Culture of Japan No.~14540245 (MK), 15-03605 (KK) and
15540247 (TM).

\appendix

\section{Photon Spectrum}
\label{app:photon}
\setcounter{equation}{0}

In this appendix, we summarize the properties of the photon spectrum
generated by the high energy photon from radiative decay of a heavy
unstable particle.  (Details of the calculation of the photon spectrum
is discussed in \cite{KawMor}.)

Once the high energy photon is injected into the thermal bath
consisting of the photon, (non-relativistic) electron, and nuclei, the
high energy photon induces cascade processes.  We have calculated the
photon spectrum taking account of effects of the following processes:
\begin{itemize}
\item Injection of the high energy photon from the radiative decay of
    $X$
\item Double photon pair creation ($\gamma+\gamma_{\rm BG}\rightarrow
    e^{+}+e^{-}$)
\item Photon-photon scattering
    ($\gamma+\gamma_{\rm BG}\rightarrow\gamma+\gamma$)
\item Compton scattering off thermal electron
    ($\gamma+e^{-}_{\rm BG}\rightarrow\gamma +e^{-}$)
\item Inverse Compton scattering off background photon
    ($e^{\pm}+\gamma_{\rm BG}\rightarrow e^{\pm}+\gamma$)
\item Pair creation in background proton (and $\alpha_{\rm BG}$)
    ($\gamma+p_{\rm BG}\rightarrow e^{+}+e^{-}+p$)
\end{itemize}
Here, the subscript ``BG'' indicates that the corresponding particles
are in the thermal bath.  In addition, in the process $\gamma+ p_{\rm
BG}\rightarrow e^{+}+e^{-}+p$, the background proton plays the role of
the source of the electric field.

The photon spectrum is determined by following the distribution
functions of the photon and electron, $f_\gamma$ and $f_e$.  Since the
expansion rate of the universe is much smaller than the scattering
rates of the electromagnetic processes, the relevant Boltzmann
equations to be solved are written in the following forms
\begin{eqnarray}
    \frac{\partial f_{\gamma}(E_\gamma)}{\partial t}
    &=&
    \left[
        \frac{\partial f_{\gamma}(E_\gamma)}{\partial t} 
    \right]_{\rm DP}
    + \left[ 
        \frac{\partial f_{\gamma}(E_\gamma)}{\partial t} 
    \right]_{\rm PP}
    + \left[ 
        \frac{\partial f_{\gamma}(E_\gamma)}{\partial t} 
    \right]_{\rm PC}
    + \left[ 
        \frac{\partial f_{\gamma}(E_\gamma)}{\partial t} 
    \right]_{\rm CS}
    \nonumber \\ &&
    + \left[ 
        \frac{\partial f_{\gamma}(E_\gamma)}{\partial t} 
    \right]_{\rm IC}
    + \left[ 
        \frac{\partial f_{\gamma}(E_\gamma)}{\partial t} 
    \right]_{\rm DE},
    \\
    \frac{\partial f_{e}(E_e)}{\partial t}
    &=&
    \left[ 
        \frac{\partial f_{e}(E_e)}{\partial t} 
    \right]_{\rm DP}
    + \left[ 
        \frac{\partial f_{e}(E_e)}{\partial t} 
    \right]_{\rm PC}
    + \left[ 
        \frac{\partial f_{e}(E_e)}{\partial t} 
    \right]_{\rm CS}
    + \left[ 
        \frac{\partial f_{e}(E_e)}{\partial t} 
    \right]_{\rm IC}
    \nonumber \\ &&
    + \left[ 
        \frac{\partial f_{e}(E_e)}{\partial t} 
    \right]_{\rm DE},
\end{eqnarray}
where the terms with the subscripts ``DP,'' ``PP,'' ``PC,'' ``CS,''
``IC'' and ``DE'' denote the contributions of double-photon pair
creation, photon-photon scattering, pair creation in nuclei, Compton
scattering, inverse Compton scattering, and the radiative decay of
$X$, respectively.

Importantly, for the energy region $E_\gamma\ll E_{\gamma,0}$ (with
$E_{\gamma,0}$ being the energy of the injected photon from the
radiative decay of $X$), the function $f_\gamma (E_\gamma)$ is
determined by the total amount of injected energy per unit time; with
this quantity being fixed, $f_\gamma (E_\gamma)$ is insensitive to
$E_{\gamma,0}$.  Thus, in the BBN with the radiatively decaying
particle $X$, resultant abundances of the light elements primarily
depend on the lifetime of $X$ and the combination $E_{\gamma,0}Y_X$,
but are insensitive to $E_{\gamma,0}$ if the combination
$E_{\gamma,0}Y_X$ is fixed.  (For the case with the hadronic decay
processes, they also depend on the hadronic branching ratio.)

\begin{figure}[t]
    \begin{center}
        \centerline{{\vbox{\epsfxsize=8.0cm
        \epsfbox{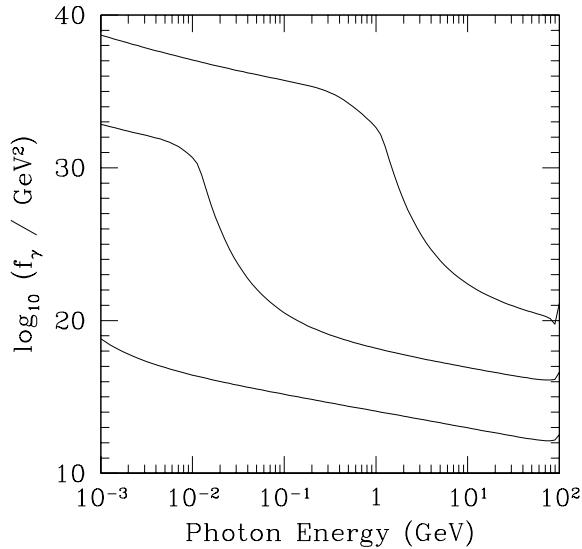}}}}
        \caption{The photon spectrum for the background 
        temperature $T=10\ {\rm eV}$, $1\ {\rm keV}$, and $100\ {\rm
        keV}$ (from above).  Here, $E_{\gamma,0}=100\ {\rm GeV}$ and
        the flux of the injected photon is $[\partial
        f_\gamma/\partial t]_{\rm DE}=\delta (E_\gamma -
        E_{\gamma,0})\ {\rm GeV^{-4}}$ while $[\partial f_e/\partial
        t]_{\rm DE}=0$.}
        \label{fig:photonspec}
    \end{center}
\end{figure}

In Fig.\ \ref{fig:photonspec}, we plot the photon spectrum
$f_\gamma(E_\gamma)$ for several values of the background temperature
$T$.  As we mentioned, first of all, $f_\gamma(E_\gamma)$ is
insensitive to the energy of the injected photon as far as the total
amount of the injected energy per unit time is fixed.  In addition,
$f_\gamma(E_\gamma)$ drastically changes its value at $E_\gamma\sim
m_e^2/22T$.  This can be understood as follows.  Photons with high
enough energy easily scatter off the background photon and create
$e^+e^-$ pairs.  Since the number density of the background photon is
much larger than those of the electron and nuclei, the pair creation
process is extremely efficient for photons with $E_\gamma\gtrsim
m_e^2/22T$ the number density of the photons with such energy is
suppressed.  As $E_\gamma$ becomes lower, however, the process
$\gamma+\gamma\rightarrow e^++e^-$ is kinematically blocked, and hence
$f_\gamma(E_\gamma)$ is no longer suppressed.

\section{Energy Loss Rates}
\label{sec:energy_loss_rates}
\setcounter{equation}{0}

In this Appendix, we summarize the energy-loss rates of high-energy
hadrons in the electromagnetic plasma.  Here, we consider the
energy-loss of the hadron with mass $m_H=Am_N$ and charge $Z$.  In
addition, $E_{\rm kin}$, $\gamma$, and $\beta$ are used for the
kinetic energy, Lorentz factor, and velocity of the hadron,
respectively.

In discussing the energy-loss processes via the electromagnetic
interactions, one of the most important processes is the scattering
with background electron (and positron).  In calculating the
scattering rate, it is necessary to determine the number density of
the background electron (and positron).  In our analysis, for the sum
of the number density of the electron and positron in the background
(called $n_e$), we use the following approximated formula which well
agrees with the exact formula:
\begin{eqnarray}
    \label{eq:nee}
    n_{e} = n_{e^{-}} + n_{e^{+}}   =\left\{
      \begin{array}{ll}
          \displaystyle{\frac32} n_{\gamma}
          &
          ~~:~~ T \ge m_{e}
          \\ \\
          4 \displaystyle{\kakko{\frac{m_{e}T}{2\pi}}}^{3/2} 
          e^{-m_{e}/T}
          &
          ~~:~~ m_{e} > T \ge m_{e}/26
          \\ \\
          \displaystyle{ \left( 1 - \frac{1}{2}Y \right) }
          \eta n_{\gamma}
          &
          ~~:~~ T < m_{e}/26
      \end{array}
    \right. .
\end{eqnarray}

\subsection{Energy loss of charged particles}

First, we consider the energy-loss rates of charged particles (i.e,
$p$, $\D$, $\T$, $\hethree$, $\hefour$, and $\pi^{\pm}$).  For those
charged particles, the Coulomb, Compton, and Bethe-Heitler scattering
are important.

First of all, we summarize the energy-loss rate through the Coulomb
scattering.  (For details, see Ref.~\cite{Reno:1987qw}.)  For the
temperature higher than the electron mass ($T \gtrsim m_{e}$), the
energy-loss rate of relativistic charged particle is given by
\begin{eqnarray}
    \label{eq:dedt_ch_rel_high}
    \left[ \frac{dE_{\rm (ch)}}{dt} \right]_{\rm Coulomb}
    = - \frac{\pi}{3}
    Z^2 \alpha_{\rm em}^{2} T^{2} \Lambda 
    ~~:~~
    T \gtrsim m_{e},\ E_{\rm kin}\gtrsim m_H,
\end{eqnarray}
where $\Lambda=\ln [2/(1-\cos\theta_{\rm min})]-1$ with $\theta_{\rm
min}$ being the minimal scattering angle in the center-of-mass frame.
In discussing the energy-loss process in the thermal plasma,
$\theta_{\rm min}$ is determined by the requirement that the energy
transfer to the electron in the comoving frame be smaller than the
plasma frequency $\omega_{\rm p}$ \cite{Jackson} which is given by
\begin{eqnarray}
    \omega_{\rm p}^2 = \frac{4\pi\alpha_{\rm em} n_e}{m_e}.
\end{eqnarray}
Thus, in our analysis, we use the approximated formula
\begin{eqnarray}
    \Lambda \simeq 
    \ln \left( 
        \frac{\tilde{\gamma}^2\tilde{\beta}^2m_e}{\omega_p} \right),
\end{eqnarray}
where $\tilde{\beta}$ is the required velocity to go to the
center-of-mass frame and $\tilde{\gamma}=(1-\tilde{\beta}^2)^{-1/2}$.
Notice that, when the incident charged particle is non-relativistic in
the center of mass frame,
$\tilde{\gamma}\tilde{\beta}\simeq\gamma\beta$.  For nonrelativistic
charged particle, we obtain
\begin{eqnarray}
    \label{eq:dedt_ch_nonrel_high}
    \left[ \frac{dE_{\rm (ch)}}{dt} \right]_{\rm Coulomb}
    = -
    \frac{4\pi}{9} 
    \frac{E_{\rm kin}}{m_H}
    Z^2 \alpha_{\rm em}^{2} T^{2} \Lambda 
    ~~:~~
    T \gtrsim m_{e},\ E_{\rm kin}\lesssim m_H.
\end{eqnarray}

For lower temperature ($T \lesssim m_{e}$), the formula of the
energy-loss rate changes.  For relativistic charged particles, we
obtain
\begin{eqnarray}
    \label{eq:dedt_ch_rel_low}
    \left[ \frac{dE_{\rm (ch)}}{dt} \right]_{\rm Coulomb}
    = - 4\pi Z^2 \alpha_{\rm em}^{2} \frac{n_{e}}{m_{e}}  \Lambda 
    ~~:~~
    T \lesssim m_{e},\ E_{\rm kin}\gtrsim m_H.
\end{eqnarray}
For non-relativistic particles, we need special care since the
energy-loss rate strongly depends on the relative velocity between the
charged hadron and the background electron.  Denoting the distribution
function of the background electron (as a function of its velocity
$\beta_e$) as $f_{e_{\rm BG}} (\beta_e)$, the energy-loss rate of the
non-relativistic hadrons is given by
\begin{eqnarray}
    \left[ \frac{dE_{\rm (ch)}}{dt} \right]_{\rm Coulomb}
    &=& 
    - \frac{4\pi Z^2 \alpha_{\rm em}^{2}}{m_e\beta} \Lambda
    \int_{\beta_e<\beta} d^3 \vec{\beta}_e     
    f_{e_{\rm BG}} (\beta_e)
    \nonumber \\ &&
    - \frac{4\pi Z^2 \alpha_{\rm em}^{2}\beta^2}{3m_e} \Lambda
    \int_{\beta_e>\beta} d^3 \vec{\beta}_e     
    \beta_e^{-1} f_{e_{\rm BG}} (\beta_e)
    ~~:~~
    T \lesssim m_{e},\ E_{\rm kin}\lesssim m_H.
    \nonumber \\
    \label{dE/dt(nr,lowT,full)}
\end{eqnarray}
Notice that, the integral in the first term of the right-hand side of
the above equation is the number density of the background electron
with velocity $\beta_e<\beta$.  Since the hadron and the electron are
both non-relativistic, contribution of the background electron with
$\beta_e>\beta$ more suppressed than the contribution of the electrons
with $\beta_e<\beta$.  Then, neglecting the second term of the
right-hand side of Eq.\ (\ref{dE/dt(nr,lowT,full)}), we obtain
\begin{eqnarray}
    \left[ \frac{dE_{\rm (ch)}}{dt} \right]_{\rm Coulomb}
    \simeq
    -\frac{4\pi\alpha_{\rm em}^{2}Z^{2}n_e}{m_{e}\beta} 
    I(\beta/\sqrt{2T/m_e}) \Lambda,
\end{eqnarray}
where
\begin{eqnarray}
    I (r) = \frac{4}{\sqrt{\pi}} \int_0^r dx x^2 e^{-x^2}.
\end{eqnarray}

Another important process is the Compton scattering.  In particular,
ultra-relativistic charged particles may lose significant amount of
energy by scattering off the background photons.  For the Compton
scattering process, the energy loss rate is approximately given by
\begin{eqnarray}
    \label{eq:dedt_compton}
    \left[ \frac{dE_{\rm (ch)}}{dt} \right]_{\rm Compton} = 
    - \frac{32 \pi^{3}}{135} \alpha_{\rm em}^{2} 
    \frac{\gamma^{2} -1}{m_H^{2}} T^{4}.
\end{eqnarray}
We use this formula in our analysis.

Next, we consider the Bethe-Heitler process (e.g., for the proton
projectile $p+\gamma\to p+e^{+}+e^{-}$) \cite{Bethe:1934za}.  This
process is important for relativistic charged nuclei with energy
$E\gtrsim A m_H m_{e} /\epsilon_{\gamma} \sim 1.6 A \gev \times
(T/0.1\ \mev)^{-1}$, where $A$ is the mass number and
$\epsilon_{\gamma}$ is the photon energy. The fitting formula of the
energy loss rate is given by \cite{Blumenthal:nn}
\begin{eqnarray}
    \label{eq:Bethe-Heitler}
    \left[ \frac{dE_{\rm (ch)}}{dt} \right]_{\rm BH} 
    = - \alpha_{\rm em} r_{0}^{2} Z^{2} m_{e}^{4}
    \int^{\infty}_{2} d\kappa 
    f_{\gamma_{\rm BG}} (\kappa/2\gamma)
    \frac{\phi(\kappa)}{\kappa^{2}},
\end{eqnarray}
where $r_{0}$ is the classical electron radius ($=\alpha_{\rm
em}/m_{e}$), and $f_{\gamma_{\rm BG}}$ is the distribution function of
the background photon which is given by
\begin{eqnarray}
    \label{eq:n_photon}
    f_{\gamma_{\rm BG}} (\epsilon) = \kakko{\frac{\epsilon}{\pi}}^{2}
    \frac1{\exp(\epsilon/T) - 1}.
\end{eqnarray}
In addition, the function $\phi(\kappa)$ in the integrand is fitted
by \cite{chodorowski:1992}
\begin{eqnarray}
    \label{eq:phi_kappa}
    \phi(\kappa) = \left\{
      \begin{array}{ll}
          \displaystyle{
          \frac{\pi}{12} \ \frac{(\kappa-2)^{4}}{1+\sum_{i=1}^{4} a_{i}
          (\kappa-2)^{i}}
          }
          &
          ~~:~~ \kappa \le 25
          \\ \\
          \displaystyle{
          \frac{\kappa \sum_{j=0}^{3} b_{j}
          \ln^{j}(\kappa)}{1-\sum_{k=1}^{3} c_{k}/\kappa^{k}}
          }
          &
          ~~:~~ \kappa > 25
      \end{array}
    \right. ,
\end{eqnarray}
where the coefficients are given by
\begin{eqnarray}
    &&
    a_{1} = 0.8048 ,~~
    a_{2}=0.1459,~~
    a_{3}= 1.137\times10^{-3},~~
    a_{4} = - 3.879\times 10^{-6}, 
    \\ &&
    b_{0} = -86.07,~~
    b_{1} = 50.96,~~
    b_{2} = -14.45,~~
    b_{3} = \frac83,
\end{eqnarray}
and
\begin{eqnarray}
    \label{eq:ci}
    c_{1} = 2.910,~~
    c_{2} = 78.35,~~
    c_{3} = 1.837 \times 10^{3}.
\end{eqnarray}

Although we include the photo-pion process ($p + \gamma \to p(n) +
\pi$) into our analysis, it is important only for highly-relativistic
nucleons with energy $E\gtrsim m_{N}\mpi/\epsilon_{\gamma}\sim
4.7\times 10^{2}\gev\times(T/0.1 \mev)^{-1}$.  The fitting formula of
the energy loss rate is approximately given by \cite{Berezinsky:wi}
\begin{eqnarray}
    \label{eq:dedt_photopion}
    \left[ \frac{dE_{\rm (ch)}}{dt}\right]_{\rm photo\mbox{-}pion}
    = \left\{
        \begin{array}{ll}
            \displaystyle{- \frac{2}{\pi^{2}}
            \sigma' \frac{\epsilon_{0}^{2}}{m_N}
            T^{3} E \exp\kakko{-\frac{\epsilon_{0} m_N}{2
            E T}}}
            & ~~:~~ E < \epsilon_{0} m_N/T
            \\
            &
            \\
            \displaystyle{- 1.8 \times 10^{-8} {\rm yr}^{-1} E
            \kakko{\frac{T}{2.7 \rm K}}^{3} }
            & ~~:~~ E \ge \epsilon_{0} m_N/T
        \end{array}
    \right. ,
\end{eqnarray}
where $\epsilon_{0}=\mpi(1+\frac{\mpi}{2m_N})$ is the (approximated)
threshold energy of the photon for the photo-pion process in the rest
frame of nucleon, and $\sigma'\simeq 6.8 \times 10^{-36} \cm^{2}/\ev$
is a constant.

\subsection{Energy loss of neutral particles}

Now, we consider the energy-loss of the neutral particles, in
particular, the neutron which is the only relevant neutral particle
for which the energy-loss rate should be discussed.  For the neutron,
scattering with the background electron with magnetic-moment
interaction is the most important process for the energy-loss.  (Even
for the neutron, we call such a process as ``Coulomb scattering.'')
For $T \gtrsim m_{e}$, we obtain
\begin{eqnarray}
    \left[ \frac{dE_n}{dt} \right]_{\rm Coulomb}
    = \left\{
        \begin{array}{ll}
            \displaystyle{
            - \frac{7\pi^{3}}{15}
            \alpha_{\rm em}^2 \frac{g_{n}^{2}}{m_N^{2}} 
            \kakko{\frac{E_{\rm kin}}{m_N}}^{2}
            T^{4}
            }
            &
            ~~:~~
            T \gtrsim m_{e},\ E_{\rm kin}\gtrsim m_N
            \\
            \\
            \displaystyle{
            -\frac{14\pi^{3}}{15}
            \alpha_{\rm em}^{2} \frac{g_{n}^{2}}{m_N^{3}} T^{4}
            E_{\rm kin}
            }
            &
            ~~:~~
            T \gtrsim m_{e},\ E_{\rm kin}\lesssim m_N
        \end{array}
    \right. ,
\end{eqnarray}
where $g_n\simeq -1.913$ is the neutron magnetic moment
\cite{Hagiwara:fs}.   Furthermore, for $T\lesssim m_e$, the
energy-loss rate is given by~\footnote{
When the temperature become so low (i.e., $T\lesssim m_e/26$) that the
number of the electron in the comoving volume becomes constant,
expression for the non-relativistic case given in Eq.\
(\ref{de/dt(n)}) becomes invalid.  For such a low temperature,
the energy-loss rate of the non-relativistic neutron is given by
\begin{eqnarray*}
    \left[ \frac{dE_n}{dt} \right]_{\rm Coulomb} =
    - \frac{4\alpha_{\rm em}^{2} g_{n}^{2}}{3\pi m_N^{3}}
    n_e \left(\frac{2\pi}{m_eT}\right)^{3/2}
    m_e^3 T E_{\rm kin}.
\end{eqnarray*}
At such low temperature, however, the energy-loss rate of the
non-relativistic neutron is negligible.  Thus, Eq.\ (\ref{de/dt(n)})
is enough for our analysis.}
\begin{eqnarray}
    \left[ \frac{dE_n}{dt} \right]_{\rm Coulomb}
    = \left\{
        \begin{array}{ll}
            \displaystyle{
            - \frac{3\pi \alpha_{\rm em}^2 g_n^2 m_e}{m_N^4} 
            n_{e} E_{\rm kin}^2
            }
            &
            ~~:~~ 
            T \lesssim m_{e},\ E_{\rm kin}\gtrsim m_N
            \\ \\
            \displaystyle{
            - \frac{16\alpha_{\rm em}^{2} g_{n}^{2}}{3\pi m_N^{3}}
            T^{4}
            e^{-x_{e}} G_{e}(x_{e}) E_{\rm kin}
            }
            &
            ~~:~~ 
            T \lesssim m_{e},\ E_{\rm kin}\lesssim m_N
        \end{array}
    \right. ,
    \label{de/dt(n)}
\end{eqnarray}
where $G_{e}(x_{e}) = x_{e}^{3} + 3  x_{e}^{2} + 6 x_{e} + 6$ with
$x_{e} = m_{e}/T$.

As well as the case of charged particles, the photo-pion process ($n +
\gamma \to n(p) + \pi$) is also important for highly-relativistic
neutrons, see Eq.~(\ref{eq:dedt_photopion}). We include this
photo-pion energy loss rate in our analysis.  Note also that, because
the energy loss rate through $n + \gamma \to n + \gamma$ is smaller
than the Coulomb energy-loss rate~\cite{Reno:1987qw,Gould:1993}, we do
not include this process in estimating the energy loss rate of the
beam neutron.

\section{Energy Transfer}
\label{sec:transfer_energy}
\setcounter{equation}{0}

In this appendix, we discuss the energy transfers into the
scattered/produced particles after the elastic and inelastic hadronic
scattering processes. There are three types of energetic nuclei after
the processes; scattered target nucleus ($p$ or $\alpha$), beam
nucleon after the scattering, and daughter nuclei which are produced
by the destruction of $\alpha$.  In order to study the evolution of
the hadronic shower, it is necessary to consider how the energy of the
initial-state particles are transferred to these final-state ones.

Throughout this appendix, $p_{i}=(E_{i},\vect{p}_{i})$ denotes the
four-dimensional momentum of $N_i$ (or $A_i$) with its mass $m_{i}$ in
the rest frame of the target particle (i.e., comoving frame of the
expanding universe), and $K_{i}=E_{i}-m_{i}$ is its kinetic energy.
The dashed quantities are those for the final state particles.  In
addition, the mass difference between the proton and neutron is not
important here so we use the approximated relation $m_{p}\simeq
m_{n}\equiv m_{N}$.  Furthermore, mass of a nucleus $A_i$ is also
approximated by the relation $m_{A_i}\simeq A_{A_i}m_N$ where
$A_{A_i}$ is the atomic number of $A_i$.

\subsection{Elastic scattering}
\label{sec:elastic}

We consider the elastic scattering of an energetic nucleon $N_i$ ($p$
or $n$) with a background nucleus $A_j$ (background proton or
$\alpha$) through the process $N_i(p_i)+A_j(p_j)\rightarrow
N_i(p'_i)+A_j(p'_j)$, where we denoted the four-momentum in the
parenthesises.  Then, $p_j=(m_{H_j},0,0,0)$ and
$p'_j=(E'_j,\vect{p'}_j)$, and the kinetic energy of the scattered
target ($p$ or $\alpha$) is given by
\begin{eqnarray}
    \label{eq:delta_E_el_bg}
     K'_j &=& E'_j - m_{A_j}.
\end{eqnarray}
Using the Mandelstam variable $t$, which is given by
\begin{eqnarray}
    \label{eq:mandelstan_t}
    t = ( p'_j - p_j )^2 = -2 m_{A_j} K'_j,
\end{eqnarray}
we obtain
\begin{eqnarray}
    \label{eq:delta_E_el_bg2}
     K'_j = \frac{-t}{2 m_{A_j}}.
\end{eqnarray}
Kinematically, the maximal possible value of $K'_j$ is given by
\begin{eqnarray}
    \label{eq:k2max}
    K'^{\rm (max)}_j =
    \frac{2m_{A_j} K_i}{(m_{N_i}+m_{H_j})^2 + 2m_{A_j} K_i}.
\end{eqnarray}

For the elastic scattering processes, the distribution of $t$ is well
approximated by the following form with the slope parameter $B_{\rm sl}$
\begin{eqnarray}
    \label{eq:dist_el}
    \frac1{ \sigma^{\rm (el)}} \frac{d \sigma^{\rm (el)}}{d t} 
    = \daikakko
    {\frac1{ \sigma^{\rm (el)}} 
      \frac{d \sigma^{\rm (el)}}{d t} }_{t = 0 }
    \exp(- B_{\rm sl} \abs{t}),
\end{eqnarray}
where $\sigma^{\rm (el)}$ is the cross section.  

\begin{figure}
    \centering
    \centerline{{\vbox{\epsfxsize=8.0cm\epsfbox{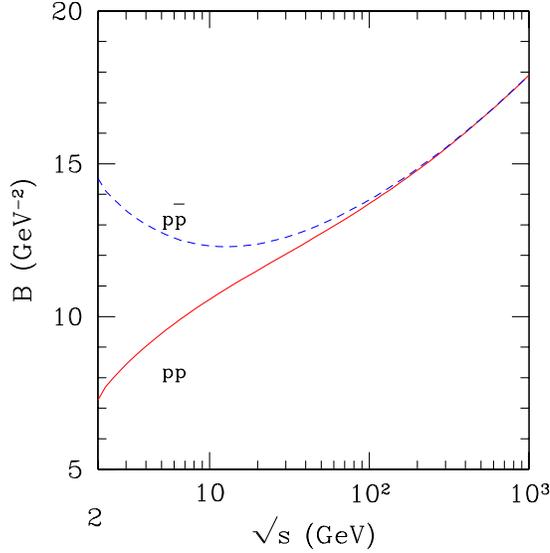}}}}
    \caption{$B_{\rm sl}$ as a function of the center-of-mass energy
    $\sqrt{s}$.  The solid (dashed) line denotes the slope parameter
    for the $pp \to pp$ ($p\bar{p} \to p\bar{p}$) process.  Here, the
    fitting formula given in \cite{Block:1985rf} is used.}
    \label{fig:bs}
\end{figure}

For the process $pp \to pp$, an accurate fitting formula of $B_{\rm
sl}$ as a function of the center-of-mass energy $\sqrt{s}$ is given in
Ref.\ \cite{Block:1985rf}.  The error of the fit is within 10 \%. In
Fig.~\ref{fig:bs} we plot $B_{\rm sl}$ for $pp \to pp$ as a function
of $\sqrt{s}$.  We use the fitting formula given in Ref.\ 
\cite{Block:1985rf} to calculate the slope parameter.~\footnote{
We use the fitting formula even for $2\ {\rm
GeV}\lesssim\sqrt{s}<4~{\rm GeV}$ although primarily the formula was
derived for $\sqrt{s}\gtrsim 4\ {\rm GeV}$ \cite{Block:1985rf}.  Note
that the experimental data of the cross sections suggest that the
energy dependence of $B_{\rm sl}$ is very mild. }
We also adopt the same formula for the elastic processes in which the
projectile is neutron (i.e., $n \pbg \to n p$).  Notice that the
Coulomb correction factor for the cross section is within 6\% for $K_n
= 100\ {\rm MeV}$ (15\% for $K_n = 10\ {\rm MeV}$). Therefore, the
error of the transfered energy $K'_p$ for $np \to np$ should be within
20\%.  With Eq.\ (\ref{eq:dist_el}), we can estimate the averaged
transfered energy as
\begin{eqnarray}
    \label{eq:averaged_kin}
    \staave{K'_j} 
    \equiv
    \int dt K'_j \frac1{ \sigma^{\rm (el)}} 
    \frac{d \sigma^{\rm (el)}}{d t}
    \sim \frac1{2 m_{A_j} B_{\rm sl}}
\end{eqnarray}
For example, for the process $p \pbg \to pp$ with $K_p=5\ {\rm GeV}$,
we can use this formula to obtain $\langle K'_p\rangle\sim 60\ {\rm
MeV}$.

For the elastic $p \alpha$ scattering, we use $B_{\rm sl} = 28\ {\rm
GeV}^{-2}$ \cite{grebenjuk:1989}, which is derived from the experiment
with the energetic proton with its kinetic energy $695-991\ {\rm
MeV}$.  Although we neglect the energy-dependence of $B_{\rm sl}$ for
the elastic scattering processes with $\alpha_{\rm BG}$, such
approximation would not significantly affect our results.  This is
because the averaged value of the transfered energy through $p \alpha
\to p \alpha$ process is $\staave{K'_\alpha} \sim 5\ {\rm MeV}$ and
the energy transfer in this process is less efficient than $p \pbg \to
p p$.  We also adopt this value in case of the elastic scattering of
high energy neutron, $n \alpha \to n \alpha$.

Then, from Eq.\ (\ref{eq:dist_el}), we derive the energy distribution
of the final-state $A_j$ as
\begin{eqnarray}
    \label{eq:f_el}
    f^{A_j}_{N_i+A_j \rightarrow N_i+A_j}
    (K_i, K'_j) =
    \frac{2 m_{A_j} B_{\rm sl}}
         {1 - \exp(-B_{\rm sl} |t|^{\rm (max)})}
         \exp (- 2 m_{A_j} B_{\rm sl} K'_j),
\end{eqnarray}
where $|t|^{\rm (max)}=2m_{A_j}K'^{\rm (max)}_j$ is the maximal
possible value of $|t|$.  In our study, we use the above formula to
evaluate the energy distribution of the scattered $A_i$.  Notice that
the energy distribution of the $N_i$ in the final state can be derived
by using the relation $K'_i=K_i-K'_j$:
\begin{eqnarray}
    \label{eq:f_el2}
    f^{N_i}_{N_i+A_j \rightarrow N_i+A_j}
    (K_i, K'_i) =
    \frac{2 m_{A_j} B_{\rm sl}}
        {1 - \exp(-B_{\rm sl} |t|^{\rm (max)})}
        \exp \left[ - 2 m_{A_j} B_{\rm sl} (K_i - K'_i) \right].
\end{eqnarray}

\subsection{Inelastic scattering}
\label{sec:inelastic}

In this subsection, we discuss the inelastic scattering process of an
energetic nucleon $N_i$ with a background nucleus $A_j$.  As we have
discussed in the previous subsection, the differences of cross
sections for the $pp$ and $np$ processes are small.  Therefore we
apply the formulae obtained for the $pp$ and $p\alpha$ collisions to
$np$ and $n\alpha$ collisions, respectively.

Since we sometimes use the Lorentz transformation from the
center-of-mass frame to the rest frame of the target particle, it is
convenient to define the Lorentz factor as a function of the kinetic
energy of the beam nucleon.  Explicit expression of the Lorentz factor 
is given by
\begin{eqnarray}
    \label{eq:gamma_cm_lab}
    \gamma^{\rm (CM)} (K_i) 
    = \frac{ (m_{N_i} + m_{A_j}) + K_i }{ E^{\rm (CM)} (K_i) },
\end{eqnarray}
where the energy in the center-of-mass frame is given by
\begin{eqnarray}
    \label{eq:Ecm}
    E^{\rm (CM)} (K_i)
    = \sqrt{ (m_{N_i} + m_{A_j})^2 + 2 m_{A_j} K_i }.
\end{eqnarray}

\subsubsection{Inelastic scattering with background proton}

First, we consider the inelastic scattering between high-energy proton
(neutron) with background proton.  The most important inelastic
processes are the ones with one pion production: $p+\pbg\to
p+p(n)+\pi$ (or $n+\pbg\to n+p(n)+\pi$).  Indeed, the cross sections
of these processes are much larger than those of other inelastic
processes.   Hereafter, we consider the process $p(p_i)+\pbg(p_j)\to
p(p'_i)+N_j(p'_j)+\pi(p'_\pi)$.

In order to study the energy distribution of $p$ in the final state
(i.e., distribution of $K'_i$), it is useful for us to introduce
Feynman's $x$-parameter.  Let us decompose the three-momentum
$\vect{p}'_i$ as
\begin{eqnarray}
    \vect{p}'_i = \vect{p}'_{i,\parallel} + \vect{p}'_{i,\perp},
\end{eqnarray}
where $\vect{p}'_{i,\parallel}$ and $\vect{p}'_{i,\perp}$ are
components of $\vect{p}'_i$ parallel to and perpendicular to the
direction of the initial-state energetic proton, respectively.  Then,
the Feynman's $x$-parameter is defined as
\begin{eqnarray}
    x \equiv \frac{p'_{i,\parallel}}{p'^{\rm (max)}_{i,\parallel}},
\end{eqnarray}
where $p'_{i,\parallel}=|\vect{p}'_{i,\parallel}|$ and $p'^{\rm
(max)}_{i,\parallel}$ is the maximal possible value of
$p'_{i,\parallel}$.  From experimental data on the high-energy
inclusive $pp$ reactions, it is known that the distribution of $x$ is
approximately independent of $\sqrt{s}$ (called ``Feynman scaling''
\cite{Feynman:ej,Perl:1974}).  Indeed, various experiments reported
that the kinetic energy of the final-state proton $K'_i$ in the
inclusive pion production process has flat distribution (see, for
e.g., \cite{Abolins:jk}).  Thus, with such experimental result, we
estimate the averaged value of $x$ as
\begin{eqnarray}
    \label{eq:inelasticity_ave}
    \langle x\rangle =
    \frac{
    \displaystyle{\int_{0}^{1} x \frac{d\sigma^{\rm (incl)}}{dx}} dx}
    {\displaystyle{\int_{0}^{1} \frac{d\sigma^{\rm (incl)}}{dx} dx }} 
    \sim 0.5.
\end{eqnarray}

Using the above result, we parameterize the energy of the beam
particle (i.e., $p$) in the final state using the ``inelasticity''
$\kappa_p$:
\begin{eqnarray}
    \label{eq:Edash1_inel}
    K'_i = (1 - \kappa_p) K_i.
\end{eqnarray}
Notice that, for high-energy pion production processes, $x\simeq
1-\kappa_p$.~\footnote{
Although, strictly speaking, the relation $x \simeq 1 - \kappa_p$
holds in the CM frame, it can be approximately satisfied even in
laboratory system for high-energy collision.}
In our analysis, we approximate that $\kappa_p$ is constant and use
$\kappa_p=0.5$ for the numerical calculations.~\footnote{
We have checked that, for the 20\ \% variation of the
$\kappa$-parameters, $\xi$-parameters change at most 15\ \%.  Such
uncertainty is relatively small compared to the total uncertainty of
the $\xi$-parameters due to the errors of the hadronic cross sections
(see section \ref{sec:results}).  Thus, in our analysis, errors from
the $\kappa$-parameters are neglected.}
That is, we approximate that the fraction of the energy-loss of the
initial-state energetic proton is always $\kappa_p$.  We also use the
above arguments in case of the energetic neutron injection.  Then, we
obtain the energy distribution of the final-state $p$ as
\begin{eqnarray}
    \label{eq:f_inel_beam}
    f^{p}_{(p,p_{\rm BG};2)}(K_i,K'_i) =
    f^{p}_{(p,p_{\rm BG};3)}(K_i,K'_i) =
    \delta \left( K'_i - (1-\kappa_p) K_i \right).
\end{eqnarray}

In order to study the energy distribution of other final-state
particles, we can use the experimental data of the momentum
distribution of the daughter particles from the $pp$ collision.  In
particular, if we consider some daughter particle $H_k$ produced by
the $pp$ collision, energy distribution is well fitted by the
following formula with the inverse slope parameter $K_{T}$
\cite{Alper:rw,Wong:jf}:
\begin{eqnarray}
    \label{eq:E_dsigma_dp}
    \frac{1}{M_{\rm T}}
    \frac{d \sigma^{\rm (inel)}}{dM_{\rm T} d y d \phi_{\rm a}} 
    \propto
    \exp (-M_{\rm T} / K_T),
\end{eqnarray}
where $M_T$ is the transverse mass which is give by
\begin{eqnarray}
    M_{\rm T}^2 = m_{H_k}^2 + |\vect{p}'_{H_k,\perp}|^2,
\end{eqnarray}
$y$ is the rapidity and $\phi_{a}$ is the azimuthal angle. Remarkably,
for any kind of the daughter particles such as protons, pions, and the
other mesons, Eq.\ (\ref{eq:E_dsigma_dp}) provides a good
approximation.  In addition, the inverse slope parameter is not
sensitive to $\sqrt{s}$.

It is notable that $K_{T}$ provides an typical kinetic energy of the
daughter particles in the CM frame since $\langle M_{\rm T}\rangle\sim
K_{T}$.  This is interpreted as the consequence of the fact that the
fragmentation scheme into hadrons is controlled by QCD and that
$K_{T}$ represents the typical energy scale of the fragmentation.
Using the results of the resent experiments of high-energy $pp$ and
$p\beseven$ collisions, we adopt the following value for $K_T$
\cite{Kaneta:ab,kaneta:1999}:
\begin{eqnarray}
    \label{eq:KT_exp}
    K_{T} =  140 \pm 15 \ \mev.
\end{eqnarray}

From the distribution of $M_{\rm T}$, we estimate the distribution of
the kinetic energy of the scattered target $N_j$.  Unfortunately,
direct information of $|\vect{p}'_{j,\parallel}|$ is not available.
Thus we assume that, at the high-energy scattering processes we are
interested in, the daughter particles are distributed isotropically in CM
system.  With this assumption, we estimate the averaged value of the
kinetic energy of $N_j$ as
\begin{eqnarray}
    \label{eq:Klab_pp}
    \langle K'_j \rangle_{pp} 
    = \gamma^{\rm (CM)} (K_i) K_T
    + \left[ \gamma^{\rm (CM)} (K_i) - 1 \right] m_N.
\end{eqnarray}
For simplicity, we approximate that the scattered target always have 
the kinetic energy $\langle K'_j \rangle_{pp}$ and hence the 
distribution of the scattered particle is written as
\begin{eqnarray}
    \label{eq:f_inel_target}
    f^{p}_{(p,p_{\rm BG};2)} (K_i,K'_j)
    = f^{n}_{(p,p_{\rm BG};3)} (K_i,K'_j)
    = \delta ( K'_j - \langle K'_j \rangle_{pp} ).
\end{eqnarray}

In a similar fashion, we can also estimate the averaged kinetic energy
of the produced pion as
\begin{eqnarray}
    \langle K'_\pi \rangle_{pp} 
    = \gamma^{\rm (CM)} (K_i) K_T
    + \left[ \gamma^{\rm (CM)} (K_i) - 1 \right] m_\pi.
\end{eqnarray}
However, for pions, the time scale for the hadronic scattering processes
is always longer than that of the electromagnetic stopping processes or
the lifetime.  Thus, we do not have to consider the energetic pions.

\subsubsection{Inelastic scattering with $\alpha_{\rm BG}$}

Next we consider the energy transfer in the inelastic processes with
background $\alpha_{\rm BG}$.  In this process, the target particle
$\alpha_{\rm BG}$ may or may not destroyed so we have to consider
various final-states.  As in the case of the inelastic scattering with
the background proton, we denote the process as $p(p_i)+\alpha_{\rm
BG}(p_j)\to p(p'_i)+N_j(p'_j)+\pi(p'_\pi)+\cdots$.

For the energy distribution of the injected particle in the final
state, we use the similar treatment as the inelastic $pp$ collision.
That is, we define
\begin{eqnarray}
    \label{eq:Edash1_inel2}
    K'_i = (1 - \kappa_\alpha) K_i,
\end{eqnarray}
where $\kappa_\alpha$ is the inelasticity related to $\hefour$.  As in
the $pp$ collision case, we approximate that $\kappa_\alpha$ is a
constant and use the distribution function
\begin{eqnarray}
    \label{eq:f_inel4_beam}
    f^{p}_{(p,\alpha_{\rm BG};2)}(K_i,K'_i) =
    f^{p}_{(p,\alpha_{\rm BG};3)}(K_i,K'_i) =
    \delta \left( K'_i - (1-\kappa_\alpha) K_i \right).
\end{eqnarray}
Since 
we do
not have sufficient data to estimate $\kappa_\alpha$, we assume that 
$\kappa_\alpha=\kappa_p$ in our analysis.~\footnote{
In the so-called ``Cascade Model,'' where a nucleus such as $\hefour$
is approximately treated as a composition of many ``independent''
nucleons with uniform density, an approximate relation
$\kappa_\alpha\sim 1 -(1-\kappa_p)^{1.3 - 1.43}$ holds (see Ref.\
\cite{Nara:1999dz} and references therein).  If we adopt $\kappa_p\sim
0.5 $, we obtain $\kappa_\alpha\sim 0.59 - 0.63$ and the error in our
treatment is about $20 \%$.}

Now we consider the daughter nuclei produced by the inelastic
scattering processes with $\alpha_{\rm BG}$.  In our study, energy
distribution of the final-state nuclei is mostly approximated as
follows.  If kinematically allowed, we assume that all the produced
debris nuclei $A_k$ (except the injected proton) have kinetic energy
\begin{eqnarray}
    \langle K'_k \rangle_{p\alpha} 
    = \gamma^{\rm (CM)} (K_i) K_T
    + \left[ \gamma^{\rm (CM)} (K_i) - 1 \right] m_{A_k},
\end{eqnarray}
where we used the transverse-mass distribution given in Eq.\ 
(\ref{eq:E_dsigma_dp}).  (The total energy is assumed to be conserved
by the pion emissions.)  If $K_i$ is not large enough, the total
kinetic energy of the final-state particles becomes larger than $K_i$
with the above approximation.  In such cases, we assume equipartition
of the total momentum in the CM system. Therefore, our approximation
is surely conservative than the case that we always assume the
equipartition at every moment.

In the study of the shower evolution and the surviving probability of
the light elements produced by the hadrodissociation of $\alpha_{\rm
BG}$, we use the energy distribution obtained by the above procedure.
For some processes, however, experimental data provides better
information about the energy distributions.  In particular for the
processes $n+\alpha_{\rm BG}\rightarrow {\rm T}+\cdots$ and
$n+\alpha_{\rm BG}\rightarrow {\rm ^3He}+\cdots$, the daughter ${\rm
T}$ and ${\rm ^3He}$ play the role of the ``spectator'' particle and
their typical energies are known to be very small ($\sim {\cal O}(1)\ 
{\rm MeV}$) \cite{meyer:1972}.  For these processes, the energies of
${\rm T}$ and ${\rm ^3He}$ are overestimated if we naively use our
shower algorithm.~\footnote{
Since the energy of ${\rm T}$ and ${\rm ^3He}$ may be overestimated in
the study of the shower evolution, the energy of the final-state $p$
and $n$ may be underestimated in some case.  This may be the case when
the $K_i$ is so small that the equipartition of the total momentum in
the CM system is adopted.  Then, such an error is expected to be small
since, in such a case, energy of the nuclei are inversely proportional
to their masses.  In addition, our treatment is also justified from
the point of view of obtaining conservative constraints.}
For the process $n+\alpha\rightarrow {\rm T}+\cdots$, the experimental
data is available and we found that the energy distribution is the
final-state ${\rm T}$ is well approximated by
\begin{eqnarray}
    \frac{1}{\sigma_{n+\alpha\rightarrow {\rm T}+\cdots}}
    \frac{d\sigma_{n+\alpha\rightarrow {\rm T}+\cdots}}{dE_{\rm T}}
    \simeq 0.09831\ {\rm MeV}^{-1} \times
    \exp \left( - \frac{E_{\rm T}}{5.789\ {\rm MeV}} \right),
    \label{dET_exp}
\end{eqnarray}
independently of the energy of the beam particle.  The energetic
non-thermally produced ${\rm T}$ and ${\rm ^3He}$ play very
significant roles in the study of the non-thermally produced ${\rm
^6Li}$.  In calculating the number of the non-thermally produced ${\rm
^6Li}$, we adopt Eq.\ (\ref{dET_exp}) as energy distributions of ${\rm
T}$ and ${\rm ^3He}$ produced by the hadrodissociation processes.

\section{Method for the Numerical Calculation}
\label{app:numerical}
\setcounter{equation}{0}

In this appendix, we explain how we study the evolution of the
hadronic shower in our numerical calculation.  In our approximation,
once a high-energy hadron $H_i$ is injected into the thermal bath with
the initial energy $E_{H_i}^{\rm (in)}$, its energy is first decreased
down to $\tilde{E}^{(R=1)}_{H_i}$, then scatters off the background
nuclei (in our case, proton or $\alpha_{\rm BG}$).  Then, the energy
distribution of the final-state particles after the hadronic
scattering process is given by the distribution function
$G_{H_i\rightarrow H_j}$ defined in Eq.\ (\ref{eq:summed_dist_nuc1})
(or $\tilde{G}_{N_i\rightarrow N_j}$ given in Eq.\ (\ref{tilde-G})
with our approximation).  The evolution of the hadronic shower can be
in principle solved by using Eq.\ (\ref{eq:eq:summed_dist_1}) once the
initial spectra of the primordial hadrons directly emitted from $X$ is
given.  Since our purpose is to count the number of the non-thermally
produced/dissociated light elements, however, there is rather a simple
method which we use in our numerical calculation.
In our numerical calculation, we first prepare the energy bins for
kinetic energy of each relevant hadrons (in our method, we prepare the
energy bins for proton, neutron and ${\rm ^4He}$).  Hereafter, we
consider the case where the $I$-th energy bin has the center value
$E_I$ and the width $\Delta E_I$.  (Thus, $E_I\pm\frac{1}{2}(\Delta
E_I+\Delta E_{I\pm 1})=E_{I\pm 1}$.)  Here, $I$ runs from $0$ to
$N_{\rm bin}$ where $0$-th bin is for stopped hadrons while $N_{\rm
bin}$-th bin is for hadrons with maximal possible energy.  Then, we
calculate the accumulated number of hadrons fallen into each energy
bin during the evolution of the hadronic shower.

As we mentioned in Section\ \ref{sec:hadrodis}, in our approximation,
evolution of the hadronic shower is studied by following the
energy-loss processes of the proton and the neutron.  Thus, the first
thing to do is to calculate the number of the nucleon $N_j$ scattered
into $J$-th bin when one nucleon $N_i$ is injected into $I$-th bin.
By using $\tilde{G}$ given in Eq.\ (\ref{tilde-G}), we can calculate
such transfer matrix, which we call $\tilde{T}^{I,J}_{H_i,H_j}$, as
\begin{eqnarray}
    \tilde{T}^{I,J}_{N_i,N_j} =
    \tilde{G}_{N_i\rightarrow N_j} (E_I, E_J; T)
    \Delta E_J,
\end{eqnarray}
where $\tilde{E}^{(R=1)}_{N_i}$ satisfies the relation
$R^{N_i}(E_I,\tilde{E}^{(R=1)}_{N_i};T)=1$.  Since, in our case, the
energy of the nucleon always decreases after the hadronic
scattering processes, $\tilde{T}^{I,J}_{N_i,N_j}$ vanishes if $I\leq
J$.~\footnote{
Strictly speaking, $\tilde{T}^{I,I}_{N_i,N_j}$ does not vanish since
the energy bin has some finite width.  In our analysis, such
correction is taken into account.}

Next, we consider the case where one $N_i$ is injected into $I$-th
bin, and estimate the accumulated number of $N_j$ fallen into $J$-th
bin after (infinite number of) multiple hadronic scattering processes.
We call such quantity as $\tilde{U}^{I,J}_{N_i,N_j}$.  Using the fact
that $\tilde{T}^{I,J}_{N_i,N_j}=0$ for $I\leq J$, the following
relation holds
\begin{eqnarray}
    \tilde{U}^{I,J}_{N_i,N_j} =  \delta_{IJ} \delta_{ij} + 
    \sum_{K=J+1}^{I-1} \sum_{N_k}
    \tilde{T}^{I,K}_{N_i,N_k} \tilde{U}^{K,J}_{N_k,N_j}.
\end{eqnarray}
Thus, using the above relation, we can recursively determine
$\tilde{U}^{I,J}_{N_i,N_j}$ from $I=J=0$ by increasing $I$ and $J$.

For the hadronic decay of $X$, then the accumulated number of $N_j$
fallen into $J$-th bin
is calculated as
\begin{eqnarray}
    \tilde{S}_{N_j} (E_J) \Delta E_J 
    \equiv
    \sum_{l=0}^\infty \tilde{F}^{(l)}_{N_j} (E_J) \Delta E_J
    = \sum_I \sum_{N_i}
    F^{(0)}_{N_i} (E_I) \Delta E_I \tilde{U}^{I,J}_{N_i,N_j},
\end{eqnarray}
where $F^{(0)}_{N_i}$ is the distribution function of $N_i$ emitted
from the hadronic decay of $X$.  Substituting this relation into Eqs.\ 
(\ref{eq:def_xi}) and (\ref{eq:xi-alpha}), we calculate the
$\xi$-parameters for the light elements.

\section{Dissociation of Non-thermally Produced Li}
\label{sec:lisurviving}
\setcounter{equation}{0}

Here, we discuss the scattering and dissociation of the non-thermally
produced ${\rm Li}$ (and ${\rm Be}$).  Non-thermally produced $^6{\rm
Li}$ and $^7{\rm Li}$ with energy of the order of $(1 - 10)\ {\rm
MeV}$ can be destroyed by scattering off background particles before
being stopped.  Thus, in order to estimate the present abundance of
$^6{\rm Li}$ and $^7{\rm Li}$, it is necessary to understand the
surviving probability of these nuclei.

As other charged nuclei, the energetic ${\rm Li}$ loses its kinetic
energy during the propagation in the thermal bath by scattering off the
background $e$ and $\gamma$.  Since the scattering rate (and the
energy-loss rate) of $^6{\rm Li}$ is much larger than the expansion rate
of the universe, the evolution of the number density of the
non-thermally produced $^6{\rm Li}$ is governed by
\begin{eqnarray}
    \frac{1}{n_{\rm Li^{(N.T.)}}}
    \frac{dn_{\rm Li^{(N.T.)}}}{dt} =
    - n_p \sigma_{{\rm Li}+p\rightarrow\cdots} \beta_{\rm Li},
\end{eqnarray}
where $\sigma_{{\rm Li}+p\rightarrow\cdots}$ is the cross section for
the ${\rm Li}$ dissociation process.  Thus, the surviving rate of
${\rm Li}$ with its initial energy $E_{\rm Li}^{\rm (in)}$ is
calculated as~\footnote{
In fact, dissociation process ${\rm ^6Li}(p,{\rm ^3He}){\rm ^4He}$ may
occur even after $^6{\rm Li}$ is thermalized.  Such an effect is taken
into account in the BBN code we use and hence the surviving rate
$P_{{\rm ^6Li}\rightarrow {\rm ^6Li}}$ parameterizes the dissociation
of $^6{\rm Li}$ before thermalization.}
\begin{eqnarray}
    P_{{\rm Li}\rightarrow {\rm Li}} (E_{\rm Li}^{\rm (in)})
    &=&
    \exp \left[ 
        - \int_0^\infty dt 
        n_p \sigma_{{\rm Li}+p\rightarrow\cdots} \beta_{\rm Li}
    \right]
    \nonumber \\ &=&
    \exp \left[ 
        - \int_{E^{\rm (th)}_{{\rm Li}
        +p\rightarrow\cdots}}^{E_{\rm Li}^{\rm (in)}}
        d E_{\rm Li} \left( \frac{d E_{\rm Li}}{dt} \right)^{-1}
        n_p \sigma_{{\rm Li}+p\rightarrow\cdots} \beta_{\rm Li}
    \right],
    \label{P_Li}
\end{eqnarray}
where $(d E_{\rm Li}/dt)$ is the energy-loss rate of ${\rm Li}$, and
$E^{\rm (th)}_{{\rm Li}+p\rightarrow\cdots}$ is the threshold energy
for the destruction process of ${\rm Li}$.

\begin{figure}
    \begin{center}
        \centerline{{\vbox{\epsfxsize=8.0cm
        \epsfbox{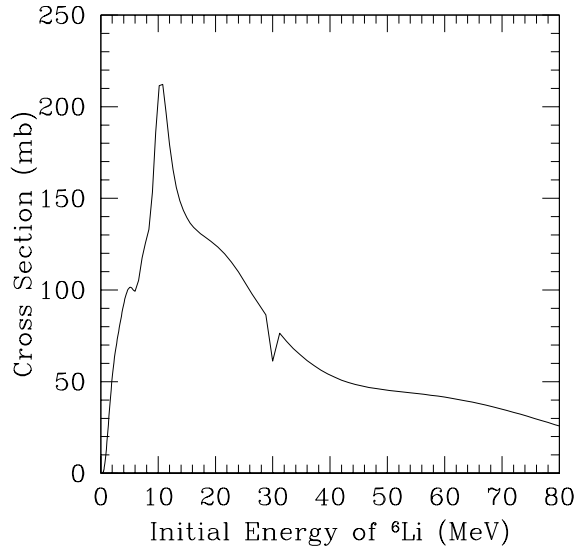}}}}
        \caption{Cross section for the process ${\rm ^6Li}(p,{\rm
        ^3He})^4{\rm He}$ as a function of the initial energy of
        $^6{\rm Li}$.  For the original data, see
        Refs.~\cite{7Li-4He,EXFOR}.}
        \label{fig:sigli6dissoc}
    \end{center}
\end{figure}

In the calculation of the surviving probability of $^6{\rm Li}$, the
dominant dissociation process is ${\rm ^6Li}(p,{\rm ^3He})^4{\rm
He}$.~\footnote{
There is another possible dissociation process of
${\rm ^6Li}(p, \gamma){\rm ^7Be}$.  Cross section for this process is,
however, much smaller than that for ${\rm ^6Li}(p,{\rm ^3He}){\rm ^4He}$
\cite{EXFOR}, and hence this process is irrelevant.}
Cross section for this process is given in Refs.~\cite{7Li-4He,EXFOR}.
For the readers' convenience, we show the plot of the cross section for
this process in Fig.\ \ref{fig:sigli6dissoc}.

\begin{figure}
    \begin{center}
        \centerline{{\vbox{\epsfxsize=8.0cm
        \epsfbox{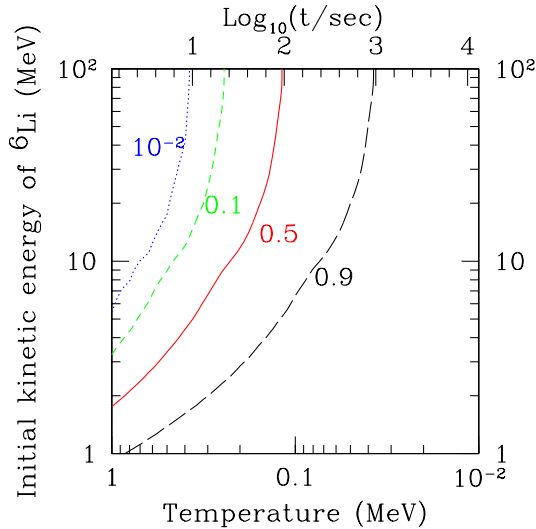}}}}
        \caption{Contours of constant $P_{{\rm ^6Li}\rightarrow {\rm
        ^6Li}}$.  The horizontal axis is the background temperature
        while the vertical axis is the initial kinetic energy of 
        ${\rm ^6Li}$.}
        \label{fig:ntli6_survival}
    \end{center}
\end{figure}

We numerically evaluate the surviving rate for ${\rm ^6Li}$ for
various values of its initial energy.  In Fig.\ 
\ref{fig:ntli6_survival}, we plot the contours of constant surviving
probability $P_{{\rm ^6Li}\rightarrow {\rm ^6Li}}$.  As one can see,
for $T\lesssim 50\ {\rm keV}$ when the thermal process ${\rm
^6Li}(p,{\rm ^4He}){\rm ^3He}$ becomes inefficient, the surviving rate
$P_{{\rm ^6Li}\rightarrow {\rm ^6Li}}$ becomes almost 1 for the
typical value of the initial energy of ${\rm ^6Li}$ (i.e., $E_{\rm
Li}^{\rm (in)}\sim (5-10)\ {\rm MeV}$).

For non-thermally produced $^7{\rm Li}$, the cross sections for the
processes $^7{\rm Li}(p,{\rm ^4He}){\rm ^4He}$ and $^7{\rm Li}(p, {\rm
^3He})X$ are available.  We use the cross sections for these processes
given in Refs.~\cite{7Li-4He_low,7Li-4He,EXFOR} and
Ref.~\cite{7Li-3He,EXFOR}, respectively.  Then, we calculate $P_{{\rm
^7Li}\rightarrow {\rm ^7Li}}$ and found that $P_{{\rm ^7Li}\rightarrow
{\rm ^7Li}}$ is also close to 1 for the cases we are interested in.

Finally, we comment on the surviving rate of non-thermally produced
$^7{\rm Be}$.  Unfortunately, the cross sections for the dissociation
processes of non-thermally produced $^7{\rm Be}$ is not available.
Thus, we use the dissociation cross sections of $^7{\rm Li}$ also for
the dissociation processes of $^7{\rm Be}$.

Although the surviving rate $P_{{\rm ^6Li}\rightarrow {\rm ^6Li}}$ is
almost 1 for the case where the non-thermal production of ${\rm ^6Li}$
may become significant, we include $P_{{\rm ^6Li}\rightarrow {\rm
^6Li}}$ in the calculation of the number of non-thermally produced
${\rm ^6Li}$.  Same is true for $^7{\rm Li}$ and $^7{\rm Be}$ when
non-thermal production of these light elements are discussed.

\section{Primordial Abundance of Gravitino}
\label{app:Ygrav}
\setcounter{equation}{0}

In this appendix, we evaluate the primordial abundance of the
gravitino in the inflationary universe.  In particular, we calculate
the gravitino abundance as a function of the well-defined reheating
temperature $T_{\rm R}$.  Compared to the old result given in Ref.\
\cite{KawMor}, there are two major improvements: (i) we have
included the effects of the gravitino production during the period
when the inflaton field is still oscillating (i.e, in the so-called
dilute plasma), and (ii) for the gravitino production cross section in
the thermal bath, formula taking account of the thermal mass of the
gauge bosons is used in order to avoid the infrared singularity.

Gravitino is the gauge field for the local supersymmetry, and hence it 
couples to the supercurrent as
\begin{eqnarray}
    {\cal L}_{\psi J} &=&
    -\frac{1}{\sqrt{2}M_*} D_{\nu} \phi^{*}
    \bar{\psi}_\mu \gamma^\nu \gamma^\mu \chi_R
    -\frac{1}{\sqrt{2}M_*} D_{\nu} \phi
    \bar{\chi}_L \gamma^\mu \gamma^\nu \psi_\mu
    \nonumber \\ &&
    -\frac{i}{8M_*} \bar{\psi}_\mu
    \left[ \gamma^\nu , \gamma^\rho \right] \gamma^\mu
    \lambda F_{\nu\rho},
    \label{L_psiJ}
\end{eqnarray}
where $\chi_R$ and $\phi$ are fermion and boson in chiral multiplets,
$\lambda$ is the gaugino, and $F_{\nu\rho}$ is the field strength of
the gauge field.  (Here, $D_\nu$ denotes the covariant derivative and
$\chi_R$ satisfies $(1-\gamma_5)\chi_R=0$.)

Importantly, the interaction of the gravitino is suppressed by the
inverse of $M_*$ and hence the gravitino interacts very weakly
compared to the standard-model particles.  In particular, if the
temperature is lower than $\sim M_*$, gravitino is not thermalized in
the expanding universe.  Although it is out of the thermal bath,
however, gravitinos are produced via the scattering processes of the
thermal particles.  In the expanding universe, evolution of the number
density of the gravitino is governed by the following Boltzmann
equation
\begin{eqnarray}
    \frac{dn_{3/2}}{dt} + 3 H n_{3/2} = 
    \langle \sigma_{\rm tot} v_{\rm rel} \rangle n_{\rm rad}^2,
    \label{dot(ngra)}
\end{eqnarray}
where $\langle\sigma_{\rm tot}v_{\rm rel}\rangle$ is the ``thermally
averaged'' total cross section (times relative velocity), and $n_{\rm
rad}=\frac{\zeta (3)}{\pi^2}T^3$.  With the Lagrangian (\ref{L_psiJ}),
the gravitino production cross sections are typically ${\cal
O}~(g_i^2/32\pi M_*^2)$, where $g_i$ denotes the gauge coupling
constant for $SU(3)_C$, $SU(2)_L$ and $U(1)_Y$ for $i=3$, $2$ and $1$,
respectively.

Approximated formula for the primordial gravitino abundance is
obtained by integrating Eq.\ (\ref{dot(ngra)}) from the highest
temperature in the radiation-dominated epoch, which approximately
corresponds to the ``reheating temperature'' $T_{\rm R}$ after the
inflation.  Using the fact that the gravitino production cross section
has very weak dependence on $T$, Eq.\ (\ref{dot(ngra)}) can be easily
solved.  In order to parameterize the primordial abundance of the
gravitino, it is convenient to define the ``yield variable''
as~\footnote{
The yield variable used in this paper differs from that in
\cite{KawMor} which defined $Y_{3/2}^{\rm (KM)}\equiv n_{3/2}/n_{\rm
rad}$.  We use $Y_{3/2}$ defined in Eq.\ (\ref{yield}) since, with
this definition, $Y_{3/2}$ is independent of time for the temperature
$T\ll T_{\rm R}$ as far as there is no entropy production.  Notice
that, for $T\ll m_e$, $Y_{3/2}^{\rm (KM)}\simeq 14Y_{3/2}$.}
\begin{eqnarray}
    Y_{3/2} \equiv \frac{n_{3/2}}{s},
    \label{yield}
\end{eqnarray}
where $s$ is the entropy density which is given by
\begin{eqnarray}
    s = \frac{2\pi^2}{45} g_{*S} T^3,
\end{eqnarray}
where $g_{*S}$ is the effective number of the massless degrees of
freedom.  For the particle content of the minimal supersymmetric
standard model (MSSM), $g_{*S}=228.75$ for temperature much higher
than the masses of the superparticles while $g_{*S}=43/11$ for $T\ll
m_e$.  As we see just below, gravitino production processes are
effective when $T\sim T_{\rm R}$ and hence, when $T\ll T_{\rm R}$,
$Y_{3/2}$ becomes constant (as far as the total entropy in the
comoving volume is conserved).  Then, integrating Eq.\ 
(\ref{dot(ngra)}) from $T=T_{\rm R}$ to $T\ll T_{\rm R}$, we obtain
\begin{eqnarray}
    Y_{3/2} \sim
    \frac{45 \zeta (3)}{2 \pi^4}
    \left[ 
        \frac{\langle\sigma_{\rm tot}v_{\rm rel}\rangle n_{\rm rad}}
        {g_{*S} H}
    \right]_{T=T_{\rm R}}.
\end{eqnarray}
The important point is that the primordial abundance of the gravitino
is (approximately) proportional to the reheating temperature after the
inflation.  Thus, gravitinos are overproduced in many cases if the
reheating temperature is too high.

For the detailed calculation of $Y_{3/2}$, however, the definition of
the ``reheating temperature'' used in the above calculation is quite
ambiguous since, in realistic models of slow-roll inflation, there
exists a period when the universe is dominated by the oscillating
energy of the inflaton field before the radiation-dominated epoch.
(We call this period as the inflaton-dominated period.)  Due to the
decay of the inflaton field, the inflaton-dominated epoch evolves into
the radiation-dominated epoch.  The transition from the
inflaton-dominated epoch to the radiation-dominated epoch occurs when
the expansion rate of the universe is comparable to the decay rate of
the inflaton.  In this paper, we define the ``reheating temperature'' as
\begin{eqnarray}
    T_{\rm R} \equiv 
    \left( 
        \frac{10}{g_* \pi^2} M_*^2 \Gamma_{\rm inf}^2 
    \right)^{1/4},
\end{eqnarray}
where $g_*=g_{*S}$.  (In estimating $T_{\rm R}$, we use
$g_*=228.75$.)  Notice that $T_{\rm R}$ given above is the same as the
reheating temperature derived with the approximation such that $X$
instantaneously decays when the relation $3H=\Gamma_{\rm inf}$ is
realized.

Then, in order to take account of the effects of the gravitino
production in the inflaton-dominated epoch, we numerically solve the
Boltzmann equations.  The Boltzmann equation for the gravitino abundance
is given in Eq.\ (\ref{dot(ngra)}), while evolutions of the energy
densities of the radiation $\rho_{\rm rad}$ and inflaton $\rho_{\rm
inf}$ are governed by the following Boltzmann equations
\begin{eqnarray}
    \frac{d \rho_{\rm rad}}{dt} + 4 H \rho_{\rm rad} &=& 
    \Gamma_{\rm inf} \rho_{\rm inf},
    \label{dot(rhorad)}
    \\
    \frac{d \rho_{\rm inf}}{dt} + 3 H \rho_{\rm inf} &=& 
    - \Gamma_{\rm inf} \rho_{\rm inf},
    \label{dot(rhoinf)}
\end{eqnarray}
where $\Gamma_{\rm inf}$ is the decay rate of the inflaton, and
$\rho_{\rm rad}$ is related to the background temperature as
\begin{eqnarray}
    \rho_{\rm rad} = \frac{\pi^2}{30} g_{*} T^4.
\end{eqnarray}

The thermally averaged cross section for the gravitino production is
calculated in Ref.~\cite{Bolz:2000fu}.~\footnote{In the gravitino
production cross section, infrared singularities arise due to the $t$-
and $u$-channel exchanges of the gauge bosons.  In the previous
calculation of $\langle\sigma_{\rm tot}v_{\rm rel}\rangle$ given in
\cite{KawMor}, such singularities were smeared by introducing a
cut-off parameter for the scattering angle while, in
\cite{Bolz:2000fu}, it is treated by properly taking account of the
thermal mass of the gauge bosons.  Numerically, difference of the
thermally averaged cross sections given in Refs.~\cite{KawMor} and
\cite{Bolz:2000fu} is at $10\ \%$ level for most of the parameter
space.}
For $SU(N)$ super Yang-Mills model with $n_{\rm f}$ pairs of
fundamental and anti-fundamental chiral superfields, we
obtain~\footnote{
For $U(1)$ gauge interactions, $(N^2-1)\rightarrow 1$, $(N+n_{\rm
f})\rightarrow n_{\rm f}$, and $n_{\rm f}$ becomes the sum of the
squared of the hypercharges of chiral multiplets.  We thank Arnd
Brandenburg for helpful correspondence on this point.}
\begin{eqnarray}
    \langle \sigma_{\rm tot} v_{\rm rel} \rangle &=& 
    \left[ 1 + \left( \frac{m_{\tilde{g}}^2}{3m_{3/2}^2} \right) 
    \right]
    \frac{3 g^2 (N^2-1)}{32\pi M_*^2}
    \nonumber \\ &&
    \times \frac{\pi^2}{\zeta(3)}
    \left\{
        \left[ \ln (T^2/m_{g,{\rm th}}^2) + 0.3224 \right]
        ( N + n_{\rm f} )
        + 0.5781 n_{\rm f}
    \right\},
\end{eqnarray}
where $m_{\tilde{g}}$ is the gaugino mass and $m_{g,{\rm th}}$ is the
thermal mass of the gauge boson which is given as
\begin{eqnarray}
    m_{g,{\rm th}}^2 = \frac{1}{6} g^2 
    ( N + n_{\rm f} ) T^2.
\end{eqnarray}

As the gaugino mass becomes larger, more gravitinos are produced.  In
our calculation, we calculate the gravitino abundance in the limit of
$m_{\tilde{g}}\rightarrow 0$ to derive a conservative constraint.
Thus, in our calculation, the gravitino abundance is underestimated by
${\cal O} (10)\ \%$ when the gaugino masses are comparable to the
gravitino mass.  If the gauginos are much heavier than the gravitino,
upper bound on $T_{\rm R}$ is approximately obtained by rescaling the
results by the factor $(m_{\tilde{g}_3}^2/3m_{3/2}^2)^{-1}$ using the
fact that the gravitino production is dominated by processes related
to the $SU(3)_C$ gauge fields.

We follow the evolutions of $n_{3/2}$, $\rho_{\rm rad}$, and
$\rho_{\rm inf}$ from the period with $H\gg\Gamma_{\rm inf}$ to
$H\ll\Gamma_{\rm inf}$ by numerically solving the Boltzmann equations
(\ref{dot(ngra)}), (\ref{dot(rhorad)}) and (\ref{dot(rhoinf)}).  In
calculating $\langle\sigma_{\rm tot}v_{\rm rel}\rangle$, we sum over
the contributions from all the MSSM gauge groups, $SU(3)_C$, $SU(2)_L$
and $U(1)_Y$.  Then, we calculate $Y_{3/2}$ at $H\ll\Gamma_{\rm inf}$
(i.e., $T\ll T_{\rm R}$).  As we mentioned, $Y_{3/2}$ is approximately
proportional to $T_{\rm R}$; more precisely, we found that the
resultant gravitino abundance is well approximated by the following
formula
\begin{eqnarray}
    Y_{3/2} &\simeq& 
    1.9 \times 10^{-12}
    \nonumber \\ &&
    \times \left( \frac{T_{\rm R}}{10^{10}\ {\rm GeV}} \right)
    \left[ 1 
        + 0.045 \ln \left( \frac{T_{\rm R}}{10^{10}\ {\rm GeV}} 
        \right) \right]
    \left[ 1 
        - 0.028 \ln \left( \frac{T_{\rm R}}{10^{10}\ {\rm GeV}} 
        \right) \right].
    \nonumber \\
\end{eqnarray}


\end{document}